\documentclass[unsortedaddress,superscriptaddress,pra,notitlepage]{revtex4-1}

\usepackage{amsmath,amssymb}
\usepackage{braket}
\usepackage{theorem}
\usepackage{color}
\usepackage{amsfonts}
\usepackage[mathscr]{eucal}
\usepackage[hidelinks]{hyperref}
\usepackage{graphicx}
\usepackage{xcolor}
\usepackage{tikz}
\usepackage{mathtools}

\usetikzlibrary{calc,decorations.pathreplacing}
\usetikzlibrary{arrows,shapes}

\newtheorem{definition}{Definition}[section]
\newtheorem{theorem}[definition]{Theorem}
\newtheorem{prop}[definition]{Proposition}
\newtheorem{lemma}[definition]{Lemma}
\newtheorem{remark}[definition]{Remark}
\newtheorem{rem}[definition]{Remark}
\newtheorem{corollary}[definition]{Corollary}
\newtheorem{cor}[definition]{Corollary}

\newenvironment{proof}[1][Proof]{\begin{trivlist}
\item[\hskip \labelsep {\bfseries #1}]}{\hfill$\Box$\end{trivlist}}

\def\A{S}
\def\B{{\mathcal B}}

\def\F{{\mathcal F}}

\def\I{{\mathcal I}}
\def\K{{\mathcal K}}

\def\M{\mathcal{M}}
\def\N{R}

\def\P{{\mathcal P}}
\def\Q{{\mathcal Q}}
\def\S{{\mathcal S}}

\def\X{{\mathcal X}}

\def\bN{\mathbb{N}}
\def\bC{\mathbb{C}}
\def\bR{\mathbb{R}}
\def\bZ{\mathbb{Z}}

\def\theta{\vartheta}
\def\ep{\varepsilon}
\def\rho{\varrho}
\def\hil{{\mathcal H}}

\def\half{\frac{1}{2}}
\def\iff{\Longleftrightarrow}
\def\imp{\Longrightarrow}

\def\bz{\left(}
\def\jz{\right)}

\def\inv{^{-1}}

\def\egy{\mathbf 1}

\def\av{\mathrm{av}}
\def\adv{\mathrm{adv}}

\def\oll{\overline}

\def\nw{^{*}}

\def\bog{^{\flat}}

\def\sc{\mathrm{sc}}

\def\sli{\mathrm{s}}

\def\cli{\mathrm{c}}

\def\dli{\mathrm{d}}

\def\e{\mathrm{e}}
\def\p{_{\ge 0}}
\def\pne{_{\gneq 0}}
\def\pp{_{>0}}
\def\sa{\mathrm{sa}}
\def\nul{0}
\def\gen{^{t}}
\def\valt{\cdot}
\def\rmax{r_{\infty}}
\def\dif{\delta}

\newcommand{\ki}[1]{\textit{\textit{#1}}}

\newcommand{\s}{\mbox{ }}
\newcommand{\ds}{\mbox{ }\mbox{ }}
\newcommand{\norm}[1]{\left\| #1\right\|}

\newcommand{\abs}[1]{\left| #1 \right|}

\newcommand{\vecc}[1]{\underline{#1}}

\newcommand{\diad}[2]{\left|#1\right\rangle\!\left\langle #2\right|}

\newcommand{\pr}[1]{\diad{#1}{#1}}

\newcommand{\floor}[1]{\lfloor #1\rfloor}
\newcommand{\ceil}[1]{\left\lceil #1\right\rceil}

\newcommand{\wtilde}[1]{\widetilde{#1}}

\makeatletter
\renewcommand{\p@enumii}{}
\makeatother

\DeclareMathOperator{\Tr}{Tr}

\DeclareMathOperator{\Exp}{\mathbb{E}}

\DeclareMathOperator{\supp}{supp}
\DeclareMathOperator{\spec}{spec}

\DeclareMathOperator{\ran}{ran}

\DeclareMathOperator{\spann}{span}

\DeclareMathOperator{\ttimes}{\times\ldots\times}

\DeclareMathOperator{\logn}{\widehat\log}

\DeclareMathOperator{\divv}{\Delta}

\DeclareMathOperator{\D}{\mathit{D}}

\DeclareMathOperator{\co}{co}
\DeclareMathOperator{\cco}{\overline{co}}
\DeclareMathOperator{\diff}{\Lambda}
\DeclareMathOperator{\ac}{\mathbb{A}}
\DeclareMathOperator{\relint}{relint}

\newcounter{szamlalo}

\usepackage{subfiles} 

\begin{document}

\title{On the error exponents of binary state discrimination with composite hypotheses}

\author{Mil\'an Mosonyi, Zsombor Szil\'agyi, Mih\'aly Weiner}

\affiliation{
MTA-BME Lend\"ulet Quantum Information Theory Research Group
}

\affiliation{
Mathematical Institute, Budapest University of Technology and Economics, \\
Egry J\'ozsef u~1., Budapest, 1111 Hungary.
}

\begin{abstract}
\centerline{\textbf{Abstract}}
\vspace{.3cm}

The trade-off between the two types of errors in binary state discrimination
may be quantified in the asymptotics by various error exponents.
In the case of simple i.i.d.~hypotheses, each of these exponents is equal to a divergence
(pseudo-distance) of the two states.
In the case of composite hypotheses, represented by sets of states $\N,\A$, 
one always has the inequality $\e(\N\|\A)\le \mathrm{E}(\N\|\A)$, 
where $\e$ is the exponent, $\mathrm{E}$ is the corresponding divergence, and the question is whether equality holds. 
This has been shown to be the case for various general settings in classical state discrimination, and for the Stein exponent with simple alternative hypothesis for 
finite-dimensional quantum state discrimination. On the other hand, examples with strict inequality have been shown for the Stein and the Chernoff exponents 
for discriminating a single finite-dimensional quantum state 
from a set of states with continuum cardinality representing the alternative hypothesis.
These results suggest that the relation between the composite exponents and the worst pairwise exponents may be influenced by a number of factors: 
the type of exponents considered; whether the problem is classical or quantum;
the cardinality and the geometric properties of the sets representing the hypotheses; and, on top of the above, possibly whether the underlying Hilbert space is finite- or 
infinite-dimensional.

Our main contribution in this paper is clarifying this landscape considerably: 
We exhibit explicit examples for hitherto unstudied cases where the above inequality fails to hold with equality, while we also prove equality for various general classes of state discrimination problems. In particular, we show that equality may fail for any of the error exponents even in the classical case, if the system is allowed to be infinite-dimensional, and the alternative hypothesis contains countably infinitely many states. 
Moreover, we show that in the quantum case strict inequality is the generic behavior in the sense that, starting from any pair of non-commuting density operators of any dimension,
and for any of the exponents, it is possible to construct an example 
with a simple null-hypothesis and an alternative hypothesis consisting of only two states,
such that strict inequality holds for the given exponent.

%
%

\end{abstract}

\maketitle

\section{Introduction}

State discrimination is one of the fundamental problems in statistics, with applications in 
many information-theoretic problems \cite{ON07,Hayashicq,Hayashibook2}. In a typical binary i.i.d.~quantum state discrimination 
problem, an experimenter is presented with several identically prepared quantum systems, 
all in the same state that either belongs to a set 
$\N\subseteq\S(\hil)$ (\ki{null-hypothesis} $H_0$), or to another set 
$\A\subseteq\S(\hil)$ (\ki{alternative hypothesis} $H_1$), where
$\S(\hil)$ is the set of all density operators on the system's Hilbert space $\hil$. The experimenter's task is to guess which hypothesis is correct, based on the result of a $2$-outcome measurement, represented by a pair of operators
$(T_n(0)=:T_n,T_n(1)=I-T_n)$, where $T_n$ is an operator on $\hil^{\otimes n}$ 
such that $0\le T_n\le I$ (also called a \ki{test}), and $n$ is the number of identically prepared systems.
If the outcome of the measurement is $k$, described by the measurement operator $T_n(k)$, the experimenter decides that hypothesis $k$ is true. 

The experimenter makes an erroneous decision by rejecting the null-hypothesis when it is true (type I error), or by accepting it when it is false (type II error). The worst-case probabilities of these events are given by 
\begin{align}\label{wc errors}
\alpha_n(\N|T_n)&:=\sup_{\rho\in\N}\Tr\rho^{\otimes n}(I-T_n),& &\text{(type I)},\\
\beta_n(\A|T_n)&:=\sup_{\sigma\in\A}\Tr\sigma^{\otimes n}T_n,& &\text{(type II)}.
\end{align}
Clearly, there is a trade-off between the two types of error probabilities, which can be
quantified in the asymptotics $n\to+\infty$ by various error exponents, depending on the way
the two types of errors are optimized with respect to each other. The most often studied ones are the \ki{Chernoff exponent} $\cli(\N\|\A)$ of symmetric state discrimination, the 
\ki{Stein exponent} $\sli(\N\|\A)$ of asymmetric
state discrimination, and the one-parameter family of \ki{direct exponents} 
$\dli_r(\N\|\A)$, $0<r<\sli(\N\|\A)$,
describing the whole trade-off curve when both errors vanish asymptotically; we will give precise definitions of these notions in Section \ref{sec:state disc}. 

Let us denote by $\e(\N\|\A)$ any of these exponents. 
It is easy to see that the exponents for the state discrimination problem  $\N$ vs.~$\A$ cannot be better than the worst pairwise exponents for the state discrimination problems 
$\rho$ vs.~$\sigma$, $\rho\in\N$, $\sigma\in\A$,
i.e, 
\begin{align}\label{exponent ineq1}
\e(\N\|\A)\le\inf_{\rho\in\N,\sigma\in\A}e(\rho\|\sigma),
\end{align} 
where we write $\e(\rho\|\sigma)$ instead of $\e(\{\rho\}\|\{\sigma\})$.
Moreover, the error exponents for simple hypotheses ($R=\{\rho\}$ vs.~$S=\{\sigma\}$) can be explicitly expressed as certain divergences of the two states \cite{HP,ON,NSz,AudChernoff,Hayashicq,Nagaoka}, and therefore if 
\eqref{exponent ineq1} holds as an equality, these results provide expressions for 
the exponents of the composite hypothesis testing problem in terms of divergence distances of the two sets $\N$ and $\A$.
Therefore, it is of fundamental importance to know when the inequality in \eqref{exponent ineq1} holds as an equality.

Such equalities have been established, or follow easily from known results, in the following cases (the Hilbert space dimension is assumed to be finite):
\begin{itemize}
\item
for the Stein and the Chernoff exponents when all states commute (classical case), and both $\N$ and $\A$ are compact convex sets \cite{AdvHyp};
\item
for the Stein exponent when $|\A|=1$ (Stein's lemma with composite null-hypothesis)
\cite{BertaBrandaoHirche2017,Sanov2005,M13,Notzel};
\item
for the Chernoff exponent when $|\N|,|\A|<+\infty$, and all states commute, or 
$\N$ consists of a single pure state \cite{AM14}.
\end{itemize}
In fact, the first result was established in \cite{AdvHyp} in a strictly stronger form, 
for the non-i.i.d.~problem of classical adversarial hypothesis testing, and the equality 
for the Stein exponent was established in \cite{Notzel} also for the 
non-i.i.d.~problem of quantum arbitrarily varying state discrimination.
We note that the statement in \cite{Sanov2005} is somewhat weaker than how we formulated 
the problem above, as in \cite{Sanov2005} the type I error is not shown to converge to zero uniformly over all candidate states in the null hypotheses, but only individually. This is a subtle 
but important difference, which will also appear at various places in our analysis of the error exponents; we will refer to the corresponding error exponents as \ki{relaxed exponents}.
On the other hand, it was shown recently in \cite{BertaBrandaoHirche2017} that 
\eqref{exponent ineq1} may hold as a strict inequality for the Stein exponent, 
in a finite-dimensional setting, 
with a simple null-hypothesis and a composite alternative hypothesis of continuum cardinality.

Instead of the worst-case error probabilities \eqref{wc errors}, it is also very natural to 
work with some weighted averages of the error probabilities, defined as
\begin{align}\label{averaged errors}
\alpha_n(p|T_n)&:=\int_{\N}dp(\rho)\,\Tr\rho^{\otimes n}(I-T_n),& &\text{(type I)},\\
\beta_n(q|T_n)&:=\int_{\A}dq(\sigma)\,\Tr\sigma^{\otimes n}T_n,& &\text{(type II)},
\end{align}
where $p,q$ are probability measures on the Borel sets of the state space, 
reflecting some prior knowledge about the likeliness of the candidate states, or
about the relative severity of misidentifying them. 
We call this the \ki{mixed i.i.d.~setting.}
It is easy to see that 
\begin{align}\label{exponent ineq2}
\e(\supp p\|\supp q)\le \e(p\|q)\le \inf_{\rho\in\supp p,\sigma\in\supp q}e(\rho\|\sigma),
\end{align}
and hence we get a refinement of the equality problem in \eqref{exponent ineq1}:
the two inequalities in \eqref{exponent ineq2} can be analyzed separately, and if both of them hold
with equality then so does the inequality in \eqref{exponent ineq1}
(with $\N=\supp p$ and $\A=\supp q$).
It is easy to see that the first equality in \eqref{exponent ineq2} holds as an equality 
whenever $\supp p$ and $\supp q$ are finite, and hence the analysis of this inequality becomes relevant when at least one of the probability distributions is supported on a set of infinite cardinality. The state discrimination problem with the weighted error probabilities
may also be interpreted as a binary state discrimination problem with simple, 
but non-i.i.d.~hypothesis, given as 
$H_0:\,(\int dp(\rho)\rho^{\otimes n})_{n\in\bN}$ vs.~$H_1:\,(\int dq(\sigma)\sigma^{\otimes n})_{n\in\bN}$, a special case of which is i.i.d.~state discrimination with 
group covariant measurements \cite{HMH09}.
In this setting, an explicit example 
demonstrating strict inequality for the 
Chernoff exponent and for the direct exponents was given in 
\cite[Example 6.2]{HMH09}, with
with $q$ being a Dirac measure (simple i.i.d.~alternative hypothesis) and 
$p$ supported on a set of continuum cardinality.

The above results suggest that the relation between the composite exponents and the worst pairwise exponents may be influenced by a number of factors: 
the type of exponents considered; whether the problem is classical or quantum;
the cardinality and the geometric properties of the sets representing the hypotheses; and, on top of the above, possibly whether the underlying Hilbert space is finite- or 
infinite-dimensional.
Our main contribution in this paper is clarifying this landscape considerably: 
We exhibit explicit examples for hitherto unstudied cases where the above inequalities fail to hold with equality, while we also prove equality for various general classes of state discrimination problems. 

The structure of the paper is as follows. In Section \ref{sec:prelim}
we give the necessary preliminaries and prove some simple general results about the
error exponents and the related divergences between sets of states, which will be used throughout the paper. 

Section \ref{sec:classical} is devoted to classical hypothesis testing. In Section 
\ref{sec:finite classical} we show that equality holds in \eqref{exponent ineq1} 
for all the error exponents (Stein, Chernoff, direct exponents)
in the most general classical setting (represented by commutative von Neumann algebras) whenever both sets
$\N$ and $\A$ are finite. In fact, we show this more generally for countable sets
and the ``relaxed'' exponents, where the errors are not required to follow the given asymptotics uniformly over all states in the two hypotheses, but only individually. 
On the other hand, in Section \ref{sec:infinite classical} we show 
that this is no longer true for any of the ``strong'' exponents (defined from the worst-case errors \eqref{wc errors}), at least if we allow the system to be infinite-dimensional; we construct an explicit example with probability distributions on 
the $[0,1]$ interval, where the null-hypothesis is simple (the uniform distribution), the alternative hypothesis consists of countably infinitely many states, and both inequalities
in \eqref{exponent ineq2} are strict for all the error exponents.
In Section \ref{sec:classical exponents} we study the finite-dimensional case, and show that 
equality holds in 
\eqref{exponent ineq1} for all the error exponents provided that both 
$\N$ and $\A$ are compact convex sets. 
In fact, we prove the stronger statement that all the error exponents 
for arbitrary $\N$ and $\A$
coincide with the respective worst pairwise error exponents over the closed convex hulls
of $\N$ and $\A$ in the
arbitrarily varying and in the adversarial settings. This extends previous results in
\cite{AdvHyp}, where the cases of the Stein and the Chernoff exponents were proved in the adversarial setting for closed convex sets, and in 
\cite{FuShen1996,FuShen1998}, where the cases of the Stein and the direct exponents 
were proved for finite sets.

In Section \ref{sec:finiteQ} we show that the quantum case is fundamentally different from what we have seen in the classical case: for any of the error exponents, 
we construct explicit examples demonstrating strict inequality in 
\eqref{exponent ineq1} 
in the simplest possible case 
with a simple null hypothesis and an alternative hypothesis consisting of two states.
In fact, we show that this phenomenon is fairly generic, in the sense that 
we can construct such examples from any two non-commuting states of arbitrary dimension. 
In particular, our examples refute some previous conjectures stated in 
\cite{HMH09} and \cite{AM14}
about the equality in \eqref{exponent ineq1} for finitely many null-hypotheses
and a simple alternative hypothesis.
In view of these, it becomes particularly relevant, and clearly a non-trivial problem, to find sufficiently general cases where 
\eqref{exponent ineq1} holds as an equality.
In Section \ref{sec:q equality}, we show 
that when both hypotheses contain finitely many states, 
equality holds for all the error exponents 
in two important general cases:
in the semi-classical case, where  $\rho\sigma=\sigma\rho$
for all $\rho\in\N$ and $\sigma\in\A$, 
and in the pure state case, where both hypotheses contain only pure states.

In Section \ref{sec:sc} we consider the one-parameter family of strong converse exponents
$\sc_r(\N\|\A)$, $r>\sli(\N\|\A)$. In this case the trivial relation is opposite to 
\eqref{exponent ineq1},
\begin{align}\label{sc triv ineq}
\sc_r(\N\|\A)\ge \sup_{\rho\in\N,\sigma\in\A}\sc_r(\rho\|\sigma),
\end{align}
and the question is again when equality holds. It was shown recently
in \cite{BunthVrana2020} that equality holds
when the alternative hypothesis is simple and the null-hypothesis 
consists of finitely many states. 
As we show in Section \ref{sec:sc countable null}, this can be easily extended to 
the relaxed version of the strong converse exponent
for a null-hypothesis of arbitrary cardinality, provided that it contains a countable set that is dense with respect to the max-relative entropy pseudo-distance;
this is satisfied, for instance, in the classical case, or if the states all have the same support. The opposite case, with a simple null-hypothesis and a composite alternative hypothesis, turns out to behave very differently.
Indeed, in Section \ref{sec:sc two alt} 
we give an explicit $2$-dimensional classical example with a simple null-hypothesis and two alternative hypotheses where \eqref{sc triv ineq}
holds with strict equality. This is in sharp contrast also with the behavior of the 
other error exponents discussed above, where equality holds in the classical case for finitely many hypotheses. 
It turns out, however, that equality still holds in the finite-dimensional classical case when the alternative hypothesis is given by a closed convex set. 
More generally, we show in Section \ref{sec:sc adv} that 
in the finite-dimensional classical case,
the strong converse exponents for arbitrary  $\N$ and $\A$
coincide with the respective worst pairwise strong converse exponents over the closed convex 
hulls of $\N$ and $\A$ in the
arbitrarily varying and in the adversarial setting. 
When combined with the result of Section \ref{sec:sc countable null}, 
this yields equality in \eqref{sc triv ineq} for the relaxed strong converse exponent
for an arbitrary null-hypothesis and a closed convex alternative hypothesis. 
Finally, in Section \ref{sec:ball test} we show equality in \eqref{sc triv ineq}
for an arbitrary null-hypothesis and a simple alternative hypothesis under some technical conditions and a restricted, but non-trivial range of parameters.

\section{Background and preliminary results}
\label{sec:prelim}

\subsection{Mathematical background}
\label{sec:Neumann algebras}

We will consider the binary state discrimination problem in a slightly more general setting 
than in the Introduction, on the one hand allowing the quantum systems to be modeled by 
general von Neumann algebras (to incorporate infinite-dimensional classical systems in the formalism), and on the other hand considering a more general notion of hypotheses. Therefore, below we collect some basic notions of such models of quantum systems. 
Note, however, that we only use this formalism for convenience of presentation, and we will only actually work with the quantum and classical models familiar to any quantum information theorists. Therefore, no familiarity with the theory of von Neumann algebras is required to follow the paper, apart from the basic notions and terminology that we collect in this section.

For a Hilbert space $\hil$, let $\B(\hil)$ denote the set of all linear operators on $\hil$. We say that $\M\subseteq\B(\hil)$ is a \ki{von Neumann algebra} on $\hil$ if it is a linear subspace of $\B(\hil)$ that is closed under the operator product, the adjoint, contains the identity operator, and is closed in the weak operator topology. (The last condition is automatically satisfied when the underlying Hilbert space is finite-dimensional.)

In this paper we will only consider the two simplest cases of von Neumann algebras: The simple quantum case where $\M=\B(\hil)$, and the classical case, where 
$\M=\{M_f:\,f\in L^{\infty}(\X,\F,\mu)\}$, with $(\X,\F,\mu)$ a measure space, and $M_f$ being the multiplication operator $L^{2}(\X,\F,\mu)\ni g\mapsto fg$. In the classical case, 
we may naturally identify $\M$ with the function algebra $L^{\infty}(\X,\F,\mu)$, in which the algebraic operations are the usual point-wise operations on functions. In particular, 
when $\X$ is finite, we will choose $\mu$ to be the counting measure $\mu(\{x\})=1$, $x\in\X$, on the full power set of $\X$, in which case the function algebra $L^{\infty}(\X,\F,\mu)$
is simply $\bC^{\X}=\{f:\,\X\to\bC\}$, and the corresponding operator algebra is the collection of all operators on $\bC^{\X}$ that are diagonal in the canonical orthonormal basis 
$(\ket{x})_{x\in\X}$ of $l^2(\X)$, i.e., can be written as $\sum_{x\in\X}a(x)\pr{x}$ with some $a\in\bC^{\X}$. The only infinite-dimensional von Neumann algebra that we will actually use in 
this paper will be the classical algebra $L^{\infty}([0,1])$, where the measure is the Lebesgue measure on the Lebesgue measurable subsets of $[0,1]$.

For a von Neumann algebra $\M\subseteq\B(\hil)$, we will use the notations 
$\M_{\sa}$, $\M\p$, $\M\pne$, and 
$\M\pp$ for the set of 
self-adjoint, positive semi-definite (PSD), non-zero positive semi-definite, and positive definite operators in $\M$, respectively, and 
\begin{align*}
\M_{[0,I]}:=\{T\in\M:\,0\le T\le I\}
\end{align*}
for the set of \ki{tests} in $\M$.
A test $T$ is \ki{projective} if $T^2=T$. In the classical case $\M=L^{\infty}(\X,\F,\mu)$,
projective tests may be identified with measurable subsets of $\X$, up to the equivalence 
$A\sim B$ if $\mu((A\setminus B)\cup(B\setminus A))=0$; the test corresponding to a subset
is the multiplication operator by the characteristic function of the given subset.

We denote by $\M_*$ the predual of $\M$, i.e., the space of normal linear functionals on $\M$. The \ki{$w^*$-topology} on $\M$ is the weak topology induced by 
$\M_*$ on $\M$.
The following is an immediate consequence of the Banach-Alaoglu theorem:
\begin{lemma}\label{lemma:tests compact}
$\M_{[0,I]}$ is $w^*$-compact.
\end{lemma}

We denote by $\M_{*}^{+}$ the set of normal positive functionals on $\M$.
A \ki{state} on $\M$ is a $\rho\in\M_*^+$ that is normalized, i.e., $\rho(I)=1$. 
We denote the set of states on $\M$ by $\S(\M)$.
In the simple quantum case, every $\rho\in\B(\hil)_*^+$ is uniquely determined by a positive trace-class operator 
$\hat\rho$ on $\hil$ via $\rho(.)=\Tr\hat\rho(.)$, 
and $\rho$ is a state if and only if $\hat\rho$ is a density operator, i.e., $\Tr\hat\rho=1$.
In this case we are going to identify normal positive functionals with their density operators, and denote the set of density operators (states) on $\hil$ by $\S(\hil)$. 

In the classical case, every non-negative function 
$\hat\rho\in L^1(\X,\F,\mu)$ defines a positive normal functional via $\rho(f):=\int_{\X}f(x)\hat\rho(x)\,d\mu(x)$, and it is a state if and only if $\hat\rho$ is a probability density function, i.e., $\int_{\X}\hat\rho(x)\,d\mu(x)=1$.
When no confusion arises, we will use the same notation $\rho$ and terminology (state) for both the function and the corresponding functional on the algebra.
For a measurable subset $A\in\F$, we will also use the notation
\begin{align*}
\rho(A):=\rho(\egy_A)=\int_A\hat\rho(x)\,d\mu(x).
\end{align*}
In particular, in a finite classical model, states of the system are simply probability density functions 
on the finite set $\X$, or equivalently, in the operator formalism, density operators on the finite-dimensional Hilbert space $l^2(\X)$ that are diagonal in the canonical basis. 
{In this case, we will denote the set of states by $\S(\X)$.}

For topological notions (e.g., compactness of subsets of $\M_*$, (semi-)continuity of functions on $\M_*$) we will always consider the weak topology induced by $\M$ on $\M_*$. In the finite-dimensional case this is just the usual topology induced by any norm on $\M_*$. 

For any $\rho\in\M_*^+$, we will denote its support projection by $\rho^0$.

\subsection{Quantum divergences for state discrimination}

The error exponents of binary i.i.d.~state discrimination may be expressed as 
certain divergences (distance-like quantities) of the two states representing the two hypotheses. The properties of these divergences are well-studied in the literature for pairs 
of states; the main purpose of this section is to extend some of this analysis to pairs of subsets of states, which we will need in our study of the error exponents of composite hypothesis testing.

For a density operator $\rho$ and a PSD operator $\sigma$ on a finite-dimensional Hilbert space 
$\hil$, their 
(Petz-type) \ki{R\'enyi $\alpha$-divergence} $D_{\alpha}(\rho\|\sigma)$ is defined as
\begin{align*}
D_{\alpha}(\rho\|\sigma):=\frac{1}{\alpha-1}\log\Tr\rho^{\alpha}\sigma^{1-\alpha}
\end{align*}
for any $\alpha\in[0,1)$. The limit $\alpha\to 1$ gives the 
\ki{Umegaki relative entropy} 
\begin{align*}
D_1(\rho\|\sigma):=\lim_{\alpha\nearrow 1}D_{\alpha}(\rho\|\sigma)=
D(\rho\|\sigma):=\begin{cases}
\Tr\rho(\logn\rho-\logn\sigma);&\rho^0\le\sigma^0,\\
+\infty,&\text{otherwise},
\end{cases}
\end{align*}
where $\log$ stands for the natural logarithm, and $\logn$ is its extension by 
$\logn 0:=0$. 

The above formulas already define the R\'enyi divergences and the relative entropy for 
a state and a positive functional on a finite-dimensional classical system, represented as diagonal operators in the same ONB. 
In the general commutative case $\M=L^{\infty}(\X,\F,\mu)$,
and non-negative functions
$\rho,\sigma\in L^1(\X,\F,\mu)$ with $\int_{\X}\rho(x)\,d\mu(x)=1$, the above divergences are defined as
\begin{align*}
D_{\alpha}(\rho\|\sigma):=\frac{1}{\alpha-1}\log\int_{\X}\rho(x)^{\alpha}\sigma(x)^{1-\alpha}\,d\mu(x),
\end{align*}
and
\begin{align*}
D_{1}(\rho\|\sigma):=D(\rho\|\sigma):=
\begin{cases}
\int_{\X}\rho(x)(\logn\rho(x)-\logn\sigma(x))\,d\mu(x);&\rho^0\le\sigma^0,\\
+\infty,&\text{otherwise}.
\end{cases}
\end{align*}
Moreover, the notions of relative entropy and R\'enyi divergences may be extended to pairs
of positive normal functionals on an arbitrary von Neumann algebra 
\cite{Araki_Relentr,HiaiStandardFdiv,Petz-quasi-Neumann}.

The \ki{Chernoff divergence} of $\rho\in\S(\M)$ and $\sigma\in\M_*^+$ is defined as
\begin{align*}
C(\rho\|\sigma):=\sup_{\alpha\in(0,1)}(1-\alpha)D_{\alpha}(\rho\|\sigma),
\end{align*}
and their \ki{Hoeffding divergence} with parameter $r\ge 0$ as
\begin{align}\label{Hoeffding def}
H_r(\rho\|\sigma):=\sup_{\alpha\in(0,1)}\frac{\alpha-1}{\alpha}\left[r-D_{\alpha}(\rho\|\sigma)\right].
\end{align}
It is immediate from their definitions that the R\'enyi- and the Hoeffding divergences satisfy the following scaling laws:
\begin{align}\label{scaling}
D_{\alpha}(t\rho\|s\sigma)=D_{\alpha}(\rho\|\sigma)-\log s+\frac{\alpha}{\alpha-1}\log t,\ds\ds\ds\ds\ds
H_r(t\rho\|s\sigma)=H_{r+\log s}(\rho\|\sigma)-\log t
\end{align}
for any $\rho\in\S(\M)$, $\sigma\in\M_*^+$, $t,s\in(0,+\infty)$,
$\alpha\in[0,1)$, and $r\ge 0$.

The above divergences may be extended to pairs of subsets $R\subseteq\S(\M)$,
$S\subseteq\M_*^+$ by
\begin{align*}
\divv(R\|S):=\inf_{\rho\in R,\sigma\in S}\divv(\rho\|\sigma),
\end{align*}
where $\divv$ stands for any of the above divergences.

\begin{lemma}\label{lemma:convex lsc}
If $\divv$ is any of the above divergences, then
the map $\S(\M)\times\M_*^+\ni(\rho,\sigma)\mapsto\divv(\rho\|\sigma)$ is convex and lower semi-continuous. Moreover, the map 
$[0,+\infty)\times\S(\M)\times\M_*^+\ni(r,\rho,\sigma)\mapsto H_r(\rho\|\sigma)$ is convex.
\end{lemma}
\begin{proof}
These statements are well-known for the relative entropy and the R\'enyi divergences; see, for instance, \cite{HiaiStandardFdiv}.
Thus, the Chernoff divergence and any Hoeffding divergence for a fixed $r$ are suprema of convex and lower semi-continuous functions, and therefore they are also convex and lower semi-continuous. 
By the above, for any fixed $\alpha\in(0,1)$, 
$\frac{\alpha-1}{\alpha}\left[r-D_{\alpha}(\rho\|\sigma)\right]$ is a convex function of the triple $(r,\rho,\sigma)\in[0,+\infty)\times\S(\M)\times\M_*^+$, and hence the same holds for its supremum over $\alpha$, i.e., $H_r(\rho\|\sigma)$.
\end{proof}

\begin{cor}\label{lemma:Hr convex in r}
For any convex subsets $R\subseteq\S(\M)$ and $S\subseteq\M_*^+$, $H_r(R\|S)$ is a convex 
function of $r\in[0,+\infty)$. 
\end{cor}
\begin{proof}
Immediate from Lemma \ref{lemma:convex lsc} and the fact that 
taking the infimum of a jointly convex function in some of its variables yields a convex function.
\end{proof}

Let us introduce the notation
\begin{align*}
\psi(R\|S|\alpha):=(\alpha-1)D_{\alpha}(R\|S)
=\sup_{\rho\in\N,\sigma\in\A}(\alpha-1)D_{\alpha}(\rho\|\sigma)
,\ds\alpha\in[0,1),\ds\ds\ds\ds\ds
\psi(R\|S|1):=\lim_{\alpha\nearrow 1}\psi(R\|S|\alpha).
\end{align*}
(The limit exists due to the convexity of $\psi$; see Lemma \ref{lemma:Renyi limits}.)
With this, the definition of the Hoeffding divergences \eqref{Hoeffding def} can be rewritten as 
\begin{align}\label{Hoeffding def2}
H_r(\rho\|\sigma)=\sup_{\alpha\in(0,1)}\left[\frac{\alpha-1}{\alpha}r-\frac{1}{\alpha}\psi(\rho\|\sigma|\alpha)\right]
=
\sup_{u\in(-\infty,0)}\left[ur-(1-u)\psi\bz\rho\Big\|\sigma\Big|\frac{1}{1-u}\jz\right].
\end{align}

\begin{rem}\label{rem:psi nonpos}
It is known \cite{HiaiStandardFdiv} that for $\rho\in\S(\M)$, $\sigma\in\M_*^+$,
\begin{align*}
\psi(\rho\|\sigma|1)=\log\rho(\sigma^0)\le 0,\ds\ds\ds\text{and hence}\ds\ds\ds
\psi(R\|S|1)\le 0
\end{align*}
for any $R\subseteq\S(\M)$, $S\subseteq\M_*^+$.
\end{rem}

For proving the $\alpha\to 1$ limit of the R\'enyi divergences in Lemma \ref{lemma:Renyi limits}, we will need the following minimax theorem from \cite[Corollary A.2]{MH}.
\begin{lemma}\label{lemma:minimax2}
Let $X$ be a compact topological space, $Y$ be an ordered set, and let $f:\,X\times Y\to \bR\cup\{-\infty,+\infty\}$ be a function. Assume that
\smallskip

\s(i) $f(.\,,\,y)$ is lower semicontinuous for every $y\in Y$ and
\smallskip

(ii) 
$f(x,.)$ is monotonic increasing for every $x\in X$, or
$f(x,.)$ is monotonic decreasing for every $x\in X$.

\noindent Then 
\begin{align}\label{minimax statement}
\inf_{x\in X}\sup_{y\in Y}f(x,y)=
\sup_{y\in Y}\inf_{x\in X}f(x,y),
\end{align}
and the infima in \eqref{minimax statement} can be replaced by minima.
\end{lemma}

\begin{lemma}\label{lemma:Renyi limits}
For any $R\subseteq\S(\M)$, $S\subseteq\M_*^+$,
$\alpha\mapsto\psi(R\|S|\alpha)$ is a 
convex function on $[0,1]$, and
\begin{align}\label{psi finite}
\forall\alpha\in[0,1]:\,\psi(R\|S|\alpha)>-\infty\ds\iff\ds \exists \rho\in R,\sigma\in S:\ds
\rho^0\not\perp\sigma^0.
\end{align}
The function
$\alpha\mapsto D_{\alpha}(R\|S)$ is monotone increasing with
\begin{align}\label{Dalpha limits}
D_0(R\|S)=\lim_{\alpha\searrow 0}D_{\alpha}(R\|S)\le
\lim_{\alpha\nearrow 1}D_{\alpha}(R\|S)=:D_{1^-}(R\|S)\le D(R\|S).
\end{align}
If, moreover, $R$ and $S$ are both compact then we also have $D_{1^-}(R\|S)= D(R\|S)$.
\end{lemma}
\begin{proof}
The above are well-known when $|R|=|S|=1$ \cite{HiaiStandardFdiv}, from which all the assertions follow immediately, except for the last one. 
Assume that $R$ and $S$ are both compact; then 
\begin{align*}
\lim_{\alpha\nearrow 1}D_{\alpha}(R\|S)&=
\sup_{\alpha\in(0,1)}\inf_{(\rho,\sigma)\in R\times S}D_{\alpha}(\rho\|\sigma)
=
\inf_{(\rho,\sigma)\in R\times S}
\underbrace{\sup_{\alpha\in(0,1)}D_{\alpha}(\rho\|\sigma)}_{=D(\rho\|\sigma)}
=
D(R\|S),
\end{align*}
where the second equality follows by Lemmas \ref{lemma:minimax2} and \ref{lemma:convex lsc}.
This proves the last assertion.
\end{proof}

\begin{cor}\label{cor:psi tilde concave}
For any fixed $\N,\A\subseteq\S(\M)$,
\begin{align*}
\tilde\psi(\N\|\A|u):=
(1-u)\psi\bz\N\|\A\Big|\frac{1}{1-u}\jz=
(1-u)\sup_{\rho\in \N,\sigma\in \A}\psi\bz\rho\Big\|\sigma\Big|\frac{1}{1-u}\jz
\end{align*}
is a non-positive 
convex function of $u$ on $(-\infty,0)$, and it is finite at every 
$u\in(-\infty,0)$ if and only if the support condition in \eqref{psi finite} holds.
\end{cor}
\begin{proof}
By Lemma \ref{lemma:Renyi limits}, $\alpha\mapsto \psi(\rho\|\sigma|\alpha)$ is convex, and therefore also continuous, on $(0,1)$. Hence, it can be written as 
$ \psi(\rho\|\sigma|\alpha)=\sup_{i\in\I}\{c_i\alpha+d_i\}$ for some index set $\I$, and 
$c_i,d_i\in\bR$. Thus, 
\begin{align*}
\tilde\psi(\rho\|\sigma|u)=(1-u)\psi\bz\rho\Big\|\sigma\Big|\frac{1}{1-u}\jz
=
(1-u)\sup_{i\in\I}\left\{c_i\frac{1}{1-u}+d_i\right\}
=
\sup_{i\in\I}\left\{c_i+d_i(1-u)\right\},
\end{align*}
which, as the supremum of convex functions on $(-\infty,0)$, is itself convex. 
Taking the supremum over $\rho\in\N,\sigma\in\A$ yields the convexity of 
$u\mapsto \tilde\psi(R\|S|u)$, and the rest of the assertions are obvious.
\end{proof}

\begin{lemma}\label{lemma:Hr identity}
For any fixed $R\subseteq\S(\M)$, $S\subseteq\M_*^+$,
\begin{align}\label{Hr lower bound}
H_r(R\|S)\ge
\sup_{\alpha\in(0,1)}\frac{\alpha-1}{\alpha}\left[r-D_{\alpha}(R\|S)\right].
\end{align}
If, moreover, both $R$ and $S$ are compact and convex then the above inequality holds as an equality.
\end{lemma}
\begin{proof}
By \eqref{Hoeffding def2}, we have 
\begin{align*}
H_r(R\|S)
&=
\inf_{\rho\in R,\sigma\in S}\sup_{u\in(-\infty,0)}
\left[ur-\tilde \psi\bz\rho\|\sigma|u\jz\right]\\
&\ge
\sup_{u\in(-\infty,0)}\inf_{\rho\in R,\sigma\in S}
\left[ur-\tilde \psi\bz\rho\|\sigma|u\jz\right]
=
\sup_{\alpha\in(0,1)}\frac{\alpha-1}{\alpha}\left[r-D_{\alpha}(R\|S)\right].
\end{align*}
If both $R$ and $S$ are compact and convex then the inequality above holds as an equality 
due to the minimax theorem in 
\cite[Theorem 5.2]{FarkasRevesz2006}, and
Lemma \ref{lemma:convex lsc} and Corollary \ref{cor:psi tilde concave}.
\end{proof}

\begin{lemma}\label{lemma:Hr limits}
For any fixed $R\subseteq\S(\M)$, $S\subseteq\M_*^+$,
the function $r\mapsto H_r(R\|S)$ is monotone non-increasing on $[0,+\infty)$, 
\begin{align}\label{Hr infty}
r<D_0(R\|S)\ds\imp\ds H_r(R\|S)=+\infty\ds\imp\ds r\le D_0(R\|S),
\end{align}
and
\begin{align}\label{Hr lower bound2}
0\le-\psi(R\|S|1)\le H_r(R\|S),\ds\ds\ds r\in[0,+\infty).
\end{align}
Moreover,
\begin{align}\label{Hr null-1}
r> D(R\|S)\ds\imp\ds H_r(R\|S)=0\ds\imp\ds r\ge D_{1^{-}}(R\|S),
\end{align}
and if $R,S$ are both compact and convex then
\begin{align}\label{Hr null}
H_r(R\|S)=0\ds\iff\ds r\ge D(R\|S).
\end{align}
\end{lemma}
\begin{proof}
It follows by its definition \eqref{Hoeffding def} that $r\mapsto H_r(R\|S)$ is monotone non-increasing.

The first implication in \eqref{Hr infty} follows as $\lim_{\alpha\searrow 0}(r-D_{\alpha}(R\|S))=r-D_0(R\|S)$ by Lemma \ref{lemma:Renyi limits}, and $\lim_{\alpha\searrow 0}\frac{\alpha-1}{\alpha}=-\infty$. To see the second implication, assume that 
$r>D_0(R\|S)$, and hence there exist $\rho\in R$, $\sigma\in S$ such that 
$r>D_0(\rho\|\sigma)$. 
Then $r>D_{\alpha}(\rho\|\sigma)$ for all small enough $\alpha>0$,
according to \eqref{Dalpha limits}, and hence
$\lim_{\alpha\searrow 0}\frac{\alpha-1}{\alpha}\left[r-D_{\alpha}(\rho\|\sigma)\right]=-\infty$. 
On the other hand, $\lim_{\alpha\nearrow 1}\frac{\alpha-1}{\alpha}\left[r-D_{\alpha}(\rho\|\sigma)\right]=-\psi(\rho\|\sigma|1)<+\infty$, where the inequality follows from the fact that 
$r>D_0(\rho\|\sigma)$ implies $\rho^0\not\perp\sigma^0$, and hence 
$\psi(\rho\|\sigma|\alpha)$ is a finite-valued function on $[0,1]$, according to 
Lemma \ref{lemma:Renyi limits}. Since $\alpha\mapsto \frac{\alpha-1}{\alpha}\left[r-D_{\alpha}(\rho\|\sigma)\right]$ is continuous on $(0,1)$, we obtain that its supremum is finite, i.e., 
$+\infty> H_r(\rho\|\sigma)\ge H_r(R\|S)$, where the last inequality is by definition.

The inequalities in \eqref{Hr lower bound2} follow from \eqref{Hr lower bound} 
and Remark \ref{rem:psi nonpos} as
\begin{align*}
\lim_{\alpha\nearrow 1}\frac{\alpha-1}{\alpha}\left[r-D_{\alpha}(R\|S)\right]
=-\psi(R\|S|1).
\end{align*}

If $r> D(R\|S)$ then there exist $\rho\in R$, $\sigma\in S$ such that 
$r>D(\rho\|\sigma)\ge D_{\alpha}(\rho\|\sigma)$, $\alpha\in(0,1)$, where the second inequality is due to 
Lemma \ref{lemma:Renyi limits}. This implies immediately that $0=H_r(\rho\|\sigma)\ge H_r(R\|S)\ge 0$, where the first inequality is by definition, and the second one is by \eqref{Hr lower bound2}. This proves the first implication in \eqref{Hr null-1}.

If $r<D_{1^{-}}(R\|S)$ then $r<D_{\alpha}(R\|S)$ for some large enough $\alpha<1$ by Lemma \ref{lemma:Renyi limits}, and 
$0<\frac{\alpha-1}{\alpha}[r-D_{\alpha}(R\|S)]\le H_r(R\|S)$, where the second inequality 
is by \eqref{Hr lower bound}. This proves \eqref{Hr null}.

Finally, if $R,S$ are both convex and compact 
then $D(R\|S)=D_{1^{-}}(R\|S)\ge D_{\alpha}(R\|S)$, $\alpha\in(0,1)$, according to 
Lemma \ref{lemma:Renyi limits}. Hence, if 
$r=D(R\|S)$ then
$\frac{\alpha-1}{\alpha}[r-D_{\alpha}(R\|S)]\le 0$, $\alpha\in(0,1)$, which implies 
$H_r(R\|S)=\sup_{\alpha\in(0,1)}\frac{\alpha-1}{\alpha}[r-D_{\alpha}(R\|S)]\le 0$, where the equality is due to Lemma \ref{lemma:Hr identity}.
\end{proof}

In Sections \ref{sec:classical exponents} and \ref{sec:classical sc}, 
we will need the following more detailed knowledge about the Hoeffding divergence 
for classical states $\rho,\sigma\in\S(\X)$ for some finite set $\X$.
The statements below are well-known and easy to prove, but 
we collect them here for ease of reference in later parts of the paper, and because it does not seem to be easy to find a single reference for all the formulas that we will need.

We will assume that $\X_{\rho,\sigma}:=(\supp\rho)\cap(\supp\sigma)\ne\emptyset$.
For every $\alpha\in\bR$ and $u\in(-\infty,1)$, let 
\begin{align}\label{psi def classical}
\psi(\alpha):=\psi(\rho\|\sigma|\alpha):=
\log\sum_{x\in\X_{\rho,\sigma}}\rho(x)^{\alpha}\sigma^{1-\alpha},\ds
\ds\ds\ds
\tilde\psi(u):=\tilde\psi(\rho\|\sigma|u):=(1-u)\psi((1-u)\inv).
\end{align}
These coincide with $\psi(\rho\|\sigma|\alpha)$ and $\tilde\psi(\rho\|\sigma|u)$ defined previously for $\alpha\in[0,1]$ and $u\in(-\infty,0)$, respectively.
A straightforward computation shows that 
\begin{align}\label{psi derivatives}
\psi'(\alpha)=\Exp_{\mu_{\alpha}}\bz\logn\rho-\logn\sigma\jz,\ds\ds\ds\ds
\psi''(\alpha)=\Exp_{\mu_{\alpha}}\bz(\logn\rho-\logn\sigma)^2\jz
-\bz\Exp_{\mu_{\alpha}}(\logn\rho-\logn\sigma)\jz^2,
\end{align}
where 
\begin{align}\label{mu alpha}
\mu_{\alpha,\rho,\sigma}(x):=\mu_{\alpha}(x):=\frac{\rho(x)^{\alpha}\sigma(x)^{1-\alpha}}{\sum_{y\in\X_{\rho,\sigma}}\rho(y)^{\alpha}\sigma(y)^{1-\alpha}}\egy_{\X_{\rho,\sigma}}(x),
\ds\ds\ds x\in\X.
\end{align}
(The family of probability distributions $(\mu_{\alpha,\rho,\sigma)})_{\alpha\in[0,1]}$ is called the \ki{Hellinger arc}, and it connects 
$\frac{\sigma}{\sigma(\X_{\rho,\sigma})}\egy_{\X_{\rho,\sigma}}$ and
$\frac{\rho}{\rho(\X_{\rho,\sigma})}\egy_{\X_{\rho,\sigma}}$.)
Moreover, 
\begin{align}\label{psi tilde derivatives}
\tilde\psi'(u)=-\psi\bz\frac{1}{1-u}\jz+\frac{1}{1-u}\psi'\bz\frac{1}{1-u}\jz,\ds\ds\ds
\tilde\psi''(u)=\bz\frac{1}{1-u}\jz^3\psi''\bz\frac{1}{1-u}\jz.
\end{align}
This immediately yields the following (see also \cite[Lemma 3.2]{HMO2} for a generalization to 
quantum states):

\begin{lemma}\label{lemma:classical strict convexity}
In the above setting, $\psi$ and $\tilde\psi$ are convex, and the following are equivalent:
\begin{enumerate}
\item
$\psi''(\alpha)=0$ for some $\alpha\in\bR$;
\item
$\logn\rho-\logn\sigma$ is constant on $\supp\rho\cap\supp\sigma$;
\item
there exists a $\kappa>0$ such that $\rho(x)=\kappa\sigma(x)$, $x\in \supp\rho\cap\supp\sigma$;
\item
$\psi(\alpha)=(\alpha-1)\log\frac{\rho(\supp\rho\cap\supp\sigma)}{\sigma(\supp\rho\cap\supp\sigma)}+\log\rho(\supp\rho\cap\supp\sigma)$, $\alpha\in\bR$; in particular, $\psi$ is affine;
\item
$\tilde\psi(u)=\log\rho(\supp\rho\cap\supp\sigma)-u\log\sigma(\supp\rho\cap\supp\sigma)$,
$u\in(-\infty,1)$;
\item
$\tilde\psi''(u)=0$ for all $u\in(-\infty,1)$;
\item
$\tilde\psi''(u)=0$ for some $u\in(-\infty,1)$.
\setcounter{szamlalo}{\value{enumi}}
\end{enumerate}
\medskip

\noindent If, moreover, $\rho^0\le\sigma^0$ then the above is further equivalent to 
\medskip

\begin{enumerate}
\setcounter{enumi}{\value{szamlalo}}
\item
$\alpha\mapsto \frac{\psi(\alpha)}{\alpha-1}$ is constant on any/some non-trivial subinterval of $(0,\infty)$,
\end{enumerate}
in which case the constant is equal to $-\log\sigma(\supp\rho)=D_0(\rho\|\sigma)$.
\end{lemma}
\medskip

Let 
\begin{align*}
\Psi(c):=\sup_{\alpha\in(0,+\infty)}\{c\alpha-\psi(\alpha)\},\ds\ds\ds
\wtilde\Psi(c):=\sup_{u\in(-\infty,1)}\{cu-\tilde\psi(u)\},\ds\ds\ds c\in\bR,
\end{align*}
be the Legendre-Fenchel transforms of $\psi$ and $\tilde\psi$, respectively. 
A straightforward computation, using \eqref{psi derivatives} and \eqref{psi tilde derivatives}, yields that for every $\alpha\in(0,+\infty)$ and $u=(\alpha-1)/\alpha$, 
\begin{align}
D(\mu_{\alpha}\|\sigma)&=
\alpha\psi'(\alpha)-\psi(\alpha)& &=\Psi(\psi'(\alpha))
& &=
\tilde\psi'(u)\label{tilde psi der1}\\
D(\mu_{\alpha}\|\rho)&=
(\alpha-1)\psi'(\alpha)-\psi(\alpha)
& &=\Psi(\psi'(\alpha))-\psi'(\alpha) & &=
u\tilde\psi'(u)-\tilde\psi(u)=
\wtilde\Psi(\tilde\psi'(u))
.\label{tilde psi der2}
\end{align}

Assume for the rest that  $\rho^0\le\sigma^0$. Let
\begin{align*}
D_{\infty}\nw(\rho\|\sigma):=\log\max\left\{\frac{\rho(x)}{\sigma(x)}:\,x\in\X\right\},
\ds\ds\ds \X_{\infty}:=\left\{x\in\X:\,\log\frac{\rho(x)}{\sigma(x)}=D_{\infty}\nw(\rho\|\sigma)\right\},
\end{align*}
where $D_{\infty}\nw(\rho\|\sigma)$ is the max-relative entropy of $\rho$ and $\sigma$; see 
Section \ref{sec:sc divergences} for details. We have 
\begin{align*}
\lim_{\alpha\searrow 0}\mu_{\alpha}(x)=\frac{\sigma(x)}{\sigma(\supp\rho)}\egy_{\supp\rho}(x),\ds\ds\ds\ds\ds
\mu_1(x)=\rho(x),\ds\ds\ds\ds\ds
\lim_{\alpha\to+\infty}\mu_{\alpha}(x)=\frac{\sigma(x)}{\sigma(\X_{\infty})}\egy_{\X_{\infty}}(x),\ds\ds\ds x\in\X,
\end{align*}
and hence, by \eqref{psi derivatives} and \eqref{tilde psi der1}--\eqref{tilde psi der2},
\begin{align}
&\lim_{u\to-\infty}\tilde\psi'(u)=-\psi(0)=-\log\sigma(\supp \rho)=D_0(\rho\|\sigma),
& &
\psi'(1)=\tilde\psi'(0)=D(\rho\|\sigma),
\label{psi derivative lim}\\
&\lim_{\alpha\to+\infty}\psi'(\alpha)=D_{\infty}\nw(\rho\|\sigma),
&  &
\lim_{u\nearrow 1}\tilde\psi'(u)=-\log\sigma(\X_{\infty})=:\rmax(\rho\|\sigma).
\label{psi tilde derivative lim}
\end{align}

\begin{rem}\label{rem:affine psi}
If $\rho^0\le\sigma^0$ and $\psi$ is affine then, by Lemma \ref{lemma:classical strict convexity} and \eqref{psi derivative lim}--\eqref{psi tilde derivative lim},
\begin{align*}
-\log\sigma(\supp \rho)=D_0(\rho\|\sigma)=D(\rho\|\sigma)=D_{\infty}\nw(\rho\|\sigma)=\rmax(\rho\|\sigma).
\end{align*}
\end{rem}

The above observations yield immediately the following:

\begin{lemma}\label{lemma:simple Hr}
Let $\X$ be a finite set, and $\rho,\sigma\in\S(\X)$ be
such that $\rho^0\le\sigma^0$.
For every $r\in(D_0(\rho\|\sigma),\rmax(\rho\|\sigma))$ there exists a unique
$u_r\in(-\infty,1)$ and corresponding $\alpha_r:=1/(1-u_r)\in(0,+\infty)$ such that 
\begin{align}\label{simple Hr}
r&=\tilde\psi'(u_r)=\alpha_r\psi'(\alpha_r)-\psi(\alpha_r)=D(\mu_{\alpha_r,\rho,\sigma}\|\sigma),
\end{align}
and for this $u_r$ and $\alpha_r$,
\begin{align}\label{simple Hr2}
\wtilde\Psi(r)=u_rr-\tilde\psi(u_r)=
\frac{\alpha_r-1}{\alpha_r}r-\frac{1}{\alpha_r}\psi(\alpha_r)=
(\alpha_r-1)\psi'(\alpha_r)-\psi(\alpha_r)=D(\mu_{\alpha_r,\rho,\sigma}\|\rho).
\end{align}
Moreover, $r\in(D_0(\rho\|\sigma),D(\rho\|\sigma))$ $\iff$ 
$u_r\in(-\infty,0)$ $\iff$ $\alpha_r\in(0,1)$, in which case
\begin{align}\label{simple Hr3}
\wtilde\Psi(r)=H_r(\rho\|\sigma),
\end{align}
and $r\in(D(\rho\|\sigma),\rmax(\rho\|\sigma))$ $\iff$ 
$u_r\in(0,1)$ $\iff$ $\alpha_r\in(1,+\infty)$, in which case
\begin{align}
\wtilde\Psi(r)=\max_{u\in[0,1)}\{ur-\tilde\psi(r)\}=
\max_{\alpha>1}\frac{\alpha-1}{\alpha}
[r-D_{\alpha}(\rho\|\sigma)]=H_r\nw(\rho\|\sigma),
\end{align}
where $H_r\nw(\rho\|\sigma):=\sup_{\alpha>1}\frac{\alpha-1}{\alpha}
[r-D_{\alpha}(\rho\|\sigma)]$ is the Hoeffding anti-divergence; see Section \ref{sec:sc divergences}. Finally, for $r\ge \rmax(\rho\|\sigma)$,
\begin{align}\label{anti Hoeffding large r}
\wtilde\Psi(r)=H_r\nw(\rho\|\sigma)=r-\tilde\psi(1)=r-D_{\infty}(\rho\|\sigma).
\end{align}
\end{lemma}
\begin{proof}
If $\psi$ is affine then $-\log\sigma(\supp\rho)=\rmax(\rho\|\sigma)$, and 
\eqref{anti Hoeffding large r} holds trivially. 
Otherwise $\psi$ is strictly convex, 
according to Lemma \ref{lemma:classical strict convexity},
and all the statements follow easily from the observations after Lemma \ref{lemma:classical strict convexity}.
\end{proof}
\medskip

Finally, let us briefly comment on different notions of R\'enyi divergences for positive definite operators $\rho,\sigma$ on a finite-dimensional Hilbert space. 
For any $\alpha\in(0,1)$, the \ki{log-Euclidean R\'enyi $\alpha$-divergence} 
of $\rho$ and $\sigma$
\cite{AD,MO-cqconv,OH,ON} is defined as
\begin{align*}
D_{\alpha}\bog(\rho\|\sigma):=\frac{1}{\alpha-1}\log\Tr e^{\alpha\log\rho+(1-\alpha)\log\sigma}-\frac{1}{\alpha-1}\log\Tr\rho,
\end{align*}
and their \ki{maximal R\'enyi $\alpha$-divergence} \cite{Matsumoto_newfdiv} as
\begin{align*}
D_{\alpha}^{\max}(\rho\|\sigma):=\frac{1}{\alpha-1}\log\Tr \sigma\#_{\alpha}\rho-\frac{1}{\alpha-1}\log\Tr\rho,
\end{align*}
where 
\begin{align}\label{alpha geom mean}
\sigma\#_{\alpha}\rho:=\sigma^{1/2}\bz\sigma^{-1/2}\rho\sigma^{-1/2}\jz^{\alpha}\sigma^{1/2}
\end{align}
is the Kubo-Ando $\alpha$-geometric mean \cite{KA}. 
The Petz-type R\'enyi $\alpha$-divergence is also defined with the extra term $-\frac{1}{\alpha-1}\log\Tr\rho$ when $\Tr\rho\ne 1$.
The following is an immediate consequence of Theorems 2.1 and 3.1 in \cite{Hiai-ALT}, 
which we will use in Section \ref{sec:finiteQ}.

\begin{lemma}\label{lemma:Renyi ordering}
For any $\alpha\in(0,1)$,
\begin{align}\label{Renyi order}
D_{\alpha}(\rho\|\sigma)\le D_{\alpha}\bog(\rho\|\sigma)
\le D_{\alpha}^{\max}(\rho\|\sigma).
\end{align}
If $\rho\sigma\ne\sigma\rho$ then all the inequalities in \eqref{Renyi order} are strict, and hence, in particular,
\begin{align*}
\log\sigma\#_{\alpha}\rho\ne \alpha\log\rho+(1-\alpha)\log\sigma.
\end{align*}
\end{lemma}

We will also use a further family of quantum R\'enyi divergences, the sandwiched R\'enyi divergences \cite{Renyi_new,WWY}, in Section \ref{sec:sc divergences}.

\subsection{Asymptotic binary quantum state discrimination}
\label{sec:state disc}

We consider various generalizations of the binary state discrimination problem
presented in the Introduction. The first level of generalization unifies the composite i.i.d.~and the (simple) mixed i.i.d.~settings discussed in the Introduction. In the unified setting, 
an experimenter is presented with several identically prepared quantum systems with von Neumann algebra $\M$, and 
the knowledge that the state of the system was either chosen according to a probability 
distribution $p$ belonging to a set of probability distributions $\P$ on the Borel sets of 
$\S(\M)$ (null-hypothesis $H_0$), or according to 
a probability distribution $q$ belonging to a set of probability distributions $\Q$ on
the Borel sets of  
$\S(\M)$ (alternative hypothesis $H_1$).

The experimenter's test on $n$ copies of the system is represented by an operator $T_n\in\M^{\otimes n}$, $0\le T_n\le I$, and the two types of error probabilities are given by 
\begin{align*}
\alpha_n(\P|T_n)&:=\sup_{p\in\P}\int_{\S(M)}dp(\rho)\,\rho^{\otimes n}(I-T_n),& &\text{(type I)}\\
\beta_n(\Q|T_n)&:=\sup_{q\in\Q}\int_{\S(\M)}dq(\sigma)\,\sigma^{\otimes n}(T_n),& &\text{(type II)}.
\end{align*}
(Recall that here $\rho^{\otimes n}(I-T_n)$ is the functional $\rho^{\otimes n}$ evaluated on the operator $(I-T_n)$.) The composite i.i.d.~setting discussed in the Introduction is obtained as a special case by choosing $\P=\{\delta_{\rho}\}_{\rho\in R}$ and  
$\Q=\{\delta_{\sigma}\}_{\sigma\in S}$, where $\delta_{\omega}$ is the Dirac measure concentrated on $\omega$. In this case, we will write $R$ and $S$ instead of 
$\P=\{\delta_{\rho}\}_{\rho\in R}$ and $\Q=\{\delta_{\sigma}\}_{\sigma\in S}$, respectively, in the error probabilities and in the error exponents introduced below. 
The mixed i.i.d.~setting, also discussed in the Introduction, may be obtained as the special case $\P=\{p\}$ and $\Q=\{q\}$, where each hypothesis is represented by a single probability distribution.

The above setting can be further generalized by dropping the i.i.d.~assumption. In the most general setting, an asymptotic binary state discrimination problem is specified by a sequence 
$(\M_n)_{n\in\bN}$ of von Neumann algebras, and two sequences of sets of states 
$(\N_n)_{n\in\bN}$ and $(\A_n)_{n\in\bN}$,
representing the null hypothesis $H_0$ and the alternative hypothesis $H_1$, respectively. We denote the two types of error probabilities corresponding to a test $T_n\in(\M_n)_{[0,I]}$ in this case by
\begin{align*}
\alpha_n(\N_n|T_n):=
\sup_{\rho_n\in\N_n}\rho_n(I-T_n),\ds\ds\ds\ds
\beta_n(\A_n|T_n):=\sup_{\sigma_n\in\A_n}\sigma_n(T_n).
\end{align*}
This level of generality is the usual setting in investigations using the information spectrum method; see, e.g., \cite{NH} for the case of binary quantum state discrimination with simple hypotheses. In this paper we will consider the non-i.i.d.~problems of arbitrarily varying and of adversarial classical state discrimination; see Section \ref{sec:classical exponents}.

A summary of the important special cases (including the ones discussed above) is given in order of decreasing generality as follows:
\begin{itemize}
\item
\ki{Consistent:} 
It is specified by a sequence of von Neumann algebras $(\M_n)_{n\in\bN}$, two index sets $I,J$, and two functions $\rho:\,\bN\times I\to\S(\M_n)$, $\sigma:\,\bN\times J\to\S(\M_n)$. The null and the alternative hypotheses are given by 
$H_0:\,(\N_n=\{\rho_{n,i}\}_{i\in I})_{n\in\bN}$ and 
$H_1:\,(\A_n=\{\sigma_{n,j}\}_{j\in J})_{n\in\bN}$.
(This is the natural setting, for instance, to consider the discrimination 
of correlated states on an infinite spin chain, where 
$\rho_{n,i}$ is the $n$-site restriction of some state 
$\rho_{\infty,i}$ on the infinite chain, and similarly for $\sigma_{n,j}$; 
see, e.g., \cite{HMO2,MO-correlated}.)
\item
\ki{Mixed i.i.d.:} $\M_n=\M^{\otimes n}$, $n\in\bN$, and 
the null and the alternative hypotheses are specified by two 
sets of probability distributions $\P,\Q$ on the Borel sets of $\S(\M)$, with 
$\N_n=\{\int_{\S(\M^{\otimes n})}dp(\rho)\rho^{\otimes n}\}_{p\in\P}$ and 
$\A_n=\{\int_{\S(\M^{\otimes n})}dq(\sigma)\sigma^{\otimes n}\}_{q\in\Q}$.
\item
\ki{I.i.d.:} In the above, $\P=\{\delta_{\rho}\}_{\rho\in\N}$ and $\Q=\{\delta_{\sigma}\}_{\sigma\in\A}$
for some sets $\N,\A\subseteq\S(\M)$. 
\end{itemize}
Note that once the sequence of von Neumann algebras is fixed (typically the tensor powers of a given algebra), we may consider mixtures of the above cases by either 
requiring only one of the hypotheses to fall in one of the above classes, or by 
requiring one of the hypotheses to fall in one class, and the other in a possibly different class; 
e.g., we may consider an i.i.d. vs. mixed i.i.d. state discrimination problem.
On top of the above, we say that the null hypothesis is \ki{simple} if $|\N_n|=1$, $n\in\bN$, and \ki{composite} otherwise, and similarly for the alternative hypothesis. The state discrimination problem is 
simple if both the null and the alternative hypotheses are simple, and composite otherwise.
We will use the notations $\alpha_n(\rho|T_n)$, $\alpha_n(\N|T_n)$, $\alpha_n(p|T_n)$,
$\alpha_n(\P|T_n)$, $\alpha_n(\N_n|T_n)$, for the type I, and similarly for the type II error probabilities, as well as for the error exponents defined later, depending on the 
precise setting that we consider. If some concept or statement is independent of the 
setting, we will also use the notations $\alpha_n(H_0|T_n)$ and $\beta_n(H_1|T_n)$. 

It is known that in the case of simple binary i.i.d.~hypothesis testing 
(i.e., $|\N|=|\A|=1$), if $\N\ne \A$ then both error probabilities can be made to vanish asymptotically with an exponential speed by a suitably chosen sequence $(T_n)_{n\in\bN}$ of tests, and the same holds in most of the more general composite settings, too.
Clearly, the faster the error probabilities decay with $n$ the better, and hence it is a natural question to ask what exponents are achievable in the following sense:

\begin{definition}\label{errexp def}
We say that $(r,r')\in[0,+\infty]^2$ is an \ki{achievable exponent pair} if there exists a sequence of tests
$T_n\in(\M_n)_{[0,I]}$, $n\in\bN$, such that 
\begin{align*}
r\le \liminf_{n\to+\infty}-\frac{1}{n}\log\beta_n(H_1|T_n),\ds\ds\ds
r'\le \liminf_{n\to+\infty}-\frac{1}{n}\log\alpha_n(H_0|T_n).
\end{align*}
We denote the set of achievable exponent pairs by $\ac(H_0\|H_1)$. 

In the case 
where the null hypothesis is consistent, given by $(\N_n=\{\rho_{n,i}\}_{i\in I})_{n\in\bN}$, 
we may introduce the relaxed set of achievable exponent pairs
$\ac^{\{0\}}\bz\bz\{\rho_{n,i}\}_{i\in I}\jz_{n\in\bN}\|H_1\jz$ as the set of all pairs 
$(r,r')\in[0,+\infty]^2$ for which there exists a sequence of tests
$T_n\in(\M_n)_{[0,I]}$, $n\in\bN$, such that 
\begin{align*}
r\le \liminf_{n\to+\infty}-\frac{1}{n}\log\beta_n(H_1|T_n),\ds\ds\ds
\forall i\in I:\ds r'\le \liminf_{n\to+\infty}-\frac{1}{n}\log\alpha_n(\rho_{n,i}|T_n).
\end{align*}

The set of achievable exponent pairs 
$\ac^{\{1\}}\bz H_0\|\bz\{\sigma_{n,j}\}_{j\in J}\jz_{n\in\bN}\jz$
and 
$\ac^{\{0,1\}}\bz \bz\{\rho_{n,i}\}_{i\in I}\jz_{n\in\bN}\|\bz\{\sigma_{n,j}\}_{j\in J}\jz_{n\in\bN}\jz$
are defined analogously when the alternative hypothesis or both hypotheses are consistent, respectively. 
\end{definition}

Note that if the index set $I$ is finite then 
$\ac^{\{0\}}\bz\bz\{\rho_{n,i}\}_{i\in I}\jz_{n\in\bN}\|H_1\jz=
\ac\bz\bz\{\rho_{n,i}\}_{i\in I}\jz_{n\in\bN}\|H_1\jz$, and similarly for the other variants, and hence these only play a role for infinite index set(s). 

We will use $\ac\gen(H_0\|H_1)$ to denote any of the above sets of achievable exponent pairs,
where $t=\emptyset$, $\{0\}$, $\{1\}$, or $\{0,1\}$.

\begin{rem}
The notation $\ac^{\{0\}}(H_0\|H_1)$ implicitly implies that the null hypothesis is consistent, and we use the analogous conventions for $t=\{1\}$ and $\{0,1\}$, as well as for the error exponents defined below. 
\end{rem}

\begin{rem}\label{rem:ac swap}
It is clear from the definition that 
\begin{align*}
(r,r')\in \ac\gen\bz\bz\N_n\jz_{n\in\bN}\|\bz\A_n\jz_{n\in\bN}\jz
\ds\iff\ds
(r',r)\in \ac^{\oll t}\bz\bz\A_n\jz_{n\in\bN}\|\bz\N_n\jz_{n\in\bN}\jz,
\end{align*}
where $\oll t:=\{1-b:\,b\in t\}$. 
\end{rem}

The following is immediate from the definition:

\begin{lemma}\label{lemma:smaller achievable}
If $(r,r')\in \ac\gen(H_0\|H_1)$ then also $(\tilde r,\tilde r')\in \ac\gen(H_0\|H_1)$ for all 
$0\le \tilde r\le r$ and $0\le \tilde r'\le r'$.
\end{lemma}

\begin{lemma}\label{lemma:achievable limit}
$\ac(H_0\|H_1)$ is closed, and 
the same holds for $\ac\gen(H_0\|H_1)$ for $t=\{0\}/\{1\}/\{0,1\}$, provided that 
$H_0/H_1/$ both is$/$are consistent with (a) countable index set(s).
\end{lemma}
\begin{proof}
The proof is standard; we give the details for readers' convenience in the case 
$t=\{0,1\}$ and $I=J=\bN$. Let $(r_m,r_m')\in\ac^{\{0,1\}}\bz\bz\{\rho_{n,i}\}_{i\in I}\jz_{n\in\bN}\|\bz\{\sigma_{n,j}\}_{j\in J}\jz_{n\in\bN}\jz$ be such that $r_m\to r$, $r_m'\to r'$. 
We assume that $r,r'<+\infty$, in which case we may also assume without loss of generality that $r_m,r_m'<+\infty$; the other cases follow in a similar way. 
For every $m\in\bN$, there exists an $N_m\in\bN$ such that for every $n\ge N_m$, 
and every $i,j\in[m]$, 
$\alpha_n(\rho_{n,i}|T_{m,n})<e^{-n(r_m'-1/m)}$,
$\beta_n(\sigma_{n,j}|T_{m,n})<e^{-n(r_m-1/m)}$ for some test $T_{m,n}$. We may also assume that $N_1<N_2<\ldots$. 
Therefore, for every $n\in\bN$, there exists a unique $m_n\in\bN$ such that 
$N_{m_n}\le n<N_{m_n+1}$.
Then with $T_n:=T_{m_n,n}$, $n\in\bN$, we have 
$\liminf_{n\to\infty}-\frac{1}{n}\log\alpha_n(\rho_{n,i}|T_n)\ge r'$,
$\liminf_{n\to\infty}-\frac{1}{n}\log\beta_n(\sigma_{n,j}|T_n)\ge r$,
for every $i,j\in\bN$. 
\end{proof}

\begin{definition}
The \ki{direct exponents} with type II rate $r> 0$ of testing $H_0$ against $H_1$ are
\begin{align}\label{direct exp def}
\dli_r\gen(H_0\|H_1):=\sup\{r':\,(r,r')\in \ac\gen(H_0\|H_1)\}.
\end{align}
\end{definition}

\begin{definition}
The \ki{Chernoff exponents} of testing $H_0$ against $H_1$ are
\begin{align*}
\cli\gen(H_0\|H_1):=\sup\{r:\,(r,r)\in \ac\gen(H_0\|H_1)\}.
\end{align*}
\end{definition}

\begin{rem}\label{rem:length}
Note that $\dli_r\gen(H_0\|H_1)$ is the length of the line segment obtained by intersecting
$\ac\gen(H_0\|H_1)$ with $\{r\}\times\bR$, while
$\cli\gen(H_0\|H_1)$ is $2^{-1/2}$ times the length of the line segment obtained by intersecting
$\ac\gen(H_0\|H_1)$ with the diagonal line $D:=\{(r,r):\,r\in\bR\}$. That is,
\begin{align*}
\dli_r\gen(H_0\|H_1)=\lambda_1\bz\bz\{r\}\times\bR\jz\cap\ac\gen(H_0\|H_1)\jz,\ds\ds\ds
\cli\gen(H_0\|H_1)=2^{-1/2}\lambda_1\bz D\cap\ac\gen(H_0\|H_1)\jz,
\end{align*}
where $\lambda_1$ is the one-dimensional Lebesgue measure (length) of the line segment in its argument, and that the intersections are indeed line segments follows from Lemma \ref{lemma:smaller achievable}.
\end{rem}

While these formulations of the error exponents are slightly different from how these quantities are usually defined, the simple geometric picture behind them is very convenient in the analysis of the error exponents of composite state discrimination problems; see, e.g.,
Lemma \ref{lemma:intersection}.
It is straightforward to rewrite the direct exponent in the familiar way, as
\begin{align*}
\dli_r(H_0\|H_1)
=
\sup\left\{\liminf_{n\to+\infty}-\frac{1}{n}\log\alpha_n(H_0|T_n):\,
\liminf_{n\to+\infty}-\frac{1}{n}\log\beta_n(H_1|T_n)\ge r\right\},
\end{align*}
where the supremum is over all sequences of tests satisfying the indicated constraint.
The relaxed variants for consistent null- and/or alternative hypotheses can be expressed similarly, e.g., as
\begin{align*}
\dli_r^{\{0\}}(H_0\|H_1)
=
\sup\left\{\inf_{i\in I}\liminf_{n\to+\infty}-\frac{1}{n}\log\alpha_n(\rho_{n,i}|T_n):\,
\liminf_{n\to+\infty}-\frac{1}{n}\log\beta_n(H_1|T_n)\ge r\right\}.
\end{align*}
For the Chernoff exponent, we have the following:

\begin{lemma}
The Chernoff exponent $\cli(H_0\|H_1)$ can be expressed as
\begin{align}\label{Chernoff liminf}
\cli(H_0\|H_1)=\liminf_{n\to+\infty}-\frac{1}{n}\log\min_{T_n\in(\M_n)_{[0,I]}}
\{\alpha_n(H_0|T_n)+\beta_n(H_1|T_n)\}.
\end{align}
\end{lemma}
\begin{proof}
If $(r,r)$ is achievable by a test sequence $(T_n)_{n\in\bN}$ then 
\begin{align*}
r\le
\liminf_{n\to+\infty}-\frac{1}{n}\log(\alpha_n(H_0|T_n)+\beta_n(H_1|T_n))
\le
\liminf_{n\to+\infty}-\frac{1}{n}\log\min_{T_n\in(\M_n)_{[0,I]}}\{\alpha_n(H_0|T_n)+\beta_n(H_1|T_n)\},
\end{align*}
and taking the supremum over all such $r$ yields LHS$\le$RHS in \eqref{Chernoff liminf}.
On the other hand, if 
LHS$<$RHS then there exists a 
$\delta>0$ and a
sequence of tests $(T_n)_{n\in\bN}$ such that for all large enough $n$,
$\alpha_n(H_0|T_n)<e^{-n(\cli(H_0\|H_1)+\delta)}$,
$\beta_n(H_1|T_n)<e^{-n(\cli(H_0\|H_1)+\delta)}$,
and hence $(\cli(H_0\|H_1)+\delta,\cli(H_0\|H_1)+\delta)$ is achievable, a contradiction.
\end{proof}

\begin{definition}
The \ki{Stein exponent} of testing $H_0$ against $H_1$ is 
\begin{align}
\sli(H_0\|H_1)&:=\sup\left\{\liminf_{n\to+\infty}-\frac{1}{n}\log\beta_n(H_1|T_n):\,
\lim_{n\to+\infty}\alpha_n(H_0|T_n)=0\right\}\label{Stein exp def1}\\
&=
\sup\left\{r\ge 0:\,\exists\,T_n\in(\M_n)_{[0,I]},\,n\in\bN,\s
\liminf_{n\to+\infty}-\frac{1}{n}\log\beta_n(H_1|T_n)\ge r,\s
\lim_{n\to+\infty}\alpha_n(H_0|T_n)=0
\right\},\label{Stein exp def2}
\end{align}
where the supremum in \eqref{Stein exp def1} is over all sequences of tests 
$T_n\in(\M_n)_{[0,1]}$, $n\in\bN$, satisfying the constraint. 
The variants $\sli^{\{0\}}(H_0\|H_1)$, etc., 
can be defined by a straightforward modification of \eqref{Stein exp def2}.
\end{definition}

It is well-known that in finite-copy state discrimination, general tests may provide an advantage over projective tests (i.e., where $T_n^2=T_n$); see, e.g., \cite[Example 2.27]{Holevobook}.
This is not the case, however, in the asymptotic study of the error exponents. Indeed, we have the following:

\begin{lemma}\label{lemma:projective enough}
All error exponents defined above
remain unchanged if in their definitions we restrict the tests to be projections.
\end{lemma}
\begin{proof}
Let $f:[0,1]\to \{0,1\}$ be the function defined by the formula
\begin{align}\label{projection from test}
f(x) = \left\{ \begin{matrix} 0 & {\rm if} \;\; x<1/2, \\ 1 & {\rm if}\;\; x\geq 1/2.\end{matrix} \right.
\end{align}
For any test $T_n\in\bz\M_n\jz_{[0,I]}$, 
$Q_n:=f(T_n)$ is a projection, $0\leq Q_n\leq 2T_n$
and $0\leq I-Q_n \leq 2(I-T_n)$ since
$f(x)\leq 2x$ and $1-f(x)\leq 2(1-x)$ for all $x\in[0,1]$.
Thus, $\alpha_n(H_0|Q_n)\le 2\alpha_n(H_0| T_n)$ and
$\beta_n(H_1|Q_n)\le 2\beta_n(H_1|T_n)$, from which the assertion follows immediately.
\end{proof}
\medskip

In the case of i.i.d.~state discrimination with simple hypotheses, i.e., when $\N=\{\rho\}$, $\A=\{\sigma\}$ for some states
$\rho,\sigma\in\S(\M)$, each of the above discussed exponents is known to be equal to a certain divergence of the two states:

\begin{lemma}\label{lemma:known exponents}
For any von Neumann algebra $\M$, and any states $\rho,\sigma\in\S(\M)$, 
\begin{align*}
\sli(\rho\|\sigma)&=D(\rho\|\sigma),
& & \text{(Stein's lemma)}\\
\cli(\rho\|\sigma)&=C(\rho\|\sigma),
& &\text{(Chernoff bound)}\\
\dli_r(\rho\|\sigma)&=H_r(\rho\|\sigma),\ds r>0.
& &\text{(Hoeffding bound)}\\
\end{align*}
\end{lemma}
See \cite{HP,ON} for the Stein's lemma, \cite{NSz,AudChernoff} for the Chernoff bound, and 
\cite{Hayashicq,Nagaoka} for the direct exponents when $\M=\B(\hil)$ with a 
finite-dimensional Hilbert space $\hil$, and \cite{JOPS} for all the error exponents in the 
case of a general von Neumann algebra. The importance of the above results is twofold: on the one hand, they give single-copy expressions for the error exponents (i.e., not involving a limit $n\to+\infty$), while on the other hand, they provide operational interpretations to the divergences appearing in them.
\medskip

In this paper we are mainly interested in the error exponents of binary i.i.d.~state discrimination, and their relations to the worst pairwise exponents.
Let us start with a few simple observations. First, 
in the most general case, it is intiuitively clear that ``larger'' hypotheses are less distinguishable; formally, if $\N_n\subseteq\wtilde\N_n$ and $\A_n\subseteq\wtilde\A_n$, $n\in\bN$, then 
\begin{align*}
\alpha_n(\wtilde\N_n|T_n)\ge\alpha_n(\N_n|T_n),\ds
\beta_n(\wtilde\A_n|T_n)\ge\beta_n(\A_n|T_n),\ds\ds\ds
T_n\in\bz\M_n\jz_{[0,I]},\,n\in\bN,
\end{align*}
and hence,
\begin{align}\label{bigger set smaller exponent}
\e\big((\wtilde\N_n)_{n\in\bN}\|(\wtilde\A_n)_{n\in\bN}\big)
\le
\e\bz\bz\N_n\jz_{n\in\bN}\|\bz\A_n\jz_{n\in\bN}\jz,
\end{align}
where $\e$ may be any of the Stein-, Chernoff, or the direct exponents.
In the consistent case we further have
\begin{align}\label{consistent exponent inequalities}
\e\bz(\{\rho_{n,i}\}_{i\in I})_{n\in\bN}\|(\{\sigma_{n,j}\}_{j\in J})_{n\in\bN}\jz
&\le
\e^{\{x\}}\bz(\{\rho_{n,i}\}_{i\in I})_{n\in\bN}\|(\{\sigma_{n,j}\}_{j\in J})_{n\in\bN}\jz
\\
&\le \e^{\{0,1\}}\bz(\{\rho_{n,i}\}_{i\in I})_{n\in\bN}\|(\{\sigma_{n,j}\}_{j\in J})_{n\in\bN}\jz
\le
\inf_{i\in I,j\in J}\e\bz(\rho_{n,i})_{n\in\bN}\|(\sigma_{n,j})_{n\in\bN}\jz,
\end{align}
where $x=0$ or $x=1$, 
and analogous inequalities may be obtained by a straightforward modification when only one of the hypotheses is assumed to be/is treated as consistent.
Specializing to the i.i.d.~case yields
\begin{align}\label{composite vs single bound1}
\e(R\|S)\le 
\left\{\begin{array}{l}\e^{\{0\}}(R\|S)\\ \e^{\{1\}}(R\|S)\end{array}\right\}
\le\e^{\{0,1\}}(R\|S)\le\inf_{\rho\in R,\,\sigma\in S}\e(\rho\|\sigma)=\mathrm{E}(\N\|\A),
\end{align}
where $\mathrm{E}$ is the divergence corresponding to the exponent $\e$ as in Lemma \ref{lemma:known exponents}.

Second, in the mixed i.i.d.~case,
\begin{align*}
\alpha_n(\P|T_n)=\sup_{p\in\P}\int_{\S(M)}dp(\rho)\,\rho^{\otimes n}(I-T_n)
\le\sup_{\rho\in\supp\P}\rho^{\otimes n}(I-T_n)=\alpha_n(\supp\P|T_n),
\end{align*}
and similarly, $\beta_n(\Q|T_n)\le\beta_n(\supp\Q|T_n)$,
where $\supp\P:=\cup_{p\in\P}\supp p$, and $\supp\Q:=\cup_{q\in\Q}\supp q$.
Combining the above, we get that in the mixed i.i.d.~case, with hypotheses 
$\P,\Q$,
\begin{align}
\e(\supp\P\|\supp\Q)
\le
\left\{\begin{array}{l}\e(\supp\P\|\Q)\\ \e(\P\|\supp\Q)\end{array}\right\}
\le 
\e(\P\|\Q)\le \inf_{p\in\P,q\in\Q}\e(p\|q).\label{composite vs single bound2}
\end{align}
In particular, for any state $\rho\in\S(\M)$, and any probability distribution $q$ on the Borel sets of $\S(\M)$, \eqref{composite vs single bound1} and 
\eqref{composite vs single bound2} give the incomparable bounds
\begin{align}\label{composite vs single bound3}
\e(\rho\|\supp q)\le\begin{cases}
\inf_{\sigma\in\supp q}\e(\rho\|\sigma),\\
\e(\rho\|q).
\end{cases}
\end{align}

\begin{rem}\label{rem:q exp vs supp q exp}
It is easy to see that for finitely supported probability distributions $p,q$ on $\S(\M)$, we have 
\begin{align*}
\e(\supp p\|\supp q)=\e(p\|q).
\end{align*}
In particular, for a finitely supported $q$, \eqref{composite vs single bound3} reduces to
\begin{align}\label{q exp vs supp q exp}
\e(\rho\|\supp q)=\e(\rho\|q)\le \inf_{\sigma\in\supp q}\e(\rho\|\sigma).
\end{align}
\end{rem}
\medskip

Our main goal in this paper is to give sufficient conditions for 
(some of) the above inequalities to hold as equalities, and exhibit 
explicit examples demonstrating strict inequality when these conditions are not satisfied.
Note that equality between the first and the last term in 
\eqref{composite vs single bound1} is equivalent to 
\begin{align*}
\e(\N\|\A)=
\mathrm{E}(\N\|\A).
\end{align*}
In particular, such an equality gives an operational interpretation of the divergence of two sets of states, extending that in 
Lemma \ref{lemma:known exponents} for pairs of single states.
\medskip

We start our investigation with some simple but very useful
necessary and sufficient conditions in terms of the set of achievable exponent pairs.
In particular, Lemmas \ref{lemma:intersection} and \ref{lemma:Stein from directexp}
below show that to prove equality for the Stein and the Chernoff exponents, it is sufficient to prove equality for the direct exponents. We will utilize this in Theorems \ref{thm:classical countable equality} and \ref{thm:classical convex hyp2}.

\begin{lemma}\label{lemma:intersection}
Let $H_0:\,(\N_n)_{n\in\bN}$, $H_1:\,(\A_n)_{n\in\bN}$ specify a binary state discrimination problem, and for every $k$ in some index set $\K$, let 
$H_0^{(k)}:\,(\N_n^{(k)})_{n\in\bN}$, $H_1^{(k)}:\,(\A_n^{(k)})_{n\in\bN}$ also specify binary state discrimination problems (on possibly different sequences of von Neumann algebras).
Moreover, let $t,t_k\in\{\emptyset,\{0\},\{1\},\{0,1\}\}$, $k\in\K$. 
Then
\begin{align}
\ac\gen(H_0\|H_1)\subseteq\bigcap_{k\in\K}\ac^{t_k}(H_0^{(k)}\|H_1^{(k)})
&\ds\ds\imp\ds\ds
\dli_r\gen(H_0\|H_1)\le\inf_{i\in\I}\dli_r^{t_k}(H_0^{(k)}\|H_1^{(k)}),\ds r>0,\label{exp bound geom1}\\
&\ds\ds\imp\ds\ds
\oll\ac\gen(H_0\|H_1)\subseteq\bigcap_{k\in\K}\oll\ac^{t_k}(H_0^{(k)}\|H_1^{(k)})
\label{exp bound geom0}\\
&\ds\ds\imp\ds\ds
\cli\gen(H_0\|H_1)\le\inf_{k\in\K}\cli^{t_k}(H_0^{(k)}\|H_1^{(k)}),\label{exp bound geom2}
\end{align}
where $\oll\ac^{t}$, $\oll\ac^{t_k}$ stand for the closures of the achievable sets, 
and
\begin{align}
\ac\gen(H_0\|H_1)\supseteq\bigcap_{k\in\K}\ac^{t_k}(H_0^{(k)}\|H_1^{(k)})
&\ds\ds\imp\ds\ds
\dli_r\gen(H_0\|H_1)\ge\inf_{k\in\K}\dli_r^{t_k}(H_0^{(k)}\|H_1^{(k)}),\ds r>0,
\label{exp bound geom3}\\
&\ds\ds\imp\ds\ds
\oll\ac\gen(H_0\|H_1)\supseteq\bigcap_{k\in\K}\oll\ac^{t_k}(H_0^{(k)}\|H_1^{(k)})\\
&\ds\ds\imp\ds\ds
\cli\gen(H_0\|H_1)\ge\inf_{k\in\K}\cli^{t_k}(H_0^{(k)}\|H_1^{(k)}).\label{exp bound geom4}
\end{align}
\end{lemma}
\begin{proof}
We only prove \eqref{exp bound geom1}--\eqref{exp bound geom2}, as 
\eqref{exp bound geom3}--\eqref{exp bound geom4} follow the same way.
We have
\begin{align*}
&\ac\gen(H_0\|H_1)\subseteq\bigcap_{k\in\K}\ac^{t_k}(H_0^{(k)}\|H_1^{(k)})\\
&\ds\imp\ds \forall r>0:\ds
\bz\{r\}\times\bR\jz\cap\ac\gen(H_0\|H_1)\subseteq
\underbrace{\bz\{r\}\times\bR\jz \cap\bz\bigcap_{k\in\K}\ac^{t_k}(H_0^{(k)}\|H_1^{(k)})\jz}_{=
\bigcap_{k\in\K}\bz \bz\{r\}\times\bR\jz \cap\ac^{t_k}(H_0^{(k)}\|H_1^{(k)})\jz}\\
&\ds\imp\ds\forall r>0:\ds
\underbrace{\lambda_1\bz\bz\{r\}\times\bR\jz\cap\ac\gen(H_0\|H_1)\jz}_{=\dli_r\gen(H_0\|H_1)}
\le
\inf_{k\in\K}\underbrace{\lambda_1\bz \bz\{r\}\times\bR\jz \cap\ac^{t_k}(H_0^{(k)}\|H_1^{(k)})\jz}_{=\dli_r^{t_k}(H_0^{(k)}\|H_1^{(k)})}\\
&\ds\imp\ds \forall r>0:\ds
\bz\{r\}\times\bR\jz\cap\oll\ac\gen(H_0\|H_1)\subseteq
\underbrace{\bz\{r\}\times\bR\jz \cap\bz\bigcap_{k\in\K}\oll\ac^{t_k}(H_0^{(k)}\|H_1^{(k)})\jz}_{=
\bigcap_{k\in\K}\bz \bz\{r\}\times\bR\jz \cap\oll\ac^{t_k}(H_0^{(k)}\|H_1^{(k)})\jz}.
\end{align*}
(The intersection $(\{0\}\times\bR)\cap\ac\gen(H_0\|H_1)$ is 
$\{0\}\times[0,+\infty]$ for any $H_0$, $H_1$, and hence it is enough to consider $r>0$ above.)
This proves \eqref{exp bound geom1}--\eqref{exp bound geom0}, and \eqref{exp bound geom2} follows in the same way as
\begin{align*}
\cli\gen(H_0\|H_1)
&=
\lambda_1\bz D\cap \oll\ac\gen(H_0\|H_1)\jz
\le
\lambda_1\bz \bigcap_{k\in\K}
\bz D\cap\oll\ac^{t_k}(H_0^{(k)}\|H_1^{(k)})\jz \jz=\inf_{k\in\K}
\underbrace{\lambda_1\bz D\cap\oll\ac^{t_k}(H_0^{(k)}\|H_1^{(k)})\jz}_{=\cli^{t_k}(H_0^{(k)}\|H_1^{(k)})},
\end{align*}
where $D:=\{(r,r):\,r\in\bR\}$, and the inequality follows from
\eqref{exp bound geom0}.
\end{proof}

\begin{cor}\label{cor:cs equality from dr equality}
In the case of binary mixed i.i.d.~hypothesis testing with null-hypothesis represented by $\P$, and the alternative hypothesis by $\Q$, we have 
\begin{align}
\ac(\P\|\Q)\supseteq\bigcap_{p\in\P,q\in\Q}\ac(p\|q)
&\ds\ds\iff\ds\ds
\dli_r(\P\|\Q)=\inf_{p\in\P,q\in\Q}\dli_r(p\|q),\ds r>0\label{c equality from dr equality1}\\
&\ds\ds\imp\ds\ds\cli(\P\|\Q)=\inf_{p\in\P,q\in\Q}\cli(p\|q).\label{c equality from dr equality2}
\end{align}
\end{cor}
\begin{proof}
Immediate from \eqref{composite vs single bound2}, Lemma \ref{lemma:intersection}, and Lemma \ref{lemma:achievable limit}.
\end{proof}

\begin{lemma}\label{lemma:Stein from directexp}
Let $H_0:\,(\N_n)_{n\in\bN}$ and $H_1:\,(\A_n)_{n\in\bN}$ be the null and the alternative hypotheses of a general binary state discrimination problem, and let $\N,\A\subseteq\S(\M)$
be compact sets for some von Neumann algebra $\M$.
\begin{align}\label{s equality from dr equality}
\text{If}\ds\ds\ds \dli_r\gen(H_0\|H_1)\ge\inf_{\rho\in\N,\sigma\in\A}\dli_r(\rho\|\sigma),
\ds r>0,\ds\ds\ds\text{then}\ds\ds\ds \sli\gen(H_0\|H_1)\ge \inf_{\rho\in\N,\sigma\in\A}\sli(\rho\|\sigma).
\end{align}
\end{lemma}
\begin{proof}
Assume that 
$\dli_r\gen(H_0\|H_1)\ge\inf_{\rho\in\N,\sigma\in\A}\dli_r(\rho\|\sigma)$,
$r>0$. Then
\begin{align*}
\sli\gen(H_0\|H_1)\ge \sup\{r>0:\,\dli_r\gen(H_0\|H_1)>0\}
&\ge
\sup\{r>0:\,\underbrace{\inf_{\rho\in\N,\sigma\in\A}\dli_r(\rho\|\sigma)}_{=H_r(\N\|\A)}>0\}\\
&=
D(\N\|\A)
=
\inf_{\rho\in\N,\sigma\in\A}\sli(\rho\|\sigma),
\end{align*}
where the inequality in the first step is obvious, the second inequality follows by assumption,
the first equality is due to Lemma \ref{lemma:known exponents} (Hoeffding bound) and 
Lemma \ref{lemma:Hr limits}, and the last equality follows again by 
Lemma \ref{lemma:known exponents} (Stein's lemma).
\end{proof}
\medskip

Finally, we note that the standard techniques used for the achievability parts 
(i.e., $\e(\rho\|\sigma)\ge E(\rho\|\sigma)$) of the equalities in Lemma \ref{lemma:known exponents}
also yield achievability bounds in the most general case; however, the resulting 
lower bounds are given in terms of regularized divergences, which are not feasible to evaluate in general. We give these bounds below with detailed proofs for completeness. We remark that an improved lower bound for the Stein exponent in the composite i.i.d.~case was given recently in 
\cite{BertaBrandaoHirche2017}.

\begin{prop}\label{prop:general lower bounds}
For the general binary state discrimination problem 
$H_0:\,(\N_n)_{n\in\bN}$ vs.~$H_1:\,(\A_n)_{n\in\bN}$, we have 
\begin{align}
\cli((\N_n)_{n\in\bN}\|(\A_n)_{n\in\bN})&\ge
\liminf_{n\to+\infty}\frac{1}{n}\sup_{\alpha\in(0,1)}(1-\alpha)D_{\alpha}(\co(\N_n)\|\co(\A_n))
\label{chernoff general lower}\\
&\ge
\liminf_{n\to+\infty}\frac{1}{n}C(\cco(\N_n)\|\cco(\A_n)),\label{chernoff general lower2}\\
\dli_r((\N_n)_{n\in\bN}\|(\A_n)_{n\in\bN})&\ge 
\liminf_{n\to+\infty}\sup_{\alpha\in(0,1)}\frac{\alpha-1}{\alpha}\left[r-\frac{1}{n}
D_{\alpha}(\co(\N_n)\|\co(\A_n))\right]\label{direct general lower1}\\
&\ge
\liminf_{n\to+\infty}\sup_{\alpha\in(0,1)}\frac{\alpha-1}{\alpha}\left[r-\frac{1}{n}D_{\alpha}(\cco(\N_n)\|\cco(\A_n))\right]\label{direct general lower2}
\\
&=
\liminf_{n\to+\infty}\frac{1}{n}H_{nr}(\cco(\N_n)\|\cco(\A_n)),\ds\ds\ds r>0,
\label{direct general lower3}\\
\sli((\N_n)_{n\in\bN}\|(\A_n)_{n\in\bN})&\ge
\sup_{\alpha\in(0,1)}\liminf_{n\to+\infty}\frac{1}{n}D_{\alpha}(\co(\N_n)\|\co(\A_n)).
\label{stein general lower}
\end{align} 
\end{prop}
\begin{proof}
The following argument is quite standard, but we explain it in detail for readers'
convenience.
For any $c\in\bR$, we have
\begin{align*}
\inf_{T\in(\M_n)_{[0,I]}}\bz\alpha_n(\N_n|T)+e^{nc}\beta_n(\A_n|T)\jz
&=
\inf_{T\in(\M_n)_{[0,I]}}\bz\alpha_n(\co(\N_n)|T)+e^{nc}\beta_n(\co(\A_n)|T)\jz\\
&=
\inf_{T\in(\M_n)_{[0,I]}}\sup_{\rho\in\co(\N_n),\sigma\in\co(\A_n)}
\bz\rho(I-T)+e^{nc}\sigma(T)\jz\\
&=
\min_{T\in(\M_n)_{[0,I]}}\sup_{\rho\in\co(\N_n),\sigma\in\co(\A_n)}
\bz\rho(I-T)+e^{nc}\sigma(T)\jz\\
&=
\sup_{\rho\in\co(\N_n),\sigma\in\co(\A_n)}\min_{T\in(\M_n)_{[0,I]}}
\bz\rho(I-T)+e^{nc}\sigma(T)\jz\\
&\le
\sup_{\rho\in\co(\N_n),\sigma\in\co(\A_n)}
e^{nc(1-\alpha)}e^{(\alpha-1)D_{\alpha}(\rho\|\sigma)},&\alpha\in(0,1),\\
&=
e^{nc(1-\alpha)+(\alpha-1)D_{\alpha}(\co(\N_n)\|\co(\A_n))},&\alpha\in(0,1).
\end{align*}
The first equality above follows as $(\rho,\sigma)\mapsto\alpha_n(\N_n|T)+e^{nc}\beta_n(\A_n|T)$ is affine, and the second equality is by definition.
Note that $T\mapsto\rho(I-T)+e^{nc}\sigma(T)$ is continuous in the 
$w^*$-topology, and hence its supremum over $\co(\N_n)\times\co(\A_n)$ is lower 
semi-continuous, and therefore it attains its infimum on the $w^*$-compact set 
$(\M_n)_{[0,I]}$ (see Lemma \ref{lemma:tests compact}). 
This proves the third equality. The fourth equality follows by the Kneser-Fan minimax theorem 
\cite{Fan,Kneser}. The inequality is due to Audenaert's inequality \cite{AudChernoff}, and its extension to 
von Neumann algebras \cite{JOPS,Ogata-PS}, and the last equality is obvious.

Taking $c=0$, we obtain the existence of a test $T_n$ such that 
\begin{align*}
\max\{\alpha_n(\N_n|T_n),\beta_n(\A_n|T_n)\}
\le
\alpha_n(\N_n|T_n)+\beta_n(\A_n|T_n)
\le
e^{\inf_{\alpha\in(0,1)}(\alpha-1)D_{\alpha}(\co(\N_n)\|\co(\A_n))},
\end{align*}
from which \eqref{chernoff general lower} follows immediately. 
Moreover, we have 
\begin{align*}
\inf_{\alpha\in(0,1)}(\alpha-1)D_{\alpha}(\co(\N_n)\|\co(\A_n))
&\le
\inf_{\alpha\in(0,1)}(\alpha-1)D_{\alpha}(\cco(\N_n)\|\cco(\A_n))\\
&=
\inf_{\alpha\in(0,1)}\sup_{\rho\in\cco(\N_n),\sigma\in\cco(\A_n)}
\underbrace{(\alpha-1)D_{\alpha}(\rho\|\sigma)}_{=\psi(\rho\|\sigma|\alpha)}\\
&=
\sup_{\rho\in\cco(\N_n),\sigma\in\cco(\A_n)}\underbrace{\inf_{\alpha\in(0,1)}
(\alpha-1)D_{\alpha}(\rho\|\sigma)}_{=-C(\rho\|\sigma)}\\
&=
-C(\cco(\N)\|\cco(\A)),
\end{align*}
where the second equality follows again from the Kneser-Fan minimax theorem, since 
$\psi(\rho\|\sigma|\alpha)$ is convex in $\alpha$, and concave in $(\rho,\sigma)$, according 
to Lemmas \ref{lemma:convex lsc} and \ref{lemma:Renyi limits}.  
This yields \eqref{chernoff general lower2}.

Let us now return to the case of a general $c$. By the above argument,
for any $c\in\bR$ and any $\alpha\in(0,1)$, there exists a test $T_{n,c,\alpha}$ such that 
\begin{align*}
\alpha_n(\N|T_{n,c,\alpha})\le e^{nc(1-\alpha)+(\alpha-1)D_{\alpha}(\co(\N_n)\|\co(\A_n))},
\ds\ds\ds\ds\ds
\beta_n(\N|T_{n,c,\alpha})\le e^{-nc\alpha+(\alpha-1)D_{\alpha}(\co(\N_n)\|\co(\A_n))}.
\end{align*}
Choosing $c(r,\alpha):=\frac{r-(1-\alpha)D_{\alpha}(\co(\N_n)\|\co(\A_n))}{\alpha}$ yields, 
with $T_{n,\alpha}:=T_{n,c(r,\alpha),\alpha}$
\begin{align*}
\alpha_n(\N|T_{n,\alpha})\le e^{-n\frac{\alpha-1}{\alpha}\left[r-\frac{1}{n}D_{\alpha}(\co(\N_n)\|\co(\A_n))\right]},
\ds\ds\ds\ds\ds
\beta_n(\N|T_{n,\alpha})\le e^{-nr}.
\end{align*}
The above argument fails only if 
$D_{\alpha}(\co(\N_n)\|\co(\A_n))=+\infty$;
however, in that case
$\rho_n\perp\sigma_n$ for all $\rho_n\in\co(\N_n)$ and $\sigma_n\in\co(\A_n)$, 
and hence we may choose $T_{n,\alpha}$ to be any test that perfectly distinguishes 
$\co(\N_n)$ and $\co(\A_n)$. 

Now, for any $\gamma$ smaller than the RHS of \eqref{direct general lower1}, and every large enough $n$, we have 
\begin{align*}
\sup_{\alpha\in(0,1)}\frac{\alpha-1}{\alpha}\left[r-\frac{1}{n}D_{\alpha}(\co(\N_n)\|\co(\A_n))\right]>\gamma,
\end{align*}
 and hence there exists an $\alpha_{\gamma,n}\in(0,1)$ such that 
$\frac{\alpha_{\gamma,n}-1}{\alpha_{\gamma,n}}\left[r-\frac{1}{n}D_{\alpha_{\gamma,n}}(\co(\N_n)\|\co(\A_n))\right]>\gamma$.
Thus, with the test sequence $T_n:=T_{n,\alpha_{\gamma,n}}$, we get
\begin{align*}
\liminf_{n\to+\infty}-\frac{1}{n}\log\beta_n(\A|T_n)\ge r,\ds\ds\ds
\liminf_{n\to+\infty}-\frac{1}{n}\log\alpha_n(\N|T_n)\ge \gamma,\ds\ds\ds\text{whence}\ds\ds\ds
\dli_r(\N\|\A)\ge \gamma.
\end{align*}
Taking the supremum over all $\gamma$ as above yields the inequality in 
\eqref{direct general lower1}. Finally, the inequality in \eqref{direct general lower2}
is trivial, and the equality in \eqref{direct general lower3} follows by Lemma \ref{lemma:Hr identity}.

Let us finally consider \eqref{stein general lower}. If 
the RHS of \eqref{stein general lower} is zero then there is nothing to prove, and hence for the rest we assume that it is strictly positive. 
Consider any $0< r<\sup_{\alpha\in(0,1)}\liminf_{n\to+\infty}\frac{1}{n}D_{\alpha}(\co(\N_n)\|\co(\A_n))$. Then 
\begin{align*}
\dli_r((\N_n)_{n\in\bN}\|(\A_n)_{n\in\bN})&\ge \liminf_{n\to+\infty}\sup_{\alpha\in(0,1)}\frac{\alpha-1}{\alpha}\left[r-\frac{1}{n}D_{\alpha}(\co(\N_n)\|\co(\A_n))\right]\\
&\ge
\sup_{\alpha\in(0,1)}\frac{\alpha-1}{\alpha}\left[r-\liminf_{n\to+\infty}\frac{1}{n}D_{\alpha}(\co(\N_n)\|\co(\A_n))\right]>0,
\end{align*}
where the first inequality is by \eqref{direct general lower1}, the second inequality is obvious,
and the third inequality is immediate from our assumption. 
Thus, $\sli((\N_n)_{n\in\bN}\|(\A_n)_{n\in\bN})>r$ for any $r$ as above, and taking the supremum over every such $r$ yields \eqref{stein general lower}.
\end{proof}

\section{Classical state discrimination}
\label{sec:classical}

\subsection{Countably many hypotheses: Equality holds for the relaxed exponents}
\label{sec:finite classical}

\begin{lemma}\label{lemma:maxmin test}
Let $\M$ be a commutative von Neumann algebra, and $T_{i,j}\in\M_{[0,I]}$ be tests for every $i\in\{1,\ldots, k\}$ and $j\in\{1,\ldots, m\}$. Then there exists a test $T\in \M_{[0,I]}$
such that 
$$
\forall i: I-T \leq \sum_{j=1}^m (I-T_{i,j}),
\quad {\rm and} \quad
\forall j:\;
T\leq \sum_{i=1}^k T_{i,j}.
$$
\end{lemma}
\begin{proof}
Without loss of generality we may assume that $\M=L^\infty(\X,\F,\mu)$ for some measure space $(\X,\F,\mu)$. Then $T_{i,j}$ are nonnegative functions on $\X$
and 
$$
T(x):=\max_i\min_j T_{i,j}(x),\quad x\in X,
$$
defines a test $T\in\M_{[0,I]}$. For any $i\in\{1,\ldots, k\}$ and $x\in \X$ we have
$T(x)\geq \min_jT_{i,j}(x)$, and hence
$$
1-T(x)\leq 1-\min_j T_{i,j}(x) = \max_j(1-T_{i,j}(x)) \leq \sum_{j=1}^m (1-T_{i,j}(x)), 
$$
implying that $I-T\leq \sum_{j=1}^m (I-T_{i,j})$. The other inequality can be shown in a similar manner. 
\end{proof}

\begin{theorem}\label{thm:classical countable equality}
Let $(\M_n)_{n\in\bN}$ be a sequence of commutative von Neumann algebras, 
$I,J$ be countable index sets, and for every $n\in\bN$, 
$\N_n,\A_n,\{\rho_{n,i}\}_{i\in I},\{\sigma_{n,j}\}_{j\in J}\subseteq\S(\M_n)$. Then 
\begin{align}
\e^{\{0\}}\bz \bz\{\rho_{n,i}\}_{i\in I}\jz_{n\in\bN}\|(\A_n)_{n\in\bN}\jz&=
\inf_i\e\bz \bz\rho_{n,i}\jz_{n\in\bN}\|(\A_n)_{n\in\bN}\jz,
\label{countable cl1}\\
\e^{\{1\}}\bz(\N_n)_{n\in\bN}\| \bz\{\sigma_{n,i}\}_{i\in I}\jz_{n\in\bN}\jz&=
\inf_j\e\bz (\N_n)_{n\in\bN}\|(\sigma_{n,j})_{n\in\bN}\jz,
\label{countable cl2}\\
\e^{\{0,1\}}\bz \bz\{\rho_{n,i}\}_{i\in I}\jz_{n\in\bN}\| \bz\{\sigma_{n,i}\}_{i\in I}\jz_{n\in\bN}\jz&=
\inf_{i,j}\e\bz  \bz\rho_{n,i}\jz_{n\in\bN}\|(\sigma_{n,j})_{n\in\bN}\jz,
\label{countable cl3}
\end{align}
where $\e$ may be any of the error exponents $\e=\sli,\cli$, or $\dli_r$, $r>0$.
Moreover, the exponents 
$\e^{\{0\}}$/$\e^{\{1\}}$/$\e^{\{0,1\}}$
on the left hand sides above coincide with $\e$ if 
$I$ is finite/$J$ is finite/both $I$ and $J$ are finite, respectively. 
\end{theorem}
\begin{proof}
Again, we may assume that for every $n\in\bN$, $\M_n=L^\infty(\X_n,\F_n,\mu_n)$ for some measure space $(\X_n,\F_n,\mu_n)$.
Let us prove \eqref{countable cl3} under the assumption that $I=J=\bN$; the rest of the 
statements follow in a similar way. Note that we have LHS$\le$RHS in 
\eqref{countable cl3}, and hence we only need to prove the converse inequality. 
Moreover, it is enough to prove the converse inequality for 
$e=d_r$, $r\in(0,+\infty)$, according to 
Lemmas \ref{lemma:intersection} and \ref{lemma:Stein from directexp}.

Thus, let $\e=\dli_r$ for some $r\in(0,+\infty)$, and let $r'$ be the RHS of \eqref{countable cl3}. 
By assumption, for every $\delta>0$ and every $i,j\in \bN\times\bN$, there exists a sequence of tests
$(T_{\delta,i,j,n})_{n\in\bN}$ such that 
\begin{align*}
\liminf_{n\to+\infty}-\frac{1}{n}\log\rho_{n,i}(I-T_{\delta,i,j,n})> r'-\delta,\ds\ds\ds
\liminf_{n\to+\infty}-\frac{1}{n}\log\sigma_{n,i}T_{\delta,i,j,n}> r-\delta.
\end{align*}
For every $n\in\bN$, let
$k_{\delta,n}:=\max\{k\in [n]:\,\rho_{m,i}(I-T_{\delta,i,j,m})<e^{-m(r'-\delta)},\ds
\sigma_{m,j}(T_{\delta,i,j,m})<e^{-m(r-\delta)},\s \s m\ge n,\s i,j\in[k]\}$, and define
$T_{\delta,n}:=\max_{i\in [k_{\delta,n}]}\min_{j\in [k_{\delta,n}]}T_{\delta,i,j,n}(x)$.
As shown in the proof of Lemma \ref{lemma:maxmin test}, 
\begin{align}\label{classical countable proof1}
\forall i\in[k_{\delta,n}]:\ds 
I-T_{\delta,n}\le \sum_{k=1}^{k_{\delta,n}}(I-T_{\delta,i,k,n}),
\ds\ds\ds\ds\ds
\forall j\in[k_{\delta,n}]:\ds 
T_{\delta,n}\le\sum_{k=1}^{k_{\delta,n}}T_{\delta,k,j,n}.
\end{align}
By assumption, $\lim_{n\to+\infty}k_{\delta,n}=+\infty$, 
and hence for every $(i,j)\in\bN\times\bN$, there exists an 
$N_{i,j}$ such that for all $n\ge N_{i,j}$, $i,j\in [k_{\delta,n}]$, and thus, by \eqref{classical countable proof1},
\begin{align*}
\rho_{n,i}(I-T_{\delta,n})\le ne^{-n(r'-\delta)},\ds\ds\ds
\sigma_{n,j}(T_{\delta,n})\le ne^{-n(r-\delta)}.
\end{align*}
This shows that $(r-\delta,r'-\delta)\in \ac^{\{0,1\}}\bz \bz\{\rho_{n,i}\}_{i\in I}\jz_{n\in\bN}\| \bz\{\sigma_{n,i}\}_{i\in I}\jz_{n\in\bN}\jz$. Since this holds for every $\delta>0$, 
Lemma \ref{lemma:achievable limit} yields that 
$(r,r')\in \ac^{\{0,1\}}\bz \bz\{\rho_{n,i}\}_{i\in I}\jz_{n\in\bN}\| \bz\{\sigma_{n,i}\}_{i\in I}\jz_{n\in\bN}\jz$, and therefore
$\dli_r^{\{0,1\}}\bz \bz\{\rho_{n,i}\}_{i\in I}\jz_{n\in\bN}\| \bz\{\sigma_{n,i}\}_{i\in I}\jz_{n\in\bN}\jz\ge r'$.
\end{proof}

\subsection{Infinite dimension with countably infinitely many alternative hypotheses: Equality may not hold}
\label{sec:infinite classical}

In this section we show that the composite hypothesis testing error exponents may be 
strictly smaller than the worst pairwise error exponents even in the classical case, provided that the system is allowed to be infinite-dimensional, and the alternative hypothesis of infinite cardinality. More precisely, we prove the following:

\begin{theorem}\label{thm:classical ce}
There exists a probability density function $\rho$ on $[0,1]$ (w.r.t.~the Lebesgue measure),
and a probability distribution $q$ supported on a countably infinite set of probability 
density functions on $[0,1]$, such that 
\begin{align}\label{classical ce1}
\e(\rho\|\supp q)\le\e(\rho\|q)<\inf_{\sigma\in\supp q}\e(\rho\|\sigma)
\end{align}
for any of the Stein-, Chernoff-, and the direct exponents. Moreover, the first inequality above is also strict for the Stein-, Chernoff-, and the direct exponents $d_r$ with $0<r<D(\rho\|q)$.
\end{theorem}

\begin{rem}
Recall that probability density functions on $[0,1]$ represent states on the commutative 
infinite-dimensional von Neumann algebra $L^{\infty}([0,1])$; see Section \ref{sec:Neumann algebras}.
\end{rem}
We start with describing our example first; Theorem \ref{thm:classical ce} will then follow from the more precise statement in Proposition \ref{prop:classical ce}.
\medskip

Assume that we receive $n$ real numbers drawn independently from the interval $[0,1]$, all with the same distribution. Our null hypothesis is that this distribution is the uniform one. The alternative hypothesis is that first a random natural number $N$ was generated with probabilities 
$$
``{\rm prob.\; of \;} N=k" = \frac{6}{\pi^2} \frac{1}{k^2} =: q_k,
$$
and then the $n$ real numbers were drawn independently from the interval $[0,1]$ with distribution $\mu_k$, where $k$ is the obtained value of $N$. The measure $\mu_k$ is given by the density function (w.r.t.\! the Lebesgue measure, i.e.\! the uniform distribution) $2\cdot \egy_{H_k}$, where $H_1=[0,\frac{1}{2}], H_2=[0,\frac{1}{4}]\cup [\frac{1}{2},\frac{3}{4}]$, and in general
$$
H_k = \bigcup_{j=0}^{2^{k-1}-1}\,  \left[\frac{j}{2^{k-1}},\frac{j}{2^{k-1}}+\frac{1}{2^k}\right],
$$
where $\egy_{H_k}$ stands for the characteristic function (or indicator function) of $H_k$.
Note that the factor ${6}/{\pi^2}$  is there to make the
sum of probabilities $\sum_{k=1}^\infty q_k = 1$.

Using the notations and conventions explained in the preliminaries, 
for a fixed number of copies $n$, our null hypothesis is 
represented by the single state $\rho^{\otimes n}=\egy_{[0,1]}^{\otimes n} = \egy_{[0,1]^n}$, whereas 
our alternative hypothesis may be represented by the convex combination of i.i.d. states 
$$
\sigma^{(n)}:=\sum_{k=1}^\infty q_k \sigma^{\otimes
n}_k = \sum_{k=1}^\infty q_k (2\cdot \egy_{H_k})^{\otimes
n} = \frac{6}{\pi^2}\sum_{k=1}^\infty \frac{2^n}{k^2} \egy_{H_k^n}
$$
where $H_k^n = H_k \times \ldots \times H_k$ is the $n$-fold Cartesian product of $H_k$ with itself. Our aim is to contrast the error-exponents of this problem with the worst one we would get if our null and alternate hypotheses were just $\rho^{\otimes n}$ and $\sigma_k^{\otimes n}$ for some $k \in \mathbb N$.
For this, we obtain the following:

\begin{prop}\label{prop:classical ce}
In the above example, we have
\begin{enumerate}
\item
\ds $\sli(\rho\|q)= \log 2<\infty =\sli(\rho\|\sigma_k)$;
\item
\ds $\dli_r(\rho\|q)= {\rm max}\{\log 2-r,0\} <\log 2 =\dli_r(\rho\|\sigma_k)$,
\ds $r>0$;
\item
\ds $\cli(\rho\|q)= \frac{1}{2}\log 2<\log 2  = \cli(\rho\|\sigma_k)$;
\end{enumerate}
for every $k\in\bN$. 
\end{prop}

\begin{proof}
Let us begin by noting that $\lambda(H_k)$ -- i.e., the Lebesgue measure of the set $H_k$ -- is always $1/2$, regardless of the value of $k$. Since $\rho =\lambda$ is the uniform distribution, this immediately implies that the ``$\rho^{\otimes n}$ {\it versus} $\sigma_k^{\otimes n}$'' cases are all equivalent in the sense that none of the exponents depend on $k$. Indeed, we have 
\begin{align*}
\psi(\rho\|\sigma_k|\alpha)=
\log\int_{[0,1]}\bz\egy_{[0,1]}(x)\jz^{\alpha}\bz 2\egy_{H_k}(x)\jz^{1-\alpha}\,d\lambda(x)
=\log \bz 2^{1-\alpha}\int_{[0,1]}\egy_{H_k}(x)\,d\lambda(x)\jz=-\alpha\log 2,
\end{align*}
and Lemma \ref{lemma:known exponents} yields
\begin{align*}
\dli_r(\rho\|\sigma_k) = \cli(\rho\|\sigma_k) ={\rm log}(2),\ds\ds\ds r>0,\ds k\in\bN,
\end{align*}
while $\supp\rho\nsubseteq\supp\sigma_k$ implies
\begin{align*}
\sli(\rho\|\sigma_k)=+\infty,\ds\ds\ds k\in\bN.
\end{align*}
We remark 
that any of these exponents remain unchanged if we replace $\rho$ and $\sigma_k$ by the $2\times 2$ diagonal density matrices
in the following manner:
$$
\rho=\egy_{[0,1]} \Longleftrightarrow  
\left(\begin{matrix}\frac{1}{2} & 0 \\ 0 & \frac{1}{2}\end{matrix}\right),
\;\;\;\;{\rm and}\;\;\;\;\sigma_k = 2\cdot \egy_{H_k} \Longleftrightarrow
2 \left(\begin{matrix}\frac{1}{2} & 0 \\ 0 & 0\end{matrix}\right).
$$

The real challenge, of course, is not the computation of the error exponents of the
``one-versus-one'' cases, but rather, the estimation of the ``one versus a convex combination'' case. 
As a first step, observe that, by construction
$$
\lambda(H_k \cap H_j) = \frac{1}{4} = \frac{1}{2} \cdot \frac{1}{2} = \lambda(H_k)\,\lambda(H_j)
$$
for all $k\neq j$. Thus, the subsets $H_1,H_2,\ldots$ (as events) are all independent 
with respect to the uniform distribution. It follows that
\begin{equation}
\label{eq:volofun}
\lambda \bz\bigcup_{k=1}^m H_k^n\jz = 1-\lambda\bz\bigcap_{k=1}^m (H_k^{n})^c\jz
= 1 - \prod_{k=1}^m \bz1-\lambda(H_k^n)\jz =
1 - \bz 1-\frac{1}{2^n}\jz^m
\end{equation}
where, with a slight abuse of notation, we use $\lambda$ for the Lebesgue measure both on the interval $[0,1]$ as well as on the hypercube $[0,1]^n$, and $(H_k^n)^c$ denotes the complement $[0,1]^n\setminus H_k^n$.

According to Lemma \ref{lemma:projective enough}, in the study of the error exponents, it is sufficient to consider projective tests only. Let $T_n$ be a projective test in 
$L^\infty([0,1])^{\otimes n} \equiv  L^\infty([0,1]^n)$; that is, $T_n=\egy_{K_n}$ where $K_n$ is a (Lebesgue) measurable subset of $[0,1]^n$. Then the type I error is 
\begin{equation}
\label{eq:type1}
\alpha_n (T_n) = \int_{[0,1]^n}\rho^{\otimes n} (1-\egy_{K_n}) d\lambda
 =
1-\lambda(K_n) = \lambda(K_n^c)
\end{equation}
is simply the {\it volume} of the complement of $K_n$. The type II error is
\begin{equation}
\label{eq:type2}
\beta_n (T_n) = \int_{[0,1]^n} \sum_{k=1}^\infty q_k \sigma_k^{\otimes n}\, \egy_{K_n}\,  d\lambda 
= 2^n \sum_{k=1}^\infty q_k\,  \lambda(H_k^n\cap K_n),
\end{equation}
which we shall {\it lower bound} by 1) replacing the summation over $k$ from $1$ to $\infty$ by a summation over $k$ from $1$ to a certain value $m$, 2) by replacing the individual coefficients $q_k$ ($k=1,\ldots,m$) by ${\rm min}\{q_k| k=1,\ldots,m\} = q_m$, and 3) by replacing a sum of volumes by the volume of the union:
\begin{equation}
\label{eq:classinfbeta}
\beta_n (T_n) \geq 2^n \sum_{k=1}^m q_k\,  \lambda(H_k^n\cap K_n)
\geq 2^n q_m \sum_{k=1}^m \lambda(H_k^n\cap K_n)
\geq 2^n q_m \lambda\bz\bigcup_{k=1}^m \bz H_k^n\cap K_n\jz\jz.
\end{equation}
Since $\bigcup_{k=1}^m (H_k^n\cap K_n)=(\cup_{k=1}^m H_k^n)\setminus K_n^c$, we can further bound the volume of the union in question as follows:
$$
\lambda\bz\bigcup_{k=1}^m (H_k^n\cap K_n)\jz
= \lambda\bz\bz\bigcup_{k=1}^m H_k^n\jz\setminus K_n^c\jz
\geq \lambda\bz\bigcup_{k=1}^m H_k^n\jz - \lambda(K_n^c).
$$
Substituting the above into (\ref{eq:classinfbeta}) together with the actual value of the coefficient $q_m$, the volume of $\cup_{k=1}^m H_k^n$ given in \ref{eq:volofun}, and the equality $\lambda(K_n^c)=\alpha_n(T_n)$, we obtain that
\begin{equation}
\label{eq:classinftradeoff1}
\beta_n (T_n) \geq 2^n \frac{6}{\pi^2}\frac{1}{m^2}\bz 1-\bz 1-\frac{1}{2^n}\jz^m -\alpha_n(T_n)\jz
\end{equation}
for any $m\in \mathbb N$. We may view this as a sequence of {\it trade-off relations}: for any given value of $m$, it 
says that a certain $m$-dependent weighted combination of the errors must be larger than some ($m$-dependent) value. 

The sequence $k\mapsto (1-1/k)^k$ is monotone increasing and converges to $1/e$. Moreover, using elementary calculus one can easily check that $1-e^{-x} \geq x/2 $ on the interval $x\in [0,1]$. So, for $m\leq 2^n$,
 the term $1-(1-1/2^n)^m$ in (\ref{eq:classinftradeoff1}) may be bounded from below as follows:
$$
1-\bz 1-\frac{1}{2^n}\jz^m = 
1-\bz\bz 1-\frac{1}{2^n}\jz^{2^n}\jz^{m 2^{-n}} 
\geq 1-e^{-m 2^{-n}} \geq \frac{1}{2} m 2^{-n}.
$$
Substituting this into (\ref{eq:classinftradeoff1}), after some reordering we get that 
\begin{align}
\label{eq:classinftradeoff2}
2^{(n+1)}\alpha_n(T_n)+\frac{\pi^2}{3} m^2 \beta_n(T_n)\geq m
\end{align}
for any $m\leq 2^n$. The choice $m:=2^n$ yields
\begin{align}
\label{eq:classinftradeoff3}
2\alpha_n(T_n)+\frac{\pi^2}{3} 2^n \beta_n(T_n)\geq 1.
\end{align}
In particular, if 
$\alpha_n(T_n)<\frac{1}{3}$, then  $\beta_n(T_n) \geq \frac{1}{\pi^2} 2^{-n}$, implying that 
$\log 2\ge \sli(\rho\|q)$. Since $\sli(\rho\|q)\ge\sup\{r>0:\,\dli_r(\rho\|q)>0\}$, we also get that for $r>\log 2$,  $\dli(\rho\|q)=0=\max\{\log 2-r,0\}$. 

Consider now a fixed $0<r<\log 2$, and assume that $(T_n)_{n\in\bN}$ is a sequence of tests such that 
\begin{align*}
\liminf_{n\to+\infty}-\frac{1}{n}\log\beta_n(T_n)\ge r.
\end{align*}
By \eqref{eq:classinftradeoff3}, if $\beta_n(T_n)\le \frac{1}{2\pi^2}2^{-n}$ for infinitely many $n$ then $\alpha_n(T_n)\ge 5/12$ for the same indeces, and hence 
$\liminf_n-\frac{1}{n}\log\alpha_n(T_n)=0$. Hence, for the rest we assume that 
$\beta_n(T_n)> \frac{1}{2\pi^2}2^{-n}$ , and hence
\begin{align*}
m:=
\left \lceil\frac{1}{3 \pi^2\beta_n(T_n)}\right\rceil
\end{align*}
is a positive integer smaller than $2^n$, 
for all sufficiently large $n$. Therefore,
 we can apply (\ref{eq:classinftradeoff2}) 
to obtain
\begin{align*}
2^{(n+1)}\alpha_n(T_n)\beta_n(T_n)+
\frac{\pi^2}{3}\underbrace{(m\beta_n(T_n))^2}_{\le 1/(2\pi^2)^2}\geq 
\underbrace{m\beta_n(T_n)}_{\ge 1/(3\pi^2)}\ds\ds\imp\ds\ds
\alpha_n(T_n)\beta_n(T_n)\ge \pi^{-2}2^{-(n+3)}.
\end{align*}
Thus,
\begin{align*}
r+\limsup_{n\to+\infty}-\frac{1}{n}\log\alpha_n(T_n)
\le
\limsup_{n\to+\infty}-\frac{1}{n}\log(\alpha_n(T_n)\beta_n(T_n))
\le
\log 2.
\end{align*}
This proves that $\dli_r(\rho\|q)\le\log 2-r$. Monotonicity of $r\mapsto \dli_r(\rho\|q)$ yields that $\dli_r(\rho\|q)\le\log 2-r=0$ for $r=\log 2$.

So far we have bounded the error exponents from above. To be able to deduce their precise value, we also need to bound them from below which we shall do by presenting an actual series of tests. Consider the projective tests $T_n:=\egy_{K_n}$ where $K_n$ is the complement of $\cup_{k\le m_n} H_k^n$ in $[0,1]^n$ and $n\mapsto m_n\in \mathbb N$ is some function of $n$.
Using that $K_n\cap H_k^n=\emptyset$ for $k\le m_n$ and the volume formula (\ref{eq:volofun}), for our specific test, the type I and II errors given by (\ref{eq:type1}) and (\ref{eq:type2}) simplify to
\begin{align*}
\alpha_n(T_n) = 1-\bz 1-\frac{1}{2^n}\jz^{m_n}\!,\;\;\;\;
{\rm and}\;\;\;\;
\beta_n(T_n) = 2^n \sum_{k=m_n+1}^\infty q_k\,  \lambda(K_n\cap H_k^n).
\end{align*}
This time we bound the errors from above. Since $\lambda(K_n\cap H_k^n)\leq \lambda(H_k^n)=2^{-n}$, the type II error can be estimated as follows:
$$
\beta_n(T_n)\leq \sum_{k=m_n+1}^\infty q_k =
\frac{6}{\pi^2}\sum_{k=m_n+1}^\infty \frac{1}{k^2}\leq
\frac{6}{\pi^2}\int_{m_n}^\infty \frac{1}{x^2}dx = 
\frac{6}{\pi^2m_n}.
$$
We now specifically choose $m_n$ to be the upper integer part of $e^{n r}$ for some $r>0$. Then, by the derived bound on $\beta_n(T_n)$, we find that 
\begin{align}\label{CL counter proof1}
\liminf_n -\frac{1}{n}\log\beta_n(T_n)\geq r.
\end{align}
On the other hand, for the type I error we have
\begin{align*}
\alpha_n(T_n)=1-(1-2^{-n})^{\ceil{e^{nr}}}
&=
2^{-n}\sum_{k=0}^{\ceil{e^{nr}}-1}(1-2^{-n})^k\le\ceil{e^{nr}}2^{-n},
\end{align*}
and hence
\begin{align}\label{CL counter proof2}
\liminf_{n\to+\infty}-\frac{1}{n}\log\alpha_n(T_n)\ge \log 2-r.
\end{align}
By \eqref{CL counter proof1} and \eqref{CL counter proof2},
$\dli_r(\rho\|q)\ge \log 2-r$ for all $r>0$.
Combining this with the previous upper bound and taking account
of the nonnegativity of $\dli_r$ yields 
$\dli_r(\rho\|q)=\max\{ \log 2-r,0\}$ for all $r>0$.
Using this expression and the derived bound on the Stein exponent, we can now also confirm the claimed values for the Stein and Chernoff exponents (see Remark \ref{rem:length}).
\end{proof}

\begin{prop}
In the above example, we have
$\e(\rho\|\supp q)=0$ for any of the exponents 
$\e=\sli,\cli,\dli_r \; (r>0)$.
\end{prop}
\begin{proof}
Assume again that $T_n=\egy_{K_n}$ for some measurable subset $K_n\subset [0,1]^n$. As in the previous proof, we have that
$\alpha_n(T_n)=1-\lambda(K_n)=\lambda(K_n^c)$. What changes is the
type two error; this time
$$
\beta_n (T_n) = \sup_k \int_{[0,1]^n} \sigma_k^{\otimes n}\, \egy_{K_n}\,  d\lambda
= \sup_k 2^n \lambda(H_k^n\cap K_n). 
$$
We shall lower bound this by 1) replacing the supremum by the average of the first $2^n$ terms and 2) replacing the sum of volumes by the volume of the union:
$$
\beta_n (T_n) \geq \frac{1}{2^n} \sum_{k=1}^{2^n} 2^n  \lambda(H_k^n\cap K_n)\geq  \lambda\bz\bigcup_{k=1}^{2^n} (H_k^n\cap K_n)\jz = \lambda\bz\bz\bigcup_{k=1}^{2^n} H_k^n\jz\setminus K_n^c\jz.
$$
Thus for the sum of the two types of errors we have
$$
\alpha_n(T_n)+\beta_n(T_n)\geq \lambda(K_n^c)+\lambda\bz\bz\bigcup_{k=1}^{2^n} H_k^n\jz\setminus K_n^c\jz \geq \lambda\bz\bigcup_{k=1}^{2^n} H_k^n\jz=1-\bz 1-\frac{1}{2^n}\jz^{2^n}.
$$
Since the expression on the right tends to $1-1/e$ as $n\to \infty$, it follows that the two types of errors cannot both converge to zero, from where our claim follows immediately.  
\end{proof}

\subsection{Finite dimension: Equality under convexity}
\label{sec:classical exponents}

Consider now the composite i.i.d.~hypothesis testing problem with the null- and the alternative hypotheses represented by 
$\N,\A\subseteq\S(\M)$.
For a fixed number of copies $n$, the two hypotheses are represented by 
\begin{align*}
H_0:\ds\N^{\otimes n}:=\{\otimes_{k=1}^n\rho:\,\rho\in\N\}\ds\ds\ds
H_1:\ds\A^{\otimes n}:=\{\otimes_{k=1}^n\sigma:\,\sigma\in\A\}.
\end{align*}
The \ki{arbitrarily varying} version of the above problem is where the hypotheses are represented by 
\begin{align*}
H_0:\ds\N^{\otimes n}_{\av}:=\{\otimes_{k=1}^n\rho_k:\,\rho_k\in\N,\,k\in[n]\}\ds\ds\ds
H_1:\ds\A^{\otimes n}_{\av}:=\{\otimes_{k=1}^n\sigma_k:\,\sigma_k\in\A,\,k\in[n]\}.
\end{align*}
That is, the problem is uniquely determined by the sets of states
$\N,\A$ on the single-copy algebra, and the state of the $n$-copy system is still a product state (independence), but the state of the $k$-th system can be an arbitrary member of the state set representing the given hypothesis. 
We denote the corresponding error probabilities and error exponents with a superscript $\av$.

In the finite-dimensional classical case, where the elements of $\N,\A$ are probability density functions on some finite set $\X$, one may also define the \ki{adversarial} version
\cite{AdvHyp}, given by 
\begin{align*}
H_0:&\ds\N^{(n)}_{\adv}:=
\{\rho\in\S(\X^n):\,\rho_1\in\N,\,
\forall k=1,\ldots,n-1,\,\forall \vecc{x}\in\X^k:\,\rho_{k+1}(\,.\,|x_1\ldots x_k)\in\N\},\\
H_1:&\ds\A^{(n)}_{\adv}:=
\{\sigma\in\S(\X^n):\,\sigma_1\in\A,\,
\forall k=1,\ldots,n-1,\,\forall \vecc{x}\in\X^k:\,\sigma_{k+1}(\,.\,|x_1\ldots x_k)\in\A\},
\end{align*}
where for a probability distribution $\omega$ on $\X^n$, 
$\omega_1(x):=\omega(\{x\}\times\X\ttimes\X)$, and for any 
$k\in\{1,\ldots,n-1\}$ and $x_1,\ldots,x_k\in\X$,
$\omega_{k+1}(\,.\,|x_1\ldots x_k)$ is the conditional distribution
\begin{align*}
\omega_{k+1}(x|x_1\ldots x_k):=
\frac{\omega(\{x_1\}\ttimes\{x_k\}\times\{x\}\times\X\ttimes\X)}
{\omega(\{x_1\}\ttimes\{x_k\}\times\X\ttimes\X)}.
\end{align*}

For a subset $A\subseteq\S(\X)$ of states, let $\co(A)$ denote the convex hull of $A$, 
and $\cco(A)$ the closure of the convex hull.

\begin{theorem}\label{thm:classical convex hyp}
Let $\X$ be a finite set and $\N,\A\subseteq\S(\X)$. Let 
$\e$ be any of the Stein-, Chernoff-, or the direct exponents, and 
$E$ be the corresponding divergence as in Lemma \ref{lemma:known exponents}. Then
\begin{align}
E(\cco(\N)\|\cco(\A)) 
&=\inf_{\rho\in\cco(\N),\sigma\in\cco(\A)}\e(\rho\|\sigma)\label{classical adversarial exponents4}\\
&
\le
\e^{\adv}(\cco(\N)\|\cco(\A))
\le
\e^{\adv}(\N\|\A)\label{classical adversarial exponents1}\\
&
\le \e^{\av}(\N\|\A)=
\e^{\av}(\cco(\N)\|\cco(\A))\label{classical adversarial exponents2}\\
&
\le
\e(\cco(\N)\|\cco(\A))
\le
\left\{\begin{array}{l}\e^{\{0\}}(\cco(\N)\|\cco(\A))\\ \e^{\{1\}}(\cco(\N)\|\cco(\A))\end{array}\right\}
\le \e^{\{01\}}(\cco(\N)\|\cco(\A))\\
&
\le
\inf_{\rho\in\cco(\N),\sigma\in\cco(\A)}\e(\rho\|\sigma),
\label{classical adversarial exponents3}
\end{align}
and hence all the inequalities above hold as equalities.
\end{theorem}
\begin{proof}
All the inequalities except for the first one in \eqref{classical adversarial exponents1}
are obvious from \eqref{bigger set smaller exponent}--\eqref{composite vs single bound1},
and the equality in \eqref{classical adversarial exponents4} is due to Lemma \ref{lemma:known exponents}.
The equality in \eqref{classical adversarial exponents2} is easily seen to follow 
as for any $\N_n,\A_n\subseteq\S(\X^n)$, and any test $T_n$,
\begin{align*}
\alpha_n(\cco(\N_n)|T_n)=\alpha_n(\N_n|T_n),\ds\ds\ds
\beta_n(\cco(\A_n)|T_n)=\beta_n(\A_n|T_n),
\end{align*}
and $(\cco(R))^{\otimes n}_{\av}\subseteq\cco(R^{\otimes n}_{\av})
\subseteq\cco\bz(\cco(R))^{\otimes n}_{\av}\jz$ implies, by taking the closed convex hull of each term, that 
\begin{align*}
\cco\bz(\cco(R))^{\otimes n}_{\av}\jz
=\cco(R^{\otimes n}_{\av}).
\end{align*}
Hence, the only thing left to be proved is the
first inequality in 
\eqref{classical adversarial exponents1}, which we establish separately below.
\end{proof}

We note that the Stein and the direct exponents 
were determined for the arbitrarily varying setting in
\cite{FuShen1996} and \cite{FuShen1998}, respectively, for finite $\N$ and $\A$.
Theorem \ref{thm:classical convex hyp} gives 
an extension of these results, on the one hand by treating $\N$ and $\A$ of arbitrary 
cardinality, and on the other hand by also determining the Chernoff exponent. We note that our proof is different from the proofs in \cite{FuShen1996} and \cite{FuShen1998}.

Regarding the first inequality in 
\eqref{classical adversarial exponents1},
we note that the cases of the Stein and the Chernoff exponent were already treated in 
\cite{AdvHyp}, and 
our proof is a straightforward adaptation of the proof for the Chernoff exponent in \cite{AdvHyp}. On the other hand, we not only extend the results of 
\cite{AdvHyp} to the case of the whole range of direct exponents, but
the proof presented below is also a simplification of that in \cite{AdvHyp}, where the statements for the Stein and the Chernoff exponents were proved separately, whereas we give a proof for the direct exponent
$\dli_r$ for every $r>0$, and obtain the cases of the Stein and the Chernoff exponents as immediate consequences. Moreover, we also utilize this proof method to show equality 
for the strong converse exponent, albeit only in the case where the null hypothesis is composite i.i.d.; see Section \ref{sec:sc adv}.

The main idea of the proof is simple: For any $0<r<D(\N\|\A)$, one takes the pair 
$\rho_r\in\N$, $\sigma_r\in\A$, that minimizes the distance between $\N$ and $\A$ in the 
$H_r$ divergence, and uses the optimal Neyman-Pearson test for discriminating the i.i.d.~extensions of these two states; the challenge is to show that this test is in fact universally good, and that it is so even in the adversarial setting.

\begin{theorem}\label{thm:classical convex hyp2}
The first inequality in 
\eqref{classical adversarial exponents1} holds.
\end{theorem}
\begin{proof}
Our aim is to prove the first inequality in \eqref{classical adversarial exponents1}, and hence we may assume without loss of generality that 
$\N$ and $\A$ are closed and convex. Hence, our aim is to prove
\begin{align}\label{adv exponents proof1}
\inf_{\rho\in\N,\sigma\in\A}\e(\rho\|\sigma)\le e^{\adv}(\N\|\A),
\end{align}
where $\e$ may be the Stein-, the Chernoff-, or any of the direct exponents. 
Moreover, it is sufficient to prove \eqref{adv exponents proof1} for the direct exponents, from which the 
inequalities for the Chernoff and the Stein exponents follow immediately
by Lemma \ref{lemma:intersection} and Lemma \ref{lemma:Stein from directexp}, respectively.

Let $r>0$ be a fixed type II rate, and 
$\e=\dli_r$ be the corresponding direct exponent, so that, by Lemma \ref{lemma:known exponents}, 
\begin{align*}
\inf_{\rho\in\N,\sigma\in\A}\e(\rho\|\sigma)
=H_r(\N\|\A).
\end{align*}
We assume that $r$ is such that $H_r(\N\|\A)>0$, since otherwise the statement is trivial. By Lemma \ref{lemma:Hr limits}, this is equivalent to assuming that $r<D(\N\|\A)$.
In particular, $D(\N\|\A)>0$, which is equivalent to $\N\cap\A=\emptyset$. 

To avoid potential difficulties arising from differing supports of the probability distributions, we introduce the ``smoothed'' sets
\begin{align}\label{smoothed hypotheses}
\N_{\theta}:=\left\{(1-\theta)\rho+\theta\frac{1}{|\X|}:\,\rho\in\N\right\},\ds\ds\ds\ds\ds
\A_{\theta}:=\left\{(1-\theta)\sigma+\theta\frac{1}{|\X|}:\,\sigma\in\N\right\},
\end{align}
for every $\theta\in[0,1)$.
Note that, by definition, for any 
$\rho_{[n]}\in\N^{(n)}_{\adv}$ there exists a 
$\tilde\rho_{[n]}\in(\N_{\theta})^{(n)}_{\adv}$ such that 
\begin{align*}
\tilde\rho_1(x)&=(1-\theta)\rho_1(x)+\theta\frac{1}{|\X|},\ds\ds\ds x\in\X,\\
\tilde\rho_{k+1}(\,.\,|x_1,\ldots,x_k)&=(1-\theta)\rho_{k+1}(\,.\,|x_1,\ldots,x_k)+\theta\frac{1}{|\X|},\ds\ds\ds k=2,\ldots,n-1,\,x_1,\ldots,x_k\in\X.
\end{align*} 
In particular, 
\begin{align}\label{smoothed upper bound}
\tilde\rho_{[n]}(x_1,\ldots,x_n)\ge (1-\theta)^n\rho_{[n]}(x_1,\ldots,x_n),\ds\ds\ds
x_1,\ldots,x_n\in\X,
\end{align}
and we have analogous inequalities for $\A$ in place of $\N$. This implies that for any 
test $T_n$, 
\begin{align*}
\alpha_n(T_n|\N^{(n)}_{\adv})\le(1-\theta)^{-n}
\alpha_n(T_n|(\N_{\theta})^{(n)}_{\adv}),\ds\ds\ds\ds\ds
\beta_n(T_n|\A^{(n)}_{\adv})\le(1-\theta)^{-n}
\beta_n(T_n|(\A_{\theta})^{(n)}_{\adv}).
\end{align*}

Clearly, $\N_{\theta}$ and $\A_{\theta}$ are again closed convex sets, 
and $\N_{\theta}\cap\A_{\theta}=\emptyset$.
Since the relative entropy is jointly lower semi-continuous, for every $\theta$ there exist
$\hat\rho_{\theta}\in\N_{\theta}$, $\hat\sigma_{\theta}\in\A_{\theta}$ such that 
$D(\hat\rho_{\theta}\|\hat\sigma_{\theta})=D(\N_{\theta}\|\A_{\theta})$. Using compactness of 
$\N$ and $\A$, we may choose a sequence $\theta_k\to0$ such that 
$\hat\rho_{\theta_k}\to\rho_0\in\N$, $\hat\sigma_{\theta_k}\to\sigma_0\in\A$ as
$k\to+\infty$.
Thus,
\begin{align}\label{relentr dist limit}
r<D(\N\|\A)\le D(\rho_0\|\sigma_0)\le
\liminf_{k\to+\infty}D(\hat\rho_{\theta_k}\|\hat\sigma_{\theta_k})
=
\liminf_{k\to+\infty}D(\N_{\theta_k}\|\A_{\theta_k}),
\end{align}
where the first inequality is by assumption, the second one is by definition, and the third 
one is due to the lower semi-continuity of the relative entropy.
Hence, $r<D(\N_{\theta_k}\|\A_{\theta_k})$ for all large enough $k$, and for the rest we assume that this is satisfied. Since $D_0(\rho\|\sigma)=0$ for all 
$\rho\in\N_{\theta}$, $\sigma\in\A_{\theta}$, Lemma \ref{lemma:Hr limits} implies that 
\begin{align*}
0<H_r(\rho\|\sigma)<+\infty,\ds\ds\ds \rho\in\N_{\theta_k},\,\sigma\in\A_{\theta_k}.
\end{align*}

Let us fix a large enough $k$ as above.
By Lemma \ref{lemma:convex lsc}, $H_r$ is lower semi-continuous on the compact set 
$\N_{\theta_k}\times\A_{\theta_k}$, and therefore there exists a pair of states 
$(\rho_r,\sigma_r)\in\N_{\theta_k}\times\A_{\theta_k}$ such that 
$H_r(\N_{\theta_k}\|\A_{\theta_k})=H_r(\rho_r\|\sigma_r)=:H_r$.
Let $\psi(\alpha):=\psi(\rho_r\|\sigma_r|\alpha)$, 
$\tilde\psi(u):=\tilde\psi(\rho_r\|\sigma_r|u)=(1-u)\psi((1-u)\inv)$, 
and let $\alpha_r$ and $u_r$ be as in 
Lemma \ref{lemma:simple Hr}.
For $\rho\in\N_{\theta_k}$, $\sigma\in\A_{\theta_k}$,
there exists a $\delta>0$ such that 
$\rho^{(t)}:=(1-t)\rho_r+t\rho$ and 
$\sigma^{(t)}:=(1-t)\sigma_r+t\sigma$ are probability density functions for every 
$t\in(-\delta,1)$. Hence, for all such $t$, we may define
\begin{align*}
f(t,u)&:=ur-\tilde\psi(\rho^{(t)}\|\sigma_r)=
ur-(1-u)\log\sum_x\rho^{(t)}(x)^{\frac{1}{1-u}}\sigma_r(x)^{1-\frac{1}{1-u}},\\
g(t,u)&:=ur-\tilde\psi(\rho_r\|\sigma^{(t)})=
ur-(1-u)\log\sum_x\rho_r(x)^{\frac{1}{1-u}}\sigma^{(t)}(x)^{1-\frac{1}{1-u}}.
\end{align*}
Then 
\begin{align*}
H_r(\rho^{(t)}\|\sigma_r)=\max_{u\in(-\infty,0)}f(t,u)=f(t,u_1(t)),\ds\ds\ds
H_r(\rho_r\|\sigma^{(t)})=\max_{u\in(-\infty,0)}g(t,u)=g(t,u_2(t)),
\end{align*}
where $u_1(t)$ and $u_2(t)$ are the unique maximizers.
Since $u_1(t)$ is the unique solution of $\partial_2 f(t,u)=0$, and 
$\partial_2^2 f(t,u)>0$ at every $u\in(-\infty,0)$, according to Lemma \ref{lemma:simple Hr},
the function $t\mapsto u_1(t)$ is differentiable due to the implicit function theorem, and similarly for $t\mapsto u_2(t)$. 
Thus,
\begin{align*}
\frac{d}{dt}H_r(\rho^{(t)}\|\sigma_r)&=
\partial_1 f(t,u_1(t))+\underbrace{\partial_2f(t,u_1(t))}_{=0}\frac{d}{dt}u_1(t)=
\partial_1 f(t,u_1(t))\\
&=
\frac{\sum_x\rho^{(t)}(x)^{\frac{1}{1-u_1(t)}-1}\sigma_r(x)^{1-\frac{1}{1-u_1(t)}}(\rho_r(x)-\rho(x))}
{\sum_x\rho^{(t)}(x)^{\frac{1}{1-u_1(t)}}\sigma_r(x)^{1-\frac{1}{1-u_1(t)}}}.
\end{align*}
By the minimizing property of $\rho_r$, we have
\begin{align}\label{LLR lower1}
0\le \frac{d}{dt} H_r(\rho^{(t)}\|\sigma_r)\Big\vert_{t=0}
\ds\imp\ds 
\sum_x\rho(x)\bz\frac{\sigma_r(x)}{\rho_r(x)}\jz^{1-\alpha_r}
\le
\sum_x\rho_r(x)\bz\frac{\sigma_r(x)}{\rho_r(x)}\jz^{1-\alpha_r},
\end{align}
where we used that $u_1(0)=u_2(0)=u_r=(\alpha_r-1)/\alpha_r$. A similar computation yields
\begin{align}
\frac{d}{dt}H_r(\rho_r\|\sigma^{(t)})&=
\partial_1 g(t,u_2(t))+\underbrace{\partial_2g(t,u_2(t))}_{=0}\frac{d}{dt}u_2(t)=
\partial_1 g(t,u_2(t))\nonumber\\
&=
\frac{u_2(t)\sum_x\rho_r(x)^{\frac{1}{1-u_2(t)}}\sigma^{(t)}(x)^{1-\frac{1}{1-u_2(t)}-1}
(\sigma(x)-\sigma_r(x))}{\sum_x\rho_r(x)^{\frac{1}{1-u_2(t)}}\sigma^{(t)}(x)^{1-\frac{1}{1-u_2(t)}}}
,\label{Hr derivative}
\end{align}
and
\begin{align}\label{LLR lower2}
0\le \frac{d}{dt}H_r(\rho_r\|\sigma^{(t)})\Big\vert_{t=0}
\ds\imp\ds 
\sum_x\sigma(x)\bz\frac{\rho_r(x)}{\sigma_r(x)}\jz^{\alpha_r}
\le
\sum_x\sigma_r(x)\bz\frac{\rho_r(x)}{\sigma_r(x)}\jz^{\alpha_r}.
\end{align}

Define $c_r:=\psi'(\alpha_r)$, and let 
\begin{align}\label{NP test}
T_{n,r}:=\left\{\vecc{x}\in\X^n:\,\frac{\rho_r^{\otimes n}(\vecc{x})}{\sigma_r^{\otimes n}(\vecc{x})}\ge e^{nc_r}\right\},\ds\text{so that}\ds
\X^n\setminus T_{n,r}=
\left\{\vecc{x}\in\X^n:\,\frac{\sigma_r^{\otimes n}(\vecc{x})}{\rho_r^{\otimes n}(\vecc{x})}> e^{-nc_r}\right\}.
\end{align}
As is usual in the classical case, we identify the subset $T_{n,r}\subseteq\X^n$ with a 
projective test
(in the algebraic formalism, this projective test is the multiplication operator corresponding to the characteristic function of $T_{n,r}$).
Then 
\begin{align}
\beta_n(T_{n,r})
&=
\sup_{\sigma_{[n]}\in\A^{(n)}_{\adv}}
\sum_{\vecc{x}\in T_{n,r}}\sigma_{[n]}(\vecc{x})\nonumber\\
&\le
(1-\theta_k)^{-n}\sup_{\sigma_{[n]}\in(\A_{\theta_k})^{(n)}_{\adv}}\sum_{\vecc{x}\in 
T_{n,r}}\sigma_{[n]}(\vecc{x})\nonumber\\
&\le
(1-\theta_k)^{-n}\sup_{\sigma_{[n]}\in(\A_{\theta_k})^{(n)}_{\adv}}
e^{-nc_r\alpha_r}\sum_{\vecc{x}\in\X^n}\sigma_{[n]}(\vecc{x})
\bz\frac{\rho_r^{\otimes n}(\vecc{x})}{\sigma_r^{\otimes n}(\vecc{x})}\jz^{\alpha_r}
\nonumber\\
&=
(1-\theta_k)^{-n}e^{-nc_r\alpha_r}\ds\times\nonumber\\
&\sup_{\sigma_{[n]}\in(\A_{\theta})^{(n)}_{\adv}}\Bigg\{
\sum_{x_1,\ldots,x_{n-1}\in\X^{n-1}}\sigma_{[n-1]}(x_1,\ldots,x_{n-1})
\bz\frac{\rho_r^{\otimes {n-1}}(x_1,\ldots,x_{n-1})}{\sigma_r^{\otimes {n-1}}(x_1,\ldots,x_{n-1})}\jz^{\alpha_r}\times\nonumber\\
& \ds\ds\ds\ds\ds\ds\ds\ds\ds\ds\ds\ds\ds\ds\ds\ds\ds\ds\ds\ds\ds\ds\ds
\underbrace{\sum_{x_n\in\X}\sigma_n(x_n|x_1,\ldots,x_{n-1})
\bz\frac{\rho_r(x_n)}{\sigma_r(x_n)}\jz^{\alpha_r}}_{\le\sum_{x\in\X}\sigma_r(x)
\bz\frac{\rho_r(x)}{\sigma_r(x)}\jz^{\alpha_r}}\Bigg\},\label{adv beta upper bound}
\end{align}
where the first inequality follows from \eqref{smoothed upper bound}, the second inequality is due to the Markov inequality, and the third one is due to 
\eqref{LLR lower2}.
Iterating the above, we get 
\begin{align}\label{adv type II upper bound}
\beta_n(T_{n,r})
&\le
(1-\theta_k)^{-n}e^{-nc_r\alpha_r}\bz\sum_{x\in\X}\sigma_r(x)
\bz\frac{\rho_r(x)}{\sigma_r(x)}\jz^{\alpha_r}\jz^n
=
(1-\theta_k)^{-n}e^{-n(c_r\alpha_r-\psi(\alpha_r))}
=e^{-n(r+\log(1-\theta_k))},
\end{align}
where the last equality is by \eqref{simple Hr}.
An exactly analogous argument yields
\begin{align*}
\alpha_n(T_{n,r})
&\le
(1-\theta_k)^{-n}e^{nc_r(1-\alpha_r)}\bz\sum_{x\in\X}\rho_r(x)
\bz\frac{\sigma_r(x)}{\rho_r(x)}\jz^{1-\alpha_r}\jz^n
=
(1-\theta_k)^{-n}e^{n(c_r(1-\alpha_r)+\psi(\alpha_r))}\\
&=e^{-n(H_r(\N_{\theta_k}\|\A_{\theta_k})+\log(1-\theta_k))},
\end{align*}
where the last equality is by \eqref{simple Hr2}--\eqref{simple Hr3}.
Thus, the pair of exponents
\begin{align}\label{achievable classical pair}
(r+\log(1-\theta_k),\,H_r(\N_{\theta_k}\|\A_{\theta_k})+\log(1-\theta_k))\ds\ds\ds\text{is achievable}
\end{align}
for any large enough $k$. 

Note that the optimal states $\rho_r, \sigma_r$ above depend on $\theta_k$, and for the rest we indicate this by denoting them as $\rho_{\theta_k,r}$, $\sigma_{\theta_k,r}$. 
Again by the compactness of $\N$ and $\A$, and by passing to a subsequence if necessary, 
we may assume that
$\rho_{\theta_k,r}\to \rho_{0,r}\in\N$, 
$\sigma_{\theta_k,r}\to\sigma_{0,r}\in\A$ as $k\to +\infty$. Thus,
\begin{align*}
H_r(\N\|\A)\le H_r(\rho_{0,r}\|\sigma_{0,r})\le
\liminf_{k\to+\infty}H_r(\rho_{\theta_k,r}\|\sigma_{\theta_k,r}),
\end{align*}
by the same argument as in \eqref{relentr dist limit}, using the lower semi-continuity of the Hoeffding divergence (Lemma \ref{lemma:convex lsc}).
By \eqref{achievable classical pair} and Lemmas \ref{lemma:achievable limit} and 
\ref{lemma:smaller achievable}, the pair of exponents
\begin{align*}
r=\lim_{k\to+\infty}(r+\log(1-\theta_k))\ds\ds\ds\text{and}\ds\ds\ds
H_r(\N\|\A)\le\liminf_{k\to+\infty}(H_r(\N_{\theta_k}\|\A_{\theta_k})+\log(1-\theta_k))
\end{align*}
is achievable, and hence 
\begin{align}\label{adv direct exp}
\dli_r^{\adv}(\N\|\A)\ge H_r(\N\|\A)=\inf_{\rho\in\N,\sigma\in\A}\dli_r (\rho\|\sigma),
\end{align}
proving \eqref{adv exponents proof1}.
\end{proof}

\section{Quantum state discrimination}
\label{sec:quantum}

\subsection{Finite dimension with two alternative hypotheses: Equality may not hold}
\label{sec:finiteQ}

In this section, for any of the error exponents 
$e = \sli, \cli, \dli_r$,
we shall construct three density operators
$\rho, \sigma_1,\sigma_2$ on a finite-dimensional Hilbert space satisfying the strict inequality 
$$
\e(\rho \| \{\sigma_1,\sigma_2\}) < 
{\rm min}\{\, \e(\rho \| \sigma_1),\,
\e(\rho \| \sigma_2)\, \}.
$$
As it turns out, in some sense, such triplets are
fairly common. So the real problem is not finding (or constructing) such triplets of density operators, but proving that we have a strict inequality between the worst-case pairwise error exponent and the composite one. Of course, the pairwise error exponents are easy to compute due to the explicit formulas given in Lemma \ref{lemma:known exponents};
the hard part is estimating
the error exponents from above in the composite case, which amounts to
estimating the two types of errors from below.

Throughout this section we shall work on finite-dimensional Hilbert spaces (so all operators appearing in our constructions, even if not explicitly mentioned, are acting on some finite-dimensional space).
Our idea is to employ the well-known operator form of the
inequality between the {\it arithmetic} and the {\it geometric means} 
to bound the type II error. Recall that the geometric mean of two positive definite operators is defined as
\begin{align}\label{geom mean def}
A\# B:=B^{1/2}\bz B^{-1/2}AB^{-1/2}\jz^{1/2} B^{1/2}
\end{align}
(corresponding to $\alpha=1/2$ in \eqref{alpha geom mean}), 
and it can be extended to pairs of positive semi-definite operators by taking decreasing limits; see, e.g., \cite{KA}. The
geometric-arithmetic mean inequality says that
\begin{align*}
A\#B\le\half(A+B)
\end{align*}
for any positive semi-definite operators $A,B$. Using this inequality, we obtain
\begin{align}\label{geometric bound on beta}
\beta_n(\{\sigma_1,\sigma_2 \}|T_n)= 
{\rm max}\{\Tr(\sigma_1^{\otimes n}T_n),
\Tr(\sigma_2^{\otimes n}T_n)\}\geq
\Tr\bz\frac{\sigma_1^{\otimes n}+\sigma_2^{\otimes n}}{2}T_n\jz  \geq \Tr ((\sigma_1^{\otimes n}\#\sigma_2^{\otimes n})T_n),
\end{align}
for any density operators $\sigma_1,\sigma_2$, and test $T_n$.
The point of replacing the arithmetic mean with the geometric mean is to replace the composite i.i.d.~problem with a simple i.i.d.~problem.
This is achieved because the geometric mean is easily seen to have the following multiplicativity property:
$$
(A_1\# B_1)\otimes (A_2\#B_2) = 
(A_1\otimes A_2)\# (B_1\otimes B_2),
$$
implying that the term $\sigma_1^{\otimes n}\# \sigma_2^{\otimes n}$ appearing in our estimation can be rewritten as $(\sigma_1\#\sigma_2)^{\otimes n}$. 

We remark that for states $\sigma_1,\sigma_2\in\S(\hil)$, their geometric mean
$\sigma_1\#\sigma_2$ is in general a subnormalized state.
Indeed, we have 
\begin{align*}
\Tr(\sigma_1\#\sigma_2)\le \Tr\bz\sigma_1^{1/2}\sigma_2\sigma_1^{1/2}\jz^{1/2}=
 F(\sigma_1,\sigma_2)\le 1,
\end{align*}
where $F$ is the fidelity \cite{NC,Uhlmann-fidelity}, the first inequality was shown in \cite{Matsumoto_newfdiv}, and the second one is well-known \cite[Chapter 9]{NC}. This also implies that 
\begin{align*}
\Tr(\sigma_1\#\sigma_2)=1 \ds\iff\ds F(\sigma_1,\sigma_2)=1\ds\iff\ds \sigma_1=\sigma_2,
\end{align*}
where the second equivalence is again well-known \cite[Chapter 9]{NC}.
Moreover, for any state $\rho\in\S(\hil)$,
\begin{align*}
D(\rho\|\sigma_1\#\sigma_2)=
\underbrace{D\bz\rho\Big\|\frac{\sigma_1\#\sigma_2}{\Tr\sigma_1\#\sigma_2}\jz}_{\ge 0}
-\log\Tr \sigma_1\#\sigma_2
\ge -\log\Tr \sigma_1\#\sigma_2\ge 0,
\end{align*}
and $D(\rho\|\sigma_1\#\sigma_2)=0$ if and only if $\rho=\sigma_1=\sigma_2$.

\begin{lemma}\label{le: finiteQ_1} 
For any density operators $\sigma_1,\sigma_2\in\S(\hil)$, 
and any set of density operators $\N\subseteq\S(\hil)$,
we have
\begin{align*}
\sli(\N \| \{\sigma_1,\sigma_2\}) \le  D(\N\|\sigma_1\#\sigma_2)
\ds\ds\ds\text{and}\ds\ds\ds 
\dli_r(\N\|\{\sigma_1,\sigma_2\}) \le H_r(\N\|\sigma_1\#\sigma_2),\ds \forall r>0.
\end{align*}
\end{lemma}
\begin{proof}
If $\sigma_1\#\sigma_2=0$ then $D(\rho\|\sigma_1\#\sigma_2)=H_r(\rho\|\sigma_1\#\sigma_2)=+\infty$, and there is nothing to prove. Hence, we may assume that $\sigma_1\#\sigma_2\neq 0$ in which case $\lambda:=\Tr(\sigma_1\#\sigma_2)>0$ and $\tilde{\sigma}:=(\sigma_1\#\sigma_2)/\lambda$ is a density operator.
By \eqref{geometric bound on beta},
\begin{align*}
\beta_n(\{\sigma_1,\sigma_2 \}|T_n) \geq \Tr((\sigma_1\#\sigma_2)^{\otimes n}T_n)=\lambda^n \beta_n(\tilde\sigma|T_n),
\end{align*}
and hence
\begin{align}\label{le: finiteQ_1 proof1}
\liminf_{n\to+\infty}-\frac{1}{n}\log \beta_n(\{\sigma_1,\sigma_2 \}|T_n)
\le-\log\lambda
+\liminf_{n\to+\infty}-\frac{1}{n}\log\beta_n(\tilde{\sigma}|T_n).
\end{align}
Thus, by the definition of the error exponents in question, we get
\begin{align*}
\sli(\N \| \{\sigma_1,\sigma_2\}) &\leq
-\log\lambda + \sli(\N\|\tilde\sigma)
\le
-\log\lambda + \inf_{\rho\in\N}\sli(\rho\|\tilde\sigma)
=
-\log\lambda + D(\N\|\tilde\sigma)=
D(\N\|\sigma_1\#\sigma_2), \\
\dli_r(\N \| \{\sigma_1,\sigma_2\}) &\leq
\dli_{r+\log\lambda}(\N\|\tilde\sigma)
\le
\inf_{\rho\in\N}\dli_{r+\log\lambda}(\N\|\tilde\sigma)
=
H_{r+\log\lambda}(\N\|\tilde\sigma)=
H_r(\N\|\sigma_1\#\sigma_2),
\end{align*}
where the first inequalities follow from \eqref{le: finiteQ_1 proof1}, 
the second inequalities from \eqref{composite vs single bound1},
the first equalities from Lemma \ref{lemma:known exponents},
and the last equalities from the scaling laws \eqref{scaling}.
\end{proof}

By the above lemma, if for some density operators $\rho,\sigma_1,\sigma_2$,
$$
D(\rho\|\sigma_1\#\sigma_2) <  \min\{D(\rho\|\sigma_1),D(\rho\|\sigma_2)\},
$$
then $\sli(\rho\|\{\sigma_1,\sigma_2\})\leq D(\rho\|\sigma_1\#\sigma_2) <  \min\{D(\rho\|\sigma_1),D(\rho\|\sigma_2)\}=
\min\{\sli(\rho\|\sigma_1),\sli(\rho\|\sigma_2)\}$, i.e.\!
we can achieve a strict inequality for the Stein exponents.
In Theorem \ref{thm:finiteQ Stein}, we will give a systematic way to construct such triplets, corresponding to any pair of non-commuting density operators. Before that, however, we will need to establish a fact regarding the relation between the geometric and 
the exponential mean.

For two positive definite operators $A,B$ let
\begin{align*}
\diff(A,B)=\log(A\#B)-\frac{\log(A)+\log(B)}{2}.
\end{align*}
Note that if $A$ and $B$ commute then $\diff(A,B)=0$,
and in fact the converse is also true, according to Lemma 
\ref{lemma:Renyi ordering}.

For a self-adjoint operator $X\in\B(\hil)_{\sa}$, let 
$\lambda_{\max}(X)$ denote its largest eigen-value.

\begin{lemma}\label{le: finiteQ_2} 
Let $A,B$ be two positive definite operators. Then $\Tr\diff(A,B)=0$. 
In particular, 
if $AB\ne BA$ then there exists a density operator $\rho$ such that 
\begin{align*}
\Tr(\rho\diff(A,B))=\dif_{A,B}:=\lambda_{\max}(\diff(A,B))>0,
\end{align*}
and there exists an invertible 
density operator $\rho$ such that 
\begin{align*}
\Tr(\rho\diff(A,B))=\dif_{A,B}/2.
\end{align*}
\end{lemma}
\begin{proof}
For a positive definite operator $X$ one has 
$\Tr({\rm log}(X))={\rm log}({\rm det}(X))$. Thus, 
$\Tr\diff(A,B)=0$ can be justified by a straightforward reordering 
relying on \eqref{geom mean def} and the multiplicative property of the determinant.
By the above, $AB\ne BA$ implies $\diff(A,B)\neq 0$, which, together with 
$\Tr\diff(A,B)=0$, yields that $\lambda_{\max}(\diff(A,B))>0$.
Choosing $\rho=\pr{\psi}$ with an eigen-vector of $\diff(A,B)$ corresponding to 
$\lambda_{\max}(\diff(A,B))$ yields $\Tr(\rho\diff(A,B))=\lambda_{\max}(\diff(A,B))$.
The existence of an invertible $\rho$ with the given property then follows from the fact that 
$\{\rho\in S(\mathcal H):\, \Tr(\rho\diff(A,B))>0\}$
is an open subset of $\S(\hil)$.
%
\end{proof}


After the above preparation, it is easy to construct triplets of density operators 
with strict inequality for the Stein exponent.

\begin{theorem}\label{thm:finiteQ Stein}
For any density operators $\rho,\sigma_1,\sigma_2\in\S(\hil)$
such that $\sigma_1,\sigma_2$ are invertible, 
\begin{align}
\min_{i=1,2}\sli(\hat{\rho}\|\hat{\sigma}_i)-
\sli(\hat{\rho}\|\{\hat{\sigma}_1,\hat{\sigma}_2\})
&=
\sli(\hat{\rho}\|\hat{\sigma}_j)
-\sli(\hat{\rho}\|\{\hat{\sigma}_1,\hat{\sigma}_2\})\nonumber\\
&\ge
D(\hat{\rho}\|\hat{\sigma}_j)-D(\hat{\rho}\|\hat{\sigma}_1\#\hat{\sigma}_2)
=
2\Tr\rho\diff(\sigma_1,\sigma_2), \ds\ds\ds j=1,2,
\label{finiteQ Stein}
\end{align}
where 
\begin{align*}
\hat{\rho} := 
\frac{1}{2}\left(\begin{matrix}\rho & 0 \\ 0 & \rho \end{matrix}\right), \quad
\hat{\sigma}_1 := 
\frac{1}{2}\left(\begin{matrix}\sigma_1 & 0 \\ 0 & \sigma_2 \end{matrix}\right), \quad
\hat{\sigma}_2 := 
\frac{1}{2}\left(\begin{matrix}\sigma_2 & 0 \\ 0 & \sigma_1 \end{matrix}\right).
\end{align*}
In particular, if $\sigma_1,\sigma_2$ are non-commuting then 
there exists an invertible density operator $\rho$ such that 
\begin{align*}
\min_{i=1,2}\sli(\hat{\rho}\|\hat{\sigma}_i)-
\sli(\hat{\rho}\|\{\hat{\sigma}_1,\hat{\sigma}_2\})
\ge
D(\hat{\rho}\|\hat{\sigma}_j)-D(\hat{\rho}\|\hat{\sigma}_1\#\hat{\sigma}_2)
=\dif_{\sigma_1,\sigma_2}>0.
\end{align*}
\end{theorem}
\begin{proof}
The inequality in \eqref{finiteQ Stein} is immediate from 
Lemma \ref{lemma:known exponents} and Lemma \ref{le: finiteQ_1}, and the equality in 
\eqref{finiteQ Stein} follows by a straightforward computation using that 
$\Tr\hat\rho\log\hat\sigma_1=\Tr\hat\rho\log\hat\sigma_2$.
The last assertion follows immediately from Lemma \ref{le: finiteQ_2}.
%
\end{proof}

\begin{rem}
Note that \eqref{finiteQ Stein} gives an explicitly computable lower bound for the gap between the composite Stein exponent and the worst pairwise Stein exponent.
\end{rem}
\medskip

We will use triplets of density operators as in Theorem \ref{thm:finiteQ Stein} to construct further triplets exhibiting strict inequality also for the Chernoff and the direct exponents. To this end, let us introduce,
for any density operators $\rho,\sigma_1,\sigma_2\in\S(\hil)$, and any 
$\lambda,\eta,\mu,\nu\in[0,1]$, 
the density operators
\begin{align*}
\rho_{\lambda,\eta}&:=\frac{\eta\lambda}{2}\rho\oplus\frac{\eta\lambda}{2}\rho\oplus\eta(1-\lambda)\oplus(1-\eta)
=(\eta\lambda)\hat\rho\oplus\eta(1-\lambda)\oplus(1-\eta),
\\
\sigma_{j,\mu,\nu}
&:=
\frac{\nu\mu}{2}\sigma_j\oplus\frac{\nu\mu}{2}\sigma_{3-j}\oplus\nu(1-\mu)\oplus\nu
=
(\nu\mu)\hat\sigma_j\oplus\nu(1-\mu)\oplus\nu,
\ds\ds j=1,2,
\end{align*}
on $\hil\oplus\hil\oplus\bC\oplus\bC$. 
Note that $\rho_{1,1}=\hat\rho\oplus 0\oplus 0$, $\sigma_{j,1,1}=\hat\sigma_j\oplus 0\oplus 0$ with the notations of 
Theorem \ref{thm:finiteQ Stein}. Clearly,
\begin{align}
\sigma_{1,\mu,\nu}\#\sigma_{2,\mu,\nu}
=\bz\sigma_1\#\sigma_2\jz_{\mu,\nu}
=
\frac{\nu\mu}{2}\sigma_1\#\sigma_2\oplus\frac{\nu\mu}{2}\sigma_1\#\sigma_2\oplus\nu(1-\mu)\oplus\nu
=
(\nu\mu)\hat\sigma_1\#\hat\sigma_2\oplus\nu(1-\mu)\oplus\nu,
\label{geommean nu}
\end{align}
and a straightforward computation shows that 
\begin{align}
D\bz\rho_{\lambda,\eta}\|\sigma_{j,\mu,\nu}\jz
&=
\lambda\eta D\bz\hat\rho\|\hat\sigma_{j}\jz+
\eta d_2(\lambda\|\mu)+d_2(\eta\|\nu),\label{scaled relentr}\\
D\bz\rho_{\lambda,\eta}\|\sigma_{1,\mu,\nu}\#\sigma_{2,\mu,\nu}\jz
&=
\lambda\eta D\bz\hat\rho\|\hat\sigma_1\#\hat\sigma_2\jz+
\eta d_2(\lambda\|\mu)+d_2(\eta\|\nu),\label{scaled relentr2}
\end{align}
where 
\begin{align*}
d_2(\alpha,\beta):=D((\alpha,1-\alpha)\|(\beta,1-\beta)),\ds\ds\ds
\alpha,\beta\in[0,1].
\end{align*}

We start with the following refined version of Theorem \ref{thm:finiteQ Stein}.

\begin{prop}\label{prop:finiteQ Stein2}
Let $\rho,\sigma_1,\sigma_2\in\S(\hil)$ be as in Theorem \ref{thm:finiteQ Stein}.
For every $r>0$ and every $0<\lambda<\min\{1,r/D(\hat\rho\|\hat\sigma_{1}\#\hat\sigma_{2})\}$,
there exists a $\mu\in(0,1)$ such that 
\begin{align*}
\sli(\rho_{\lambda,1}\|\{\sigma_{1,\mu,1},\sigma_{2,\mu,1}\}) 
\le r
&=D(\rho_{\lambda,1}\|\sigma_{1,\mu,1}\#\sigma_{2,\mu,1})\\
&=D(\rho_{\lambda,1}\|\sigma_{j,\mu,1})-\lambda\delta_{\sigma_1,\sigma_2}\\
&=
\sli(\rho_{\lambda,1}\|\sigma_{j,\mu,1})-\lambda\delta_{\sigma_1,\sigma_2},
\ds\ds\ds j=1,2.
\end{align*}
\end{prop}
\begin{proof}
By Theorem \ref{thm:finiteQ Stein},
\begin{align*}
r_0
:=D(\hat\rho\|\hat\sigma_{1}\#\hat\sigma_{2})
< r_0+\dif_{\sigma_1,\sigma_2}
=
D(\hat\rho\|\hat\sigma_{j}),\ds\ds j=1,2.
\end{align*}
Let $\lambda$ be as in the statement, so that
$\lambda\in(0,1)$ and $r-\lambda r_0>0$, and let   
$\mu\in(0,1)$ be such that $r-\lambda r_0=d_2(\lambda\|\mu)$.
Such a $\mu$ exists, since $\mu\mapsto d_2(\lambda\|\mu)$ is convex, 
it is finite-valued on $(0,1)$, 
$d_2(\lambda\|\lambda)=0$, and $\lim_{\mu\to 0}d_2(\lambda\|\mu)=\lim_{\mu\to 1}d_2(\lambda\|\mu)=+\infty$.
By \eqref{scaled relentr}--\eqref{scaled relentr2},
\begin{align}
D(\rho_{\lambda,1}\|\sigma_{j,\mu,1})
&=
\underbrace{\lambda D(\hat\rho\|\hat\sigma_j)}_{=\lambda(r_0+\dif_{\sigma_1,\sigma_2})}+
\underbrace{d_2(\lambda\|\mu)}_{=r-\lambda r_{\nul}}
=r+\lambda\dif_{\sigma_1,\sigma_2},\ds\ds\ds j=1,2,
\label{refined Stein proof1}\\
D(\rho_{\lambda,1}\|\sigma_{1,\mu,1}\#\sigma_{2,\mu,1})
&=
\underbrace{\lambda D(\hat\rho\|\hat\sigma_1\#\hat\sigma_2)}_{=\lambda r_0}
+
\underbrace{d_2(\lambda\|\mu)}_{=r-\lambda r_{\nul}}
=r.\label{refined Stein proof2}
\end{align}
%
Thus,
\begin{align*}
\sli(\rho_{\lambda,1}\|\{\sigma_{1,\mu,1},\sigma_{2,\mu,1}\}) \leq 
D(\rho_{\lambda,1}\|\sigma_{1,\mu,1}\#\sigma_{2,\mu,1})=
r=  D(\rho_{\lambda,1}\|\sigma_{j,\mu,1})-\lambda\dif_{\sigma_1,\sigma_2}=
 \sli(\rho_{\lambda,1}\|\sigma_{j,\mu,1})-\lambda\dif_{\sigma_1,\sigma_2},\ds\ds j=1,2,
\end{align*}
where the first inequality is due to Lemma \ref{le: finiteQ_1}, and the last equality 
is by Lemma \ref{lemma:known exponents}.
This proves the assertion.
\end{proof}

\begin{rem}
Note that in Proposition \ref{prop:finiteQ Stein2}, 
\begin{align*}
\supp\rho_{\lambda,1}=\hil\oplus\hil\oplus\bC\oplus 0=\supp\sigma_{j,\mu,1}.
\end{align*}
\end{rem}

Using the freedom in the choice of the parameters $\lambda,\eta,\mu,\nu$, we can also get examples with strict inequality for the direct exponents.
Our construction below also provides examples with strict inequality for the 
direct exponents in the setting where the null-hypothesis is composite (consisting of two hypotheses), and the alternative hypothesis is simple.

\begin{theorem}\label{thm:finiteQ}
Let $\rho,\sigma_1,\sigma_2\in\S(\hil)$ be as in Theorem \ref{thm:finiteQ Stein}.
For every $r,t>0$, there exist $\lambda,\eta,\mu,\nu\in(0,1)$ and $\gamma>0$ such that 
for every $r',r''\in(r-\gamma,r+\gamma)$, 
\begin{align}
\dli_{r'}(\rho_{\lambda,\eta}\|\{\sigma_{1,\mu,\nu},\sigma_{2,\mu,\nu}\}) 
< t <
\min_{i\in\{1,2\}} \dli_{r''}(\rho_{\lambda,\eta}\|\sigma_{i,\mu,\nu})=
\dli_{r''}(\rho_{\lambda,\eta}\|\sigma_{j,\mu,\nu}),
\ds\ds j=1,2.
\label{Hoeffding separation}
\end{align}
Moreover, if $r'<r''$ then 
\begin{align}
\dli_{t}(\{\sigma_{1,\mu,\nu},\sigma_{2,\mu,\nu}\}\|\rho_{\lambda,\eta}) 
< r'<r'' \le
\min_{i\in\{1,2\}} \dli_{t}(\sigma_{i,\mu,\nu}\|\rho_{\lambda,\eta})=
\dli_{t}(\sigma_{j,\mu,\nu}\|\rho_{\lambda,\eta}),
\ds\ds j=1,2.
\label{Hoeffding separation2}
\end{align}
\end{theorem}
\begin{proof}
Let us fix $r$, and choose $\lambda,\mu$ as in Proposition \ref{prop:finiteQ Stein2}.
Note that 
\begin{align*}
\rho_{\lambda,\eta}=\eta(\underbrace{\lambda\hat\rho\oplus(1-\lambda)\oplus 0}_{=\rho_{\lambda,1}})+0\oplus 0\oplus 0\oplus (1-\eta),\ds\ds\ds
\sigma_{j,\mu,\nu}=\nu(\underbrace{\mu\hat\sigma_j\oplus(1-\mu)\oplus 0}_{=\sigma_{j,\mu,1}})+0\oplus 0\oplus 0\oplus (1-\nu).
\end{align*}
Thus, for any $\eta\in(0,1]$,
\begin{align}
H_r(\rho_{\lambda,\eta}\|\sigma_{1,\mu,1}\#\sigma_{2,\mu,1})&=-\log\eta+
H_r(\rho_{\lambda,1}\|\sigma_{1,\mu,1}\#\sigma_{2,\mu,1})\\
&=-\log\eta
<-\log\eta+\underbrace{H_r(\rho_{\lambda,1}\|\sigma_{j,\mu,1})}_{=:\kappa>0}=
H_r(\rho_{\lambda,\eta}\|\sigma_{j,\mu,1}),\ds j=1,2,\label{Hoeffding separation proof2}
\end{align}
where the first and the last equalities follow by straightforward computations
using the scaling laws \eqref{scaling},
The second equality follows by Lemma \ref{lemma:Hr limits}, since 
$r=D(\rho_{\lambda,1}\|\sigma_{1,\mu,1}\#\sigma_{2,\mu,1})$ 
according to \eqref{refined Stein proof2}, and 
the inequality follows again 
by Lemma \ref{lemma:Hr limits}, since $r<D(\rho_{\lambda,1}\|\sigma_{j,\mu,1})$
according to \eqref{refined Stein proof1}.

Let $s\in(0,1/3)$ be such that $s\kappa<t$. Set $\eta:=e^{s\kappa-t}\in(0,1)$. By 
\eqref{Hoeffding separation proof2}, we have
\begin{align}
H_r(\rho_{\lambda,\eta}\|\sigma_{1,\mu,1}\#\sigma_{2,\mu,1})=t-s\kappa<t,\ds\ds\ds
H_r(\rho_{\lambda,\eta}\|\sigma_{j,\mu,1})=-\log\eta+\kappa=t+\kappa(1-s)>t+2\kappa/3.
\label{Hoeffding separation proof3}
\end{align}
By Lemma \ref{lemma:convex lsc}, 
$\nu\mapsto H_r(\rho_{\lambda,\eta}\|\sigma_{j,\mu,\nu})$ 
is a convex lower semi-continuous function on $[0,1]$.
Moreover, it is finite-valued; for $\nu<1$ this follows from 
\begin{align}\label{quantum Hoeffding support}
\supp\rho_{\lambda,\eta}=\hil\oplus\hil\oplus\bC\oplus\bC=\supp\sigma_{j,\mu,\nu},
\end{align}
while for $\nu=1$ it follows from \eqref{Hoeffding separation proof2}.
By \eqref{geommean nu}, $\sigma_{1,\mu,\nu}\#\sigma_{2,\mu,\nu}$ is affine in $\nu$, 
and hence, again by Lemma \ref{lemma:convex lsc},
$\nu\mapsto H_r(\rho_{\lambda,\eta}\|\sigma_{1,\mu,\nu}\#\sigma_{2,\mu,\nu})$
is convex and lower semi-continuous, and it is also finite-valued, by the same argument as above. Hence, both functions are continuous in $\nu$, and thus there exists a 
$\nu_0\in[0,1)$ such that for all $\nu\in(\nu_0,1)$, 
\begin{align}\label{Hoeffding separation proof4}
H_r(\rho_{\lambda,\eta}\|\sigma_{1,\mu,\nu}\#\sigma_{2,\mu,\nu})<t,\ds\ds\ds
H_r(\rho_{\lambda,\eta}\|\sigma_{j,\mu,\nu})>t+2\kappa/3,
\end{align}
according to \eqref{Hoeffding separation proof3}.
Let us fix any such $\nu$. 

By \eqref{quantum Hoeffding support}, $D_0(\rho_{\lambda,\eta}\|\sigma_{1,\mu,\nu}\#\sigma_{2,\mu,\nu})=0=D_0(\rho_{\lambda,\eta}\|\sigma_{j,\mu,\nu})$, whence the functions 
$r\mapsto H_r(\rho_{\lambda,\eta}\|\sigma_{1,\mu,\nu}\#\sigma_{2,\mu,\nu})$
and $r\mapsto H_r(\rho_{\lambda,\eta}\|\sigma_{j,\mu,\nu})$
are finite-valued on $(0,+\infty)$ according to Lemma \ref{lemma:Hr limits}, and, by 
Corollary \ref{lemma:Hr convex in r}, they are also convex, and hence continuous. Thus, 
by \eqref{Hoeffding separation proof4}, there exists a $\gamma\in(0,r)$ such that 
\begin{align*}
H_{r'}(\rho_{\lambda,\eta}\|\sigma_{1,\mu,\nu}\#\sigma_{2,\mu,\nu})<t,\ds\ds\ds
H_{r''}(\rho_{\lambda,\eta}\|\sigma_{j,\mu,\nu})>t+2\kappa/3,
\end{align*}
for any $r'\in(r-\gamma,+\infty)$ and $r''\in(0,r+\gamma)$, where we also took into account that $H_r$ is monotone non-increasing in $r$.
Finally, for any $r',r''\in(r-\gamma,r+\gamma)$,
\begin{align}
\dli_{r'}(\rho_{\lambda,\eta}\|\{\sigma_{1,\mu,\nu},\sigma_{2,\mu,\nu}\})
\le
H_{r'}(\rho_{\lambda,\eta}\|\sigma_{1,\mu,\nu}\#\sigma_{2,\mu,\nu})
 < t <
H_{r''}(\rho_{\lambda,\eta}\|\sigma_{j,\mu,\nu})
=
\dli_{r''}(\rho_{\lambda,\eta}\|\sigma_{j,\mu,\nu}),
\ds\ds j=1,2,
\end{align}
where the first inequality is due to Lemma \ref{le: finiteQ_1}, and the 
equality is due to Lemma \ref{lemma:known exponents}. 
This proves \eqref{Hoeffding separation}.

For $r'$ as above, we have
\begin{align}
\dli_{r'}(\rho_{\lambda,\eta}\|\{\sigma_{1,\mu,\nu},\sigma_{2,\mu,\nu}\}) < t
&\ds\iff\ds
(r',t)\notin\ac(\rho_{\lambda,\eta}\|\{\sigma_{1,\mu,\nu},\sigma_{2,\mu,\nu}\})
\nonumber\\
&\ds\iff\ds
(t,r')\notin\ac(\{\sigma_{1,\mu,\nu},\sigma_{2,\mu,\nu}\}\|\rho_{\lambda,\eta})
\nonumber\\
&\ds\iff\ds
\dli_{t}(\{\sigma_{1,\mu,\nu},\sigma_{2,\mu,\nu}\}\|\rho_{\lambda,\eta}) < r',
\label{Hoeffding separation proof5}
\end{align}
where the first inequality is by \eqref{Hoeffding separation},
the first and the last equivalences are by definition, and the second equivalence is by 
Remark \ref{rem:ac swap}. Similarly, for $r''$ as above,
\begin{align}
t<\dli_{r''}(\rho_{\lambda,\eta}\|\sigma_{j,\mu,\nu})
&\ds\iff\ds
(r'',t)\in\ac(\rho_{\lambda,\eta}\|\sigma_{j,\mu,\nu})\nonumber\\
&\ds\iff\ds
(t,r'')\in\ac(\sigma_{j,\mu,\nu}\|\rho_{\lambda,\eta})\nonumber\\
&\ds\iff\ds
r''\le \dli_t(\sigma_{j,\mu,\nu}\|\rho_{\lambda,\eta}).\label{Hoeffding separation proof6}
\end{align}
Finally, if $r'<r''$ then \eqref{Hoeffding separation proof5} and 
\eqref{Hoeffding separation proof6} yield \eqref{Hoeffding separation2}.
\end{proof}

Theorem \ref{thm:finiteQ} yields immediately a construction with strict inequality for the Chernoff exponent.

\begin{theorem}\label{thm:Chernoff Q separation}
Let $\rho,\sigma_1,\sigma_2\in\S(\hil)$ be as in Theorem \ref{thm:finiteQ Stein}.
For every $r>0$, there exist $\lambda,\eta,\mu,\nu\in(0,1)$ such that 
\begin{align}\label{Chernoff Q separation}
\cli(\rho_{\lambda,\eta}\|\{\sigma_{1,\mu,\nu},\sigma_{2,\mu,\nu}\})\leq r < 
\min_{i\in\{1,2\}} \cli(\rho_{\lambda,\eta}\|\sigma_{i,\mu,\nu})=
\cli(\rho_{\lambda,\eta}\|\sigma_{j,\mu,\nu}), \ds j=1,2.
\end{align}
\end{theorem}
\begin{proof}
Let $t:=r$, and choose $\lambda,\eta,\mu,\nu$ as in Theorem \ref{thm:finiteQ}.
By \eqref{Hoeffding separation},
$\dli_{r}(\rho_{\lambda,\eta}\|\{\sigma_{1,\mu,\nu},\sigma_{2,\mu,\nu}\}) \le r$, and hence
$\cli(\rho_{\lambda,\eta}\|\{\sigma_{1,\mu,\nu},\sigma_{2,\mu,\nu}\})\leq r$. 

On the other hand, let $r''>r$ be such that \eqref{Hoeffding separation} holds, and let 
$\hat r:=\dli_{r''}(\rho_{\lambda,\eta}\|\sigma_{j,\mu,\nu})$. 
Then $\hat r>r$, according to \eqref{Hoeffding separation}. Finally, with 
$\tilde r:=\min\{r'',\hat r\}$, we have 
$d_{\tilde r}(\rho_{\lambda,\eta}\|\sigma_{j,\mu,\nu})\ge 
\dli_{r''}(\rho_{\lambda,\eta}\|\sigma_{j,\mu,\nu})=\hat r\ge \tilde r$, whence 
$\cli(\rho_{\lambda,\eta}\|\sigma_{j,\mu,\nu})\ge \tilde r>r$, proving 
\eqref{Chernoff Q separation}.
\end{proof}


\begin{rem}
Note that the unitaries $U_0:=I$, $U_1:\,X\oplus Y\oplus t\oplus s\mapsto 
Y\oplus X\oplus t\oplus s$ give a unitary representation of $\bZ_2$ on 
$\hil\oplus\hil\oplus\bC\oplus\bC$, and we have 
\begin{align*}
\rho_{\lambda,\eta}=U_k\,\rho_{\lambda,\eta}\, U_k^*,\ds k=0,1,
\ds\ds\ds\ds\ds
U_1\,\sigma_{1,\mu,\nu}\,U_1^*=\sigma_{2,\mu,\nu}.
\end{align*}
Thus, the examples in Theorem \ref{thm:finiteQ Stein}, Proposition \ref{prop:finiteQ Stein2}, 
and Theorems \ref{thm:finiteQ}--\ref{thm:Chernoff Q separation}
fit into the setting of the group symmetric state discrimination problem considered in \cite{HMH09}, and they refute the conjecture formulated in 
\cite[Section VII]{HMH09}, which, if true, would have implied the equality 
$\e(\{\sigma_{1,\mu,\nu},\sigma_{2,\mu,\nu}\}\|\rho_{\lambda,\eta})
=
\e(\sigma_{j,\mu,\nu}\|\rho_{\lambda,\eta})$, $j=1,2$, for the Chernoff and the direct exponents.
\end{rem}

\begin{rem}
Our example for the strict inequality in the direct exponents works for a non-trivial parameter range, which depends on the states constructed.  
It would be interesting to find an example of similarly simple structure, for which strict inequality holds simultaneously over the whole 
trade-off curve. 
An example with this property, but with continuum many null-hypotheses, was given in
\cite[Example 6.2]{HMH09}.
\end{rem}

\begin{remark}
\label{remark:minimal}
The above constructed examples are minimal in the sense that their null-hypothesis is always a singleton while their alternative hypothesis contain only two elements. However, regarding the dimension of the Hilbert space, they are not necessarily minimal. For example, to have a strict inequality for the asymmetric error exponent, our construction uses matrices of size at least $4\times 4$. 
This can in fact be improved: by an explicit computation of geometric means and relative entropies, one can verify that the $2\times 2$ matrices  
$$
\rho := 
\frac{1}{2}\left(\begin{matrix} 1 & -1 \\ -1 & 1 \end{matrix}\right), \quad
\sigma_1 := 
\frac{1}{4}\left(\begin{matrix} 3 & 1 \\ 1 & 1 \end{matrix}\right), \quad
\sigma_2 := 
\frac{1}{4}\left(\begin{matrix} 1 & 1 \\ 1 & 3 \end{matrix}\right)
$$
also give an example for the strict inequality 
$\sli(\rho\|\{\sigma_1,\sigma_2\}) < \min\{\sli(\rho\|\sigma_1), 
\sli(\rho\|\sigma_2)\}$, thereby providing an example that is also minimal in dimension.
On the other hand, for the strict inequality in the symmetric case, we were unable to find an example using $2\times 2$ matrices; numerical attempts to compute the asymptotic behaviour using the first $\sim\!40$ tensor powers\footnote{We thank Sloan Nietert whose clever script made computations with such high tensor powers feasible.}
 seemed to 
indicate that for that, one needs at least $3\times 3$ matrices.
\end{remark}

\subsection{Equality cases in quantum state discrimination}  
\label{sec:q equality}

In this section we consider the case where 
both the null and the alternative hypotheses
are given by finite sets of density operators $\N=\{\rho_1,\ldots, \rho_k\}$ and 
$\A=\{\sigma_1,\ldots, \sigma_m\}$ on some (possibly infinite-dimensional) Hilbert space 
$\hil$.

As we have seen in Section \ref{sec:finiteQ}, the optimal error exponents need not coincide with the worst pairwise error exponents even in this simple case, and therefore it becomes important to identify classes of states for which such an equality can nevertheless be established, for some or for all of the error exponents discussed in Section 
\ref{sec:state disc}. As we have already mentioned in Section \ref{sec:state disc}, equality holds in the classical case, i.e., when all states commute with each other. Our first result, in Section \ref{sec:semi-classical}, shows that this condition can be relaxed, and equality 
holds even in the \ki{semi-classical} case, where we only require that $\rho\sigma=\sigma\rho$ 
for all $\rho\in\N$ and $\sigma\in\A$. 
In Section \ref{sec:pure} we show that equality holds also 
when all states $\rho,\sigma$ are pure. 

Note that the above two special cases are disjoint in the sense that if $\N$ and $\A$ satisfy both conditions then for any 
$\rho\in\N$ and $\sigma\in\A$, they are either equal or orthogonal to each other, a trivial 
situation.

We mention that, on top of the equality cases discussed in this section
and the ones mentioned in the Introduction, 
further special cases of equality are given in \cite{Sz20}
for the Chernoff exponent. 
Moreover, it is easy to see that the main result of \cite{KeLi} on symmeric multi-state discrimination implies that we also have equality for the Chernoff exponent 
when the Hilbert space is finite-dimensional and 
the null and alternative hypotheses  
$\N,\A$ are such that 
\begin{align*}
\min_{\rho\in\N,\sigma\in\A}C(\rho\|\sigma)=\min_{\tau_1,\tau_2\in\N\cup\A,\tau_1\ne\tau_2}
C(\tau_1\|\tau_2).
\end{align*}

\subsubsection{Semi-classical case}
\label{sec:semi-classical}

For a PSD operator $A\in\B(\hil)\p$ and a test $T\in\B(\hil)_{[0,I]}$, let 
\begin{align*}
\alpha(A|T):=\Tr A(I-T),\ds\ds\ds
\beta(A|T):=\Tr AT.
\end{align*}

\begin{lemma}\label{le:semi-classical_1}
Let $A_1,\ldots, A_k,B\in\B(\hil)\p$ be PSD trace-class operators, 
and $T_1,\ldots, T_k\in\B(\hil)_{[0,I]}$ be tests. If $[A_i,B]=0$ for every $i$, then there exists a projective test $Q$ such that
\begin{align}\label{sc error bounds1}
\max_{1\le i\le k}\alpha(A_i|Q) \leq 2\sum_{i=1}^k \alpha(A_i|T_i),
\ds\ds\ds\text{and}\ds\ds\ds
\beta(B|Q) \, \leq \, 2\sum_{i=1}^k \beta(B|T_i).
\end{align}
\end{lemma}

\begin{proof}
Note that replacing each $T_i$ with its diagonal in an orthonormal basis in which 
both $A_i$ and $B$ are diagonal does not change the $\alpha$ and $\beta$ quantities,
and therefore we may assume that $T_i$ commutes with both $A_i$ and $B$.
Let $Q_i:=f(T_i)$ with the function $f$ in \eqref{projection from test}. Then 
$Q_i$ is a projective test that also commutes with $A_i$ and $B$, and,  
by Lemma  \ref{lemma:projective enough}, 
\begin{align*}
\alpha(A_i|Q_i) \leq 2 \alpha(A_i|T_i), 
\ds\ds\ds\text{and}\ds\ds\ds
\beta(B|Q_i) \leq 2 \beta(B|T_i).
\end{align*}
Let $Q:=\oll Q^0$ be the projection onto the support of $\oll Q:=\sum_{i=1}^k Q_i$.
Then
\begin{align*}
\max_{1\le i\le k}\alpha(A_i|Q)\le
\max_{1\le i\le k}\alpha(A_i|Q_i)\le
\max_{1\le i\le k}2\alpha(A_i|T_i)\le
2\sum_{i=1}^k \alpha(A_i|T_i),
\end{align*}
where the first inequality follows from $Q\ge Q_i$, $i\in[k]$.
This proves the first inequality in \eqref{sc error bounds1}.

Let $P_{\lambda}$ denote the spectral projection of $B$ corresponding to a 
$\lambda\in\spec(B)$, and for an arbitrary PSD operator $X\in\B(\hil)\p$, let 
$E^X(H)$ denote its spectral projection corresponding to a Borel set $H\subseteq\bR$.
Note that $B$ commutes with $\oll Q$, and hence all $P_{\lambda}$ commute with the spectral projections $E^{\oll Q}(\cdot)$ of $\oll Q$. In particular,
$E^{P_{\lambda}\oll Q}(H)=P_{\lambda}E^{\oll Q}(H)$ for any Borel set $H\subseteq(0,+\infty)$, and 
\begin{align*}
(P_{\lambda}\oll Q)^0=
\int_{(0,+\infty)}\,E^{P_{\lambda}\oll Q}(dt)=
\int_{(0,+\infty)}\,P_{\lambda}E^{\oll Q}(dt)=
P_{\lambda}\int_{(0,+\infty)}\,E^{\oll Q}(dt)=P_{\lambda}\oll Q^0=P_{\lambda}Q.
\end{align*}
Thus,
\begin{align*}
\Tr(P_\lambda Q) =
\Tr \bz \sum_{i=1}^k P_\lambda Q_i\jz^0 
=
\dim\bz\spann\cup_{i=1}^k\bz\ran  P_{\lambda}\cap\ran Q_i\jz\jz
\le
\sum_{i=1}^k\dim\bz\ran  P_{\lambda}\cap\ran Q_i\jz
=
\sum_{i=1}^k\Tr P_{\lambda}Q_i.
\end{align*}
Finally,
\begin{align*}
\beta(B|Q)  =   
\sum_{\lambda \in \spec(B)\setminus \{0\}}\lambda\Tr( P_\lambda Q) \le
\sum_{i=1}^k \sum_{\lambda \in \spec(B)\setminus \{0\}} \lambda \Tr (P_\lambda Q_i) =
\sum_{i=1}^k\beta(B|Q_i) \leq 2\sum_{i=1}^k \beta(B|T_i),
\end{align*}
proving the second inequality in \eqref{sc error bounds1}.
\end{proof}

\begin{lemma}\label{le:semi-classical_2}
Let $A_i,\ldots,A_k,B_1,\ldots,B_m\in\B(\hil)\p$ be PSD trace-class operators, 
and $T_{i,j}\in\B(\hil)_{[0,I]}$, $i\in[k]$, $j\in[m]$, be tests.
If $[A_i,B_j]=0$ for all $i,j$, then there exists a projective test $Q$ such that
\begin{align}\label{sc lemma 2}
\max_{1\le i\le k}\alpha(A_i|Q)\le 4k\sum_{i=1}^{k}\sum_{j=1}^{m}\alpha(A_i|T_{ij}),
\ds\ds\ds\text{and}\ds\ds\ds
\max_{1\le j\le m}\beta(B_j|Q)\le 4\sum_{i=1}^{k}\sum_{j=1}^{m}\beta(B_j|T_{ij}).
\end{align}
\end{lemma}
\begin{proof}
The proof goes by a double application of Lemma \ref{le:semi-classical_1}. 
First, we fix $j\in[m]$ and apply Lemma \ref{le:semi-classical_1} with 
$B:=B_j$ and $T_i:=T_{i,j}$ to obtain a projective test $Q_j$ such that 
\begin{align}\label{sc lemma2 proof1}
\max_{1\le i\le k}\alpha(A_i|Q_j)\leq 2\sum_{i=1}^k \alpha(A_i|T_{i,j}),
\ds\ds\ds\text{and}\ds\ds\ds
\beta(B_j|Q_j)\le 2\sum_{i=1}^k \beta(B_j|T_{i,j}).
\end{align}
Next, we apply Lemma \ref{le:semi-classical_1} 
with $\wtilde A_j:=B_j$, $\wtilde B:=\bar A:=\sum_{i=1}^k A_i$, and $\wtilde T_j:=I-Q_j$,
to obtain a projective test $\wtilde Q$ such that 
\begin{align}
&\max_{1\le j\le m}\underbrace{\alpha(\wtilde A_j|\wtilde Q)}_{=\Tr B_j(I-\wtilde Q)}\le 2\sum_{j=1}^m\underbrace{\alpha(\wtilde A_j|\wtilde T_j)}_{=\Tr B_jQ_j}
=
2\sum_{j=1}^m\beta(B_j|Q_j)
\le 2\sum_{j=1}^m 2\sum_{i=1}^k \beta(B_j|T_{i,j}),\ds\ds\text{and}\label{sc lemma2 proof2}\\
&\beta(\wtilde B|\wtilde Q)\le 
2\sum_{j=1}^m\beta(\wtilde B|\wtilde T_j)=
2\sum_{j=1}^m\underbrace{\Tr\bar A(I-Q_j)}_{=\sum_l\Tr A_l(I-Q_j)}
=
2\sum_{j=1}^m\sum_{l=1}^k\alpha(A_l|Q_j)
\le
2\sum_{j=1}^m\sum_{l=1}^k2\sum_{i=1}^k \alpha(A_i|T_{i,j}).\label{sc lemma2 proof3}
\end{align}
The first inequality in \eqref{sc lemma2 proof2} is due to Lemma \ref{le:semi-classical_1}, and the second one follows from the second inequality in \eqref{sc lemma2 proof1}.
Similarly, the first inequality in \eqref{sc lemma2 proof3} is due to 
Lemma \ref{le:semi-classical_1}, and the second 
inequality follows from the first inequality in \eqref{sc lemma2 proof1}.
Defining $Q:=I-\wtilde Q$, the LHS in \eqref{sc lemma2 proof2} becomes $\max_{1\le j\le m}\beta(B_j|Q)$, and we
obtain the second inequality in \eqref{sc lemma 2} from \eqref{sc lemma2 proof2}, while the LHS in 
\eqref{sc lemma2 proof3} becomes
\begin{align*}
\beta(\wtilde B|\wtilde Q)=\Tr\bar A(I-Q)=\sum_{i=1}^k\Tr A_i(I-Q)=
\sum_{i=1}^k\alpha(A_i|Q)\ge \max_{1\le i\le k}\alpha(A_i|Q),
\end{align*}
and combining it with the rest of \eqref{sc lemma2 proof3} yields 
the first inequality in \eqref{sc lemma 2}.
\end{proof}

\begin{corollary}\label{cor:semi-classical} Consider the composite hypothesis testing problem given by the finite sets of 
density operators $\N=\{\varrho_1,...,\varrho_k\}$  (null-hypothesis)
and  $\A=\{\sigma_1,...,\sigma_m\}$ (alternative hypothesis).
If $[\varrho_i, \sigma_j]=0$ for all $i,j$, then 
$\e(\N\|\S) = \min_{i,j}\, \e(\varrho_i\| \sigma_j)$
for any of the exponents $\e=\sli, \cli$, and $\dli_r$ with $r>0$.
\end{corollary}

\begin{proof}
By \eqref{composite vs single bound1}, it is enough to show that   $\e(R\|S)\geq \min_{ij}\e(\rho_i\|\sigma_j)$, which follows immediately from the inequalities in Lemma \ref{le:semi-classical_2}.
\end{proof}

\subsubsection{Pure state case}
\label{sec:pure}

\begin{theorem}\label{th:pure}
Let $\hil$ be a Hilbert space, and 
$\N,\A \subset S(\mathcal H)$. If both $\N$ and $\A$ are finite collections of rank-one 
projections (pure states), then 
\begin{align*}
\e(\N\|\A)=\min_{\rho\in\N,\sigma\in \A}\e(\rho\|\sigma)
\end{align*}
for any of the error exponents $\e=\sli,\cli$, and $\dli_r$, $r>0$.
\end{theorem}

\begin{proof}
Let $\N =\{\rho_1,\ldots,\rho_k\}$ and $\A=\{\sigma_1,\ldots, \sigma_m\}$
where $\rho_i=\pr{\Psi_i}$ and $\sigma_j=\pr{\Phi_j}$ are rank-one projections for every $i,j$. We may assume without loss of generality that $\rho_i\ne\rho_j$, $i\ne j$, and 
$\sigma_i\ne\sigma_j$, $i\ne j$.
Note that for every $i\in[k]$, $j\in[m]$,
\begin{align*}
\psi(\rho_i\|\sigma_j|\alpha)=\log\Tr \rho_i\sigma_j=-C(\rho_i\|\sigma_j)=:-C_{i,j}
\end{align*} 
for every $\alpha\in(0,1)$, and thus
\begin{align*}
\dli_r(\rho_i\|\sigma_j)=H_r(\rho_i\|\sigma_j)
=
\sup_{\alpha\in(0,1)}\frac{\alpha-1}{\alpha}\left[r-\frac{1}{\alpha-1}\log\Tr \rho_i\sigma_j\right]
=
r-\inf_{\alpha\in(0,1)}\frac{1}{\alpha}\left[r-C_{i,j}\right]
=
\begin{cases}
+\infty,&r<C_{i,j},\\
C_{i,j},&r\ge C_{i,j},
\end{cases}
\end{align*}
where the first equality is by Lemma \ref{lemma:known exponents}, and the rest are immediate 
from the definition of $H_r$. Hence,
\begin{align*}
\ac(\rho_i\|\sigma_j)=[0,+\infty]^2\setminus(C_{i,j},+\infty]^2,\ds\ds\text{and thus}\ds\ds
\bigcap_{i,j}\ac(\rho_i\|\sigma_j)=[0,+\infty]^2\setminus(C_{\min},+\infty]^2,
\end{align*}
where $C_{\min}:=\min_{i,j}C_{i,j}$.
By Corollary \ref{cor:cs equality from dr equality} and 
Lemma \ref{lemma:Stein from directexp}, the assertion will follow 
for all the exponents if we can show that 
$[0,+\infty]^2\setminus(C_{\min},+\infty]^2\subseteq\ac(\N\|\A)$. 
For this it is sufficient to show that $(C_{\min},\infty)$ and $(\infty,C_{\min})$ are both in $\ac(\N\|\A)$, according to Lemma \ref{lemma:smaller achievable}. 
Due to the symmetry of the problem with respect to interchanging $\N$ and $\A$, it is sufficient to show the achievability of the first pair. To this end, let us define
\begin{align*}
K_n:=\sum_{i=1}^k \rho_i^{\otimes n},\ds\ds\ds
T_n:=K_n^0,
\ds\ds\ds n\in\bN,
\end{align*}
so that 
\begin{align*}
\alpha_n(\N|T_n)=\max_{1\le i\le k}\Tr\rho_i^{\otimes n}(I-T_n)=0,\ds n\in\bN\ds\ds\imp\ds\ds
\liminf_{n\to+\infty}-\frac{1}{n}\log\alpha_n(\N|T_n)=+\infty.
\end{align*}

Note that $\supp K_n=\spann\{\Psi_i^{\otimes n}:\,i\in[k]\}$ is a $k$-dimensional subspace in $\hil^{\otimes n}$, and
the nonzero eigenvalues of $K_n$ coincide with those of the $k\times k$ Gram matrix $G^{(n)}$ with entries
$$
G^{(n)}_{i,l} = 
\langle \Psi_i^{\otimes n},\Psi_l^{\otimes n}\rangle = 
\langle \Psi_i,\Psi_l \rangle ^n.
$$
Clearly, $G^{(n)}$ converges to the $k\times k$ identity matrix as $n\to \infty$. Hence, for every large enough $n$, the smallest non-zero eigenvalue of $K_n$ is larger than $1/2$,
and thus $2K_n\ge K_n^0=T_n$. Therefore,
\begin{align*}
\beta(\N|T_n) 
= 
\max_{1\le j\le m}\Tr\bz \sigma_j^{\otimes n}T_n\jz 
\le
2 \max_{1\le j\le m}\Tr\bz \sigma_j^{\otimes n}K_n\jz 
=
2 \max_{1\le j\le m}\sum_{i=1}^k\Tr\sigma_j^{\otimes n}\rho_i^{\otimes n}
\le
2 \bz\max_{1\le j\le m} \max_{1\le i\le k} \Tr(\sigma_j \rho_i)\jz^n.
\end{align*}
Hence,
\begin{align*}
\liminf_{n\to+\infty}\frac{1}{n}\log\beta(\N|T_n) \ge
-\max_{i,j}\log\Tr(\rho_i \sigma_j) =C_{\min},
\end{align*}
completing the proof.
\end{proof}

\section{The strong converse exponent}
\label{sec:sc}

By the definition of the Stein exponent $\sli(H_0\|H_1)$, if the type II error exponent 
$r>\sli(H_0\|H_1)$ then the type I error cannot converge to $0$. In typical scenarios
it even converges to $1$ (this is called the strong converse property), and 
hence in this case the exponent of interest 
is the following quantity:

\begin{definition}
The \ki{strong converse exponent} of the hypothesis testing problem $H_0$ vs.~$H_1$ with type II error exponent $r$ is defined as
\begin{align*}
\sc_r(H_0\|H_1):=
\inf\left\{\limsup_{n\to+\infty}-\frac{1}{n}\log(1-\alpha_n(H_0|T_n)):\,
\liminf_{n\to+\infty}-\frac{1}{n}\log\beta_n(H_1|T_n)\ge r
\right\},
\end{align*}
where the infimum is taken along all test sequences $(T_n)_{n\in\bN}$
satisfying the indicated condition. The relaxed exponents 
$\sc_r\gen(H_0\|H_1)$ for $t=\{0\}$, $\{1\}$, and $\{0,1\}$ are defined analogously to Definition 
\ref{errexp def}.
\end{definition}

Clearly, we have
\begin{align*}
\sc_r\gen(H_0\|H_1)=0,\ds\ds\ds r\le \sli\gen(H_0\|H_1),
\end{align*}
and hence we will be interested in the strong converse exponent only for 
type II error exponents $r>\sli\gen(H_0\|H_1)$.

Similarly to the case of the direct exponents, 
it follows immediately from the definitions that 
in the case of a consistent null hypothesis,
\begin{align}
\sc_r\bz(\{\rho_{n,i}\}_{i\in I})_{n\in\bN}\|(\A_n)_{n\in\bN}\jz
&\ge
\sc_r^{\{0\}}\bz(\{\rho_{n,i}\}_{i\in I})_{n\in\bN}\|(\A_n)_{n\in\bN}\jz
\ge
\sup_{i\in I}\sc_r\bz(\rho_{n,i})_{n\in\bN}\|(\A_n)_{n\in\bN}\jz,
\label{sc consistent inequalities}
\end{align}
and similar inequalities hold when the alternative hypothesis is consistent,
and when both hypotheses are consistent. 

In this section we analyze cases where the inequalities 
in \eqref{sc consistent inequalities} do/do not hold as equalities. 
Of course, 
\begin{align*}
I\ds\text{is finite}\ds\imp\ds
\sc_r\bz(\{\rho_{n,i}\}_{i\in I})_{n\in\bN}\|(\A_n)_{n\in\bN}\jz
=
\sc_r^{\{0\}}\bz(\{\rho_{n,i}\}_{i\in I})_{n\in\bN}\|(\A_n)_{n\in\bN}\jz,
\end{align*}
and therefore in this case there is only one inequality to consider.

\subsection{Composite null hypothesis and arbitrary alternative hypothesis}
\label{sec:sc countable null}

It was shown in \cite{BunthVrana2020} that 
$\sc_r\bz\{\rho_i\}_{i\in I}\|\sigma\jz=\max_i\sc_r(\rho_i\|\sigma)$
in the composite i.i.d.~vs.~simple i.i.d.~setting if the index set $I$ is finite. The same
proof method gives the equality in the more general setting of Proposition 
\ref{prop:sc general equality finite} below. We give a proof for completeness, and as a 
preparation for the proof of Proposition \ref{prop:sc general equality}.

\begin{prop}\label{prop:sc general equality finite}
Let $H_0:\,(\{\rho_{n,i}\}_{i\in I})_{n\in\bN}$ vs.~$H_1:\,(\A_n)_{n\in\bN}$ be a 
binary state discrimination problem with a consistent null hypothesis. If $I$ is finite then 
\begin{align}\label{sc general equality finite}
\sc_r\bz(\{\rho_{n,i}\}_{i\in I})_{n\in\bN}\|H_1\jz
=
\max_{i\in I}\sc_r\bz(\rho_{n,i})_{n\in\bN}\|H_1\jz.
\end{align}
\end{prop}
\begin{proof}
We have LHS$\ge$RHS in \eqref{sc general equality finite} due to the general inequalities in 
\eqref{sc consistent inequalities}, and hence we only have to prove the converse inequality. 
Let $s>\max_{i\in I}\sc_r\bz(\rho_{n,i})_{n\in\bN}\|H_1\jz$, so that  
for every $i\in I$ there exists a sequence of tests $T_{n,i}$, $n\in\bN$, satisfying 
\begin{align*}
\limsup_{n\to+\infty}-\frac{1}{n}\log(1-\alpha_n(\rho_{n,i}|T_{n,i}))<s,\ds\ds\ds
\liminf_{n\to+\infty}-\frac{1}{n}\log\beta_n(\N_n|T_{n,i})\ge r.
\end{align*}
Therefore there exists an $N\in\bN$ such that
$\min_i\rho_{n,i}(T_{n,i})>e^{-ns}$, $n\ge N$. 
Let $T_n:=|I|\inv\sum_{i\in I}T_{n,i}$, $n\in\bN$. Then 
\begin{align*}
\min_{i\in I}\rho_{n,i}(T_{n})>|I|\inv e^{-ns},\ds\ds\ds
\sup_{\sigma_n\in\A_n}\sigma_n(T_{n})
\le|I|\inv\sum_{i\in I}\sup_{\sigma_n\in\A_n}\sigma_n(T_{n,i}),\ds\ds\ds
n\ge N,
\end{align*}
from which the assertion follows immediately.
\end{proof}


Next, we give some general converses to the inequalities in 
\eqref{sc consistent inequalities}. For this, we will need the following:

\begin{definition}
For $\rho,\sigma\in\S(\M)$, their \ki{max-relative entropy} is defined as \cite{Datta,RennerPhD}
\begin{align*}
D_{\infty}\nw(\rho\|\sigma):=\inf\{\lambda\in\bR:\,\rho\le e^{\lambda}\sigma\}.
\end{align*}
\end{definition}

The following proposition gives a generalization of 
Proposition \ref{prop:sc general equality finite}. The main idea of the proof is due to 
\cite{Petiprivate}. 

\begin{prop}\label{prop:sc general equality}
Let $H_0:\,(\{\rho_{n,i}\}_{i\in I})_{n\in\bN}$ vs.~$H_1:\,(\A_n)_{n\in\bN}$ be a 
binary state discrimination problem with a consistent null hypothesis, and assume that there exists a countable max-relative entropy dense set for the null hypothesis in the following sense: there exists 
a countable $\tilde I\subseteq I$ such that for every $i\in I$ and every $\ep>0$ there exists 
a $k_{i,\ep}\in I$ and $N_{i,\ep}$ such that $\frac{1}{n}D_{\infty}\nw(\rho_{n,k_{i,\ep}}\|\rho_{n,i})\le\ep$ for every $n\ge N_{i,\ep}$.
Then 
\begin{align}\label{sc general equality}
\inf_{r'>r}\sup_{i\in \tilde I}\sc_{r'}\bz(\rho_{n,i})_{n\in\bN}\|H_1\jz
\ge
\sc_r^{\{0\}}\bz(\{\rho_{n,i}\}_{i\in I})_{n\in\bN}\|H_1\jz
&\ge
\sup_{i\in I}\sc_r\bz(\rho_{n,i})_{n\in\bN}\|H_1\jz
\end{align}
\end{prop}
\begin{proof}
When $I$ is finite, we have the stronger statement given in Proposition \ref{prop:sc general equality finite}, and hence for the rest we assume that 
$I$ is of infinite cardinality. In particular, $\tilde I$ is of countably infinite cardinality, and hence we may assume without loss of generality that 
$\tilde I=\bN$. 

Clearly, \eqref{sc general equality} is equivalent to 
\begin{align}\label{sc general equality1}
\sc_r^{\{0\}}\bz(\{\rho_{n,i}\}_{i\in I})_{n\in\bN}\|H_1\jz
\ge
\sup_{i\in I}\sc_r\bz(\rho_{n,i})_{n\in\bN}\|H_1\jz
&\ge
\sup_{i\in \tilde I}\sc_r\bz(\rho_{n,i})_{n\in\bN}\|H_1\jz\\
&\ge 
\sup_{r'<r}\sc_{r'}^{\{0\}}\bz(\{\rho_{n,i}\}_{i\in I})_{n\in\bN}\|H_1\jz,
\label{sc general equality2}
\end{align}
and we will prove the statement in this form.
The inequalities in \eqref{sc general equality1} are immediate from the general inequalities in \eqref{sc consistent inequalities}, and hence we only have to prove the inequality in 
\eqref{sc general equality2}.

Let $s> \sup_{i\in \tilde I}\sc_r\bz(\rho_{n,i})_{n\in\bN}\|H_1\jz$.
Then for every $k\in \tilde I$ there exists a sequence of tests $T_{n,k}$, $n\in\bN$, such that 
\begin{align}\label{sc general equality proof}
\limsup_{n\to+\infty}-\frac{1}{n}\log(1-\alpha_n(\rho_{n,k}|T_{n,k}))<s,\ds\ds\ds
\liminf_{n\to+\infty}-\frac{1}{n}\log\beta_n(\N_n|T_{n,k})\ge r.
\end{align}
Let us fix a probability distribution $(q_k)_{k\in\bN}$ with all $q_k>0$.
For every $\delta>0$ and every $n\in\bN$, let 
\begin{align*}
\tilde I_{n,\delta}:=\left\{k\in\tilde I\cap [n]:\,\rho_{m,k}(T_{m,k})>e^{-ms},\ds\ds\sup_{\sigma_m\in\A_m}\sigma_m(T_{m,k})<e^{-m(r-\delta)},\ds m\ge n\right\},\ds\ds\ds
T_n:=\sum_{k\in \tilde I_n}q_kT_{n,k}.
\end{align*}
Clearly, the sequence $(\tilde I_{n,\delta})_{n\in\bN}$ is increasing, and 
by \eqref{sc general equality proof},
$\cup_{n\in\bN}\tilde I_{n,\delta}=\tilde I$, i.e., for every $k\in\tilde I$ there exists an
$M_{k,\delta}\in\bN$ such that $k\in\tilde I_{n,\delta}$ for every $n\ge M_{k,\delta}$. 
For arbitrary $i\in I$ and $\ep>0$, let $k_{i,\ep}$ and $N_{i,\ep}$ be as in the assumption. 
Then for every $n\ge \max\{N_{i,\ep},M_{k_{i,\ep},\delta}\}$,
\begin{align*}
\rho_{n,i}(T_n)\ge e^{-n\ep}\rho_{n,k_{i,\ep}}(T_n)
\ge
e^{-n\ep}q_{k_{i,\ep}}\rho_{n,k_{i,\ep}}(T_{n,k_{i,\ep}})
>
q_{k_{i,\ep}}e^{-n(s+\ep)},\ds\ds\ds
\sup_{\sigma_n\in\A_n}\sigma_n(T_{n})<e^{-n(r-\delta)}.
\end{align*} 
This shows that 
$\sc_{r-\delta}^{\{0\}}\bz(\{\rho_{n,i}\}_{i\in I})_{n\in\bN}\|H_1\jz\le s+\ep$
for all $\ep>0$ and $s$ as above. Taking the infimum in $\ep$ and $s$, and then 
in $\delta$, yields \eqref{sc general equality2}.
\end{proof}

Let us now consider the case where the null hypothesis is composite i.i.d., given by a subset 
$\N\subseteq\S(\M)$. We say that a subset $\tilde\N\subseteq\N$ is 
\ki{max-relative entropy dense} in $\N$, if for every $\rho\in\N$ and every $\ep>0$ there exists a $\tilde\rho_{\ep}\in\tilde\N$ such that $D_{\infty}\nw(\tilde\rho_{\ep}\|\rho)<\ep$,
or equivalently, $e^{-\ep}\tilde \rho_{\ep}\le\rho$. Clearly then
$D_{\infty}\nw(\tilde\rho_{\ep}^{\otimes n}\|\rho^{\otimes n})<n\ep$ for every $n\in\bN$.
Hence, if a countable $\tilde\N$ exists with the above properties then 
the sufficient condition in Proposition \ref{prop:sc general equality} is satisfied, and therefore the inequalities in \eqref{sc general equality} hold.
Hence, our next goal is to find sufficient conditions for the existence of a 
countable max-relative entropy dense set.

\begin{lemma}\label{lemma:maxrelentr bound}
Let $\rho,\sigma$ be non-zero PSD operators on a finite-dimensional Hilbert space $\hil$. 
If $\rho^0\le\sigma^0$ then  
\begin{align}\label{Dinfty bounds}
\half\norm{\rho-\sigma}_1^2\le
D_{\infty}\nw(\rho\|\sigma)\le
\log\bz 1+\frac{1}{\lambda_{\min}(\sigma)}\norm{\rho-\sigma}_1\jz\le
\frac{1}{\lambda_{\min}(\sigma)}\norm{\rho-\sigma}_1,
\end{align}
where $\lambda_{\min}(\sigma)$ is the smallest non-zero eigen-value of $\sigma$.
\end{lemma}
\begin{proof}
The first inequality is an immediate consequence of the quantum Csisz\'ar-Pinsker inequality
\cite{HiaiOhyaTsukada}.
The duality of linear programming yields that 
\begin{align*}
D_{\infty}\nw(\rho\|\sigma)&=
\log\max\left\{\frac{\Tr \rho T}{\Tr\sigma T}:\,T\in\B(\hil)_{[0,I]},\,T^0\le\sigma^0\right\}
=
\log\max\left\{\frac{\Tr \rho\, \tau}{\Tr\sigma \tau}:\,\tau\in\S(\hil),\,\tau^0\le\sigma^0\right\}.
\end{align*}
By the H\"older inequality, $|\Tr\rho\, \tau-\Tr\sigma \tau|\le\norm{\rho-\sigma}_1$, and hence
\begin{align*}
\abs{\frac{\Tr \rho\, \tau}{\Tr\sigma \tau}-1}\le\frac{1}{\Tr\sigma\tau}\norm{\rho-\sigma}_1
\le
\frac{1}{\lambda_{\min}(\sigma)}\norm{\rho-\sigma}_1,
\end{align*}
from which the statement follows.
\end{proof}

\begin{lemma}\label{lemma:maxrelentr dense set}
Let $\hil$ be a finite-dimensional Hilbert space and $\N\subseteq\S(\hil)$.
If only countably many of the sets
\begin{align*}
\N_P:=\{\rho\in\N:\,\rho^0=P\},\ds\ds\ds P\ds\text{projection on }\hil,
\end{align*} 
are non-empty then 
there exists a countable max-relative entropy dense set in $\N$. This is satisfied, for instance, if $\N$ is a convex set and (i)  $\N\setminus\relint\N$ is countable, or (ii)
$\N$ is classical, i.e., any two elements in $\N$ commute. 
\end{lemma}
\begin{proof}
Finite-dimensionality implies that for each $P$, there exists a countable dense set 
$\tilde\N_P$ in $\N_P$ with respect to the 
trace-norm, and therefore the main claim follows immediately from Lemma 
\ref{lemma:maxrelentr bound} due to 
$\N=\bigcup_{P\text{ projection}}\N_P$.

Assume now that $\N$ is convex. It is easy to see that any two
$\rho,\sigma\in\relint\N$ have the same support, from which the assertion in case (i) follows. 
In case (ii), we may represent the elements on $\N$ as probability density functions on a 
finite set $\X$. Hence, there are only finitely many possible support sets, 
from which the assertion in case (ii) follows. 
\end{proof}

\begin{cor}\label{cor:general sc equality}
Consider a binary state discrimination problem with composite i.i.d.~null-hypothesis
$H_0:\,(\N^{\otimes n})_{n\in\bN}$ vs.~$H_1:\,(\A_n)_{n\in\bN}$.
If $\N$ is as in Lemma \ref{lemma:maxrelentr dense set} then 
\eqref{sc general equality} holds (with $I=\N$).
\end{cor}

We will utilize the above corollary in the proof of Corollary \ref{cor:adv sc}.

\begin{rem}
Let $\N\subseteq\S(\bC^2)$ be a set obtained by slicing off a part of the Bloch ball by a 
hyperplane, such that $|\N|>1$. Then $\N$ does not contain a countable max-relative entropy
dense set, because for every pure state $\pr{\psi}\in\N$ and every $\pr{\psi}\ne\rho\in\N$, 
$D_{\infty}\nw(\rho\|\pr{\psi})=+\infty$. 
This shows that max-relative entropy density is a strictly stronger notion than 
density in any norm, and also demonstrates how the sufficient conditions in 
Lemma \ref{lemma:maxrelentr dense set} might fail.
\end{rem}

\subsection{Divergences for the strong converse exponent}
\label{sec:sc divergences}

In the case of finite-dimensional simple i.i.d.~quantum state discrimination, the 
list of equalities in Lemma \ref{lemma:known exponents} can be continued with $\sc_r(\rho\|\sigma)=H_r\nw(\rho\|\sigma)$, where the quantity on right-hand side is the Hoeffding anti-divergence \cite{MO}.
In this section we discuss some important properties of the extension of these quantities to 
pairs of sets of states. While we will only use these later in the classical case, 
in Section \ref{sec:Hanti basics} we consider the more general quantum case, as this does not 
make the discussion more complicated. 

Similarly to the Hoeffding divergence, the Hoeffding anti-divergence is also a certain transform of a function defined from a family of R\'enyi $\alpha$-divergences
(denoted by $\psi$ and $\psi\nw$ in the respective cases),
which is convex in $\alpha$ for pairs of states. However, 
for a pair of sets of states $\N,\A$, the function is given by 
$\psi(\N\|\A|\alpha)=\sup_{\rho\in\N,\sigma\in\A}\psi(\rho\|\sigma|\alpha)$
in the first case, and hence it is still convex in $\alpha$, 
while in the second case it is 
$\psi\nw(\N\|\A|\alpha)=\inf_{\rho\in\N,\sigma\in\A}\psi\nw(\rho\|\sigma|\alpha)$,
and it is not clear whether convexity in $\alpha$ still holds. 
For this reason, establishing some of the relevant properties is more complicated for the 
Hoeffding anti-divergences, and in Section \ref{sec:Hanti LF} we give an analysis of these properties
under minimal assumptions on the properties of $\psi$.

\subsubsection{The Hoeffding anti-divergence}
\label{sec:Hanti basics}

For a density operator $\rho\in\S(\hil)$, a non-zero PSD operator $\sigma\in\B(\hil)\pne$, 
and $\alpha\in(0,+\infty)$, let 
\begin{align*}
Q_{\alpha}\nw(\rho\|\sigma):=
\begin{cases}
\Tr\bz\rho^{1/2}\sigma^{\frac{1-\alpha}{\alpha}}\rho^{1/2}\jz^{\alpha},
&\alpha\in(0,1)\text{ or }\rho^0\le\sigma^0,\\
+\infty,&\text{otherwise},
\end{cases}
\end{align*}
and
\begin{align*}
\psi\nw(\rho\|\sigma|\alpha):=\log Q_{\alpha}\nw(\rho\|\sigma).
\end{align*}
The \ki{sandwiched R\'enyi $\alpha$-divergences} of $\rho$ and $\sigma$ 
are defined as \cite{Renyi_new,WWY} 
\begin{align*}
D_{\alpha}\nw(\rho\|\sigma):=\frac{1}{\alpha-1}\psi\nw(\rho\|\sigma|\alpha),\ds\ds\ds
\alpha\in(0,+\infty)\setminus\{1\}.
\end{align*}
These notions have also been extended to the setting of general von Neumann algebras
in \cite{Sandwiched_Masuda,Jencova_NCLp,Jencova_NCLpII}, but here we only consider the finite-dimensional case.
It is known that 
$\alpha\mapsto D_{\alpha}\nw(\rho\|\sigma)$ is monotone non-decreasing, with
\begin{align}
D_{1}\nw(\rho\|\sigma):=\lim_{\alpha\to 1} D_{\alpha}\nw(\rho\|\sigma)=D(\rho\|\sigma),\ds\ds\ds
\lim_{\alpha\to +\infty} D_{\alpha}\nw(\rho\|\sigma)=D_{\infty}\nw(\rho\|\sigma)
:=\inf\{\lambda\in\bR:\,\rho\le e^{\lambda}\sigma\};
\label{sandwiched limits}
\end{align}
see \cite{Renyi_new}.

In the above setting, let us introduce the \ki{Hoeffding anti-divergence} of $\rho$ and $\sigma$ as 
\begin{align}
H_r\nw(\rho\|\sigma)
&:=
\max_{u\in[0,1]}\left[ur-\tilde\psi\nw(\rho\|\sigma|u)\right],\label{Hoeffding anti def2}
\end{align}
where $r>0$ is a parameter, and 
\begin{align*}
\tilde\psi\nw(\rho\|\sigma|u):=(1-u)\psi\nw(\rho\|\sigma|(1-u)\inv).
\end{align*}
(The existence of the $\max$ follows from properties of the $\psi\nw$ function that we establish below.)
In the above, we use the conventions 
\begin{align*}
\frac{+\infty-1}{+\infty}:=\lim_{\alpha\to+\infty}\frac{\alpha-1}{\alpha}=1,\ds\ds
0\cdot(\pm\infty):=0,\ds\ds
\tilde\psi\nw(\rho\|\sigma|1):=\lim_{u\nearrow 1}\tilde\psi\nw(\rho\|\sigma|u)=D_{\infty}\nw(\rho\|\sigma).
\end{align*}
The equality in \eqref{Hoeffding anti def2} follows by the simple change of variables $u:=(\alpha-1)/\alpha$. 

\begin{rem}
Note that for commuting states $\rho,\sigma$ (which can be assumed to be 
probability distributions on a finite set $\X$), 
$\psi\nw(\rho\|\sigma|\alpha)=\psi(\rho\|\sigma|\alpha)$ and
$\tilde\psi\nw(\rho\|\sigma|u)=\tilde\psi(\rho\|\sigma|u)$, 
where the latter were defined in 
\eqref{psi def classical}. However, $H_r(\rho\|\sigma)\ne H_r\nw(\rho\|\sigma)$ in general, even if the states commute, because in their definitions not only the types of R\'enyi divergences differ, but also the ranges of optimization.
\end{rem}

\begin{rem}
$H_r\nw$ is called an anti-divergence because it is monotone non-decreasing under CPTP maps; 
this is immediate from the monotonicity of $D_{\alpha}\nw$ under CPTP maps for $\alpha\ge 1$
\cite{Beigi,FL}. 
\end{rem}

The importance of the Hoeffding anti-divergence stems from the following fact \cite{MO}:
\begin{lemma}\label{lemma:simple sc exp}
In the simple binary i.i.d.~state discrimination problem $\rho$ vs.~$\sigma$,
\begin{align*}
\sc_r(\rho\|\sigma)=H_r\nw(\rho\|\sigma),\ds\ds\ds r>0.
\end{align*}
\end{lemma}
\medskip

For two sets $\N\subseteq\S(\hil)$ and $\A\subseteq\B(\hil)\pne$, let
their Hoeffding anti-divergence be defined as
\begin{align*}
H_r\nw(\N\|\A)&:=\sup_{\rho\in\N,\sigma\in\A}H_r\nw(\rho\|\sigma)=
\sup_{\alpha\in[1,+\infty]}\frac{\alpha-1}{\alpha}
\Big[r-\underbrace{\inf_{\rho\in\N,\sigma\in\A}D_{\alpha}\nw(\rho\|\sigma)}_{=:D_{\alpha}\nw(\N\|\A)}\Big]
=
\sup_{\alpha\in[1,+\infty]}
\frac{\alpha-1}{\alpha}\left[r-D_{\alpha}\nw(\N\|\A)\right].
\end{align*}
Note that 
\begin{align}
\lim_{\alpha\searrow 1}D_{\alpha}\nw(\N\|\A)
=
\inf_{\alpha>1}D_{\alpha}\nw(\N\|\A)
=
\inf_{\alpha>1}\inf_{\rho\in\N,\sigma\in\A}D_{\alpha}\nw(\rho\|\sigma)
=
\inf_{\rho\in\N,\sigma\in\A}\inf_{\alpha>1}D_{\alpha}\nw(\rho\|\sigma)
=
\inf_{\rho\in\N,\sigma\in\A}D(\rho\|\sigma)
=
D(\N\|\A),\label{composite D1}
\end{align}
and thus
\begin{align}\label{Hanti composite pos}
H_r\nw(\N\|\A)\ge 0,\ds\ds\ds\text{and}\ds\ds\ds
H_r\nw(\N\|\A)= 0\ds\iff\ds r\le D(\N\|\A).
\end{align}

\begin{rem}\label{rem:composite D infty}
Note that 
\begin{align*}
D_{\infty}\nw(\N\|\A)
&=
\inf_{\rho\in\N,\sigma\in\A}D_{\infty}\nw(\rho\|\sigma)
=
\inf_{\rho\in\N,\sigma\in\A}\sup_{\alpha>1}D_{\alpha}\nw(\rho\|\sigma)\\
&\ge
\sup_{\alpha>1}\inf_{\rho\in\N,\sigma\in\A}D_{\alpha}\nw(\rho\|\sigma)
=
\sup_{\alpha>1}D_{\alpha}\nw(\N\|\A)
=
\lim_{\alpha\to+\infty}D_{\alpha}\nw(\N\|\A).
\end{align*}
If $\N$ and $\A$ are compact then the inequality holds with equality; this follows from 
Lemma \ref{lemma:minimax2} and the lower semi-continuity of the sandwiched R\'enyi divergences; see Lemma \ref{lemma:Hanti usc} below.
\end{rem}

The following is immediate from the definition:
\begin{lemma}
For any $\N,\A\subseteq\S(\hil)$, the function $(0,+\infty)\ni r\mapsto H_r\nw(\N\|\A)$ is 
finite-valued, monotone increasing, and convex; in particular, it is continuous. 
\end{lemma}
\begin{proof}
Monotonicity is trivial from the definition, and convexity follows as 
$H_r\nw(\N\|\A)$ is the supremum of convex functions. For
$r\le\D(\N\|\A)$, $H_r(\N\|\A)=0$. For $r>D(\N\|\A)$, we have
\begin{align}\label{Hanti upper bound}
\underbrace{\frac{\alpha-1}{\alpha}}_{\le 1}\big[ r-\underbrace{D_{\alpha}\nw(\N\|\A)}_{\ge D(\N\|\A)}\big]\le r-D(\N\|\A)
\end{align}
for any $\alpha\ge 1$, and hence $H_r\nw(\N\|\A)\le r-D(\N\|\A)$, showing finiteness. 
Continuity then follows from convexity and finiteness.
\end{proof}

We will need certain further continuity properties of the above quantities.
Recall that an extended real-valued function $f$ on a topological space $X$ 
is \ki{upper semi-contunous}, if $\{x\in X:\,f(x)\ge c\}$ is closed for every $c\in\bR$. 
The following is easy to show from the definition:

\begin{lemma}\label{lemma:usc}
Let $X$ be a topological space, $Y$ be an arbitrary set, and 
$f:\,X\times Y\to\bR\cup\{\pm\infty\}$ be a function.
\begin{enumerate}
\item\label{usc1}
If $f(.,y)$ is upper semi-continuous for every $y\in Y$ then $\inf_{y\in Y}f(.,y)$ is upper semi-continuous.

\item\label{usc2}
If $Y$ is a compact topological space, and $f$ is upper semi-continuous on $X\times Y$ w.r.t.~the product topology, then 
$\sup_{y\in Y}f(.,y)$ is upper semi-continuous.
\end{enumerate}
\end{lemma}

For $\alpha\in(1,+\infty]$, the map $(0,+\infty)\ni\ep\mapsto (\sigma+\ep)^{\frac{1-\alpha}{\alpha}}$ is monotone non-increasing, and hence so is $\ep\mapsto Q_{\alpha}\nw(\rho\|\sigma+\ep)$. Moreover, it is easy to verify that 
\begin{align}
Q_{\alpha}\nw(\rho\|\sigma)=\sup_{\ep>0}Q_{\alpha}\nw(\rho\|\sigma+\ep),\ds\ds\ds
\psi\nw(\rho\|\sigma|\alpha)=\sup_{\ep>0}\psi\nw(\rho\|\sigma+\ep|\alpha),\ds\ds\ds
D_{\alpha}\nw(\rho\|\sigma)=\sup_{\ep>0}D_{\alpha}\nw(\rho\|\sigma+\ep),
\label{sandwiched smoothing}
\end{align}
where the last equality also holds for $\alpha=1$.

\begin{lemma}\label{lemma:Hanti usc}
For any fixed $r>0$, the map
\begin{align*}
[1,+\infty]\times\S(\hil)\times\B(\hil)\pne\ni (\alpha,\rho,\sigma)\mapsto
\frac{\alpha-1}{\alpha}\left[r-D_{\alpha}\nw(\rho\|\sigma)\right]
\end{align*}
is upper semi-continuous.
\end{lemma}
\begin{proof}
For any fixed $\ep>0$, the map
\begin{align*}
[1,+\infty]\times\S(\hil)\times\B(\hil)\pne\ni (\alpha,\rho,\sigma)\mapsto
\frac{\alpha-1}{\alpha}\left[r-D_{\alpha}\nw(\rho\|\sigma+\ep)\right]
\end{align*}
is continuous, (where $(+\infty-1)/+\infty:=1$), and hence 
\begin{align*}
[1,+\infty]\times\S(\hil)\times\B(\hil)\pne\ni (\alpha,\rho,\sigma)\mapsto
\frac{\alpha-1}{\alpha}\left[r-D_{\alpha}\nw(\rho\|\sigma)\right]
=
\inf_{\ep>0}\frac{\alpha-1}{\alpha}\left[r-D_{\alpha}\nw(\rho\|\sigma+\ep)\right],
\end{align*}
where the equality is due to \eqref{sandwiched smoothing},
is upper semi-continuous according to \ref{usc1} of Lemma \ref{lemma:usc}.
\end{proof}

\begin{cor}\label{cor:Hanti usc1}
For any fixed $r>0$, the map $\S(\hil)\times\B(\hil)\pne\ni(\rho,\sigma)\mapsto H_r\nw(\rho\|\sigma)$ is upper semi-continuous.
\end{cor}
\begin{proof}
Immediate from Lemma \ref{lemma:Hanti usc} and  
\ref{usc2} of Lemma \ref{lemma:usc}.
\end{proof}

\begin{cor}\label{cor:Hanti usc2}
For any fixed $r>0$ and compact sets $\N\subseteq\S(\hil)$, $\A\subseteq\B(\hil)\pne$,
there exist $\rho_r\in\N$, $\sigma_r\in\A$, and $\alpha_r\in[1,+\infty]$ such that 
\begin{align*}
H_r\nw(\N\|\A)=
\frac{\alpha_r-1}{\alpha_r}\left[r-D_{\alpha_r}\nw(\N\|\A)\right]=
\frac{\alpha_r-1}{\alpha_r}\left[r-D_{\alpha_r}\nw(\rho_r\|\sigma_r)\right]=H_r\nw(\rho_r\|\sigma_r).
\end{align*}
In particular, $D_{\alpha_r}\nw(\rho_r\|\sigma_r)=D_{\alpha_r}\nw(\N\|\A)$.
\end{cor}
\begin{proof}
Immediate from Lemma \ref{lemma:Hanti usc} and the fact that an upper semi-continuous function attains its supremum on a compact set.
\end{proof}


Some of the important properties of the Hoeffding anti-divergence, which we will need in 
Section \ref{sec:sc adv}, can be obtained from very general properties of the function 
$\psi(\alpha):=\inf_{\rho\in\N,\sigma\in\A}\psi\nw(\rho\|\sigma|\alpha)$. We discuss these in the following separate section, which essentially gives a generalization of Lemmas 15 and 16 in \cite{MO-correlated} to the case where $\psi$ is not assumed to be convex or differentiable. 

\subsubsection{Legendre-Fenchel representation of the Hoeffding anti-divergence}
\label{sec:Hanti LF}


Let $\psi:\,[1,+\infty)\to\bR$ be a function such that 
\begin{align*}
\psi(1)=0,\ds\ds\ds\text{and}\ds\ds\ds
\alpha\mapsto\frac{\psi(\alpha)}{\alpha-1}\ds\ds\text{is monotone increasing}.
\end{align*}
This implies that for the function $\tilde\psi(u):=(1-u)\psi\bz\frac{1}{1-u}\jz$, $u\in[0,1)$,
\begin{align*}
\tilde\psi(0)=0,\ds\ds\ds\text{and}\ds\ds\ds u\mapsto \frac{\tilde\psi(u)}{u}\ds\ds\text{is monotone increasing}.
\end{align*}
(Note that $\tilde\psi$ is the so-called transpose function of $\psi$ evaluated at $1-u$.)
The monotonicity assumption yields the existence of the 
half-sided limits
\begin{align*}
\psi(\alpha^-):=\lim_{\beta\nearrow \alpha}\psi(\beta),\ds\ds\ds
\psi(\alpha^+):=\lim_{\beta\searrow \alpha}\psi(\beta),\ds\ds\ds
\tilde\psi(u^-):=\lim_{v\nearrow u}\tilde\psi(v),\ds\ds\ds
\tilde\psi(u^+):=\lim_{v\searrow u}\tilde\psi(v)
\end{align*}
for every $\alpha\in(1,+\infty)$ and $u\in(0,1)$, and the existence of 
\begin{align}
D_1^+&:=\partial^+\psi(1):=
\lim_{\alpha\searrow 1}\frac{\psi(\alpha)}{\alpha-1}=
\lim_{u\searrow 0}\frac{\tilde\psi(u)}{u}=:\partial^+\tilde\psi(0)
\label{D1+}\\
D_{\infty}&:=\lim_{\alpha\to+\infty}\frac{\psi(\alpha)}{\alpha-1}=
\lim_{u\nearrow 1}\frac{\tilde\psi(u)}{u}=\lim_{u\nearrow 1}\tilde\psi(u)=:
\tilde\psi(1).\label{Dinfty}
\end{align}
(It is easy to see that if $\psi$ is convex then 
$D_{\infty}=\lim_{\alpha\to+\infty}\partial^-\psi(\alpha)=\lim_{\alpha\to+\infty}\partial^+\psi(\alpha)$.)
%
%
For every $c\in\bR$, let 
\begin{align*}
\Psi(c):=\sup_{1\le\alpha<+\infty}\{c\alpha-\psi(\alpha)\},\ds\ds\ds\ds\ds
\Psi^-(c):=\sup_{1\le\alpha<+\infty}\{c(\alpha-1)-\psi(\alpha)\}=\Psi(c)-c,
\end{align*}
where $\Psi$ is the Legendre-Fenchel transform of $\psi$.

\begin{lemma}\label{lemma:Psi prop}
\begin{enumerate}
\item\label{Psi prop1}
$\Psi$ and $\Psi^-$ are convex, increasing, and lower semi-continuous on $\bR$.
\item\label{Psi prop2}
For every $c\in\bR$, $\Psi(c)\ge c$, $\Psi^-(c)\ge 0$, and the inequalities are strict if and only if $c>D_1^+$. 
\item\label{Psi prop3}
$\Psi(c)=\Psi^-(c)=+\infty$ for $c>D_{\infty}$.
\setcounter{szamlalo}{\value{enumi}}
\end{enumerate}
\bigskip

\noindent Assume for the rest that $D_1^+>-\infty$.

\begin{enumerate}
\setcounter{enumi}{\value{szamlalo}}

\item\label{Psi prop4}
$\Psi(c)$ and $\Psi^-(c)$ are finite for $c<D_{\infty}$.

\item\label{Psi prop7}
$\Psi$ and $\Psi^-$ are continuous on $(-\infty,D_{\infty}]$, and $\Psi$ is strictly 
increasing on $(-\infty,D_{\infty}]$.

\item\label{Psi prop5}
For every $c\in(D_1^+,D_{\infty})$ there exist 
$1<\alpha_{c,\min}<\alpha_c\le\alpha_{c,max}<+\infty$ such that
\begin{align}\label{Psi restricted opt2}
\sup_{\alpha\in[1,+\infty)\setminus[\alpha_{c,\min},\alpha_{c,\max}]}
\{(\alpha-1)c-\psi(\alpha)\}
<\frac{\Psi^-(c)}{2}<\Psi^-(c),\ds\ds\ds\text{and}\ds\ds\ds\ds
\Psi^-(c)=c(\alpha_c-1)-\psi(\alpha_c^-).
\end{align}

\item\label{Psi prop6}
$\Psi^-$ is strictly increasing on 
$(D_1^+,D_{\infty}]$.
\end{enumerate}
\end{lemma}
\begin{proof}
The properties listed in \ref{Psi prop1} are obvious from the definitions of  
$\Psi$ and $\Psi^-$.

The inequalities in \ref{Psi prop2} are obvious from taking $\alpha=1$, for which 
$c(\alpha-1)-\psi(\alpha)=0$. We have 
\begin{align*}
\Psi^-(c)=
\max\left\{0,\sup_{\alpha>1}\left\{(\alpha-1)\left[c-\frac{\psi(\alpha)}{\alpha-1}\right]\right\}\right\}>0
\ds\iff\ds
\exists\, \alpha>1:\,c-\frac{\psi(\alpha)}{\alpha-1}>0\ds\iff\ds
c>\inf_{\alpha>1}\frac{\psi(\alpha)}{\alpha-1}=D_1^+,
\end{align*}
proving the assertion about the strict inequalities in \ref{Psi prop2}. 
If $D_{\infty}<+\infty$ then 
\begin{align*}
\Psi^-(c)\ge\limsup_{\alpha\to+\infty}(\alpha-1)\Bigg[c-D_{\infty}+\underbrace{D_{\infty}-\frac{\psi(\alpha)}{\alpha-1}}_{\ge 0}\Bigg]
\ge 
\lim_{\alpha\to+\infty}(\alpha-1)\left[c-D_{\infty}\right],
\end{align*}
and the last expression is equal to $+\infty$ for $c>D_{\infty}$, proving \ref{Psi prop3}.

Assume for the rest that $D_1^+>-\infty$.
If $c<D_{\infty}$ then $(\alpha-1)\left[c-\frac{\psi(\alpha)}{\alpha-1}\right]\le 0$ for all $\alpha>\alpha_{c,\max}:=\sup\left\{\alpha>1:\,\frac{\psi(\alpha)}{\alpha-1}<c\right\}<+\infty$ .
As a consequence, $\Psi^-(c)\le(\alpha_{c,\max}-1)\left[c-D_1^+\right]$, proving
\ref{Psi prop4}.

Since $\Psi$ and $\Psi^-$ are convex and lower semi-continuous, they are also continuous on the closure of the interval where they take finite values; by 
\ref{Psi prop3} and \ref{Psi prop4}, this is $(-\infty,D_{\infty}]$. Since $\Psi^-$ is increasing on $(-\infty,D_{\infty}]$,
and $\Psi(c)=\Psi^-(c)+c$, it is clear that $\Psi$ is strictly increasing on the same interval. This proves \ref{Psi prop7}.

$D_1^+>-\infty$ implies that $\lim_{\alpha\searrow 1}(\alpha-1)\left[c-\frac{\psi(\alpha)}{\alpha-1}\right]=0$
for every $c\in\bR$. 
If $c>D_1^+$ then $\Psi^-(c)>0$ by \ref{Psi prop2}, and by the observation in the previous sentence, $(\alpha-1)c-\psi(\alpha)<\Psi^-(c)/2$ for $\alpha\in[1,\alpha_{c,\min})$ with some $\alpha_{c,\min}>1$. 
Hence, if $c\in(D_1^+,D_{\infty})$ then the inequalities in 
\eqref{Psi restricted opt2} hold. 
Therefore,
\begin{align}\label{Psi restricted opt}
\Psi^-(c)=\sup_{\alpha_{c,\min}\le\alpha\le \alpha_{c,\max}}\left\{(\alpha-1)\left[c-\frac{\psi(\alpha)}{\alpha-1}\right]\right\}.
\end{align}
Consequently, for every $n\in\bN$ there exists an $\alpha_{c,n}\in [\alpha_{c,\min}, \alpha_{c,\max}]$
such that $c(\alpha_{c,n}-1)-\psi(\alpha_{c,n})>\Psi^-(c)-1/n$. Thus, there exists a strictly increasing function $k:\,\bN\to\bN$ such that $(\alpha_{c,k(n)})_{n\in\bN}$ is monotone increasing or decreasing, and converges to some $\alpha_c\in [\alpha_{c,\min}, \alpha_{c,\max}]$. Hence,
\begin{align}\label{Psi approximate optimizer proof}
\Psi^-(c)\ge 
\lim_{n\to+\infty}(\alpha_{c,k(n)}-1)\left[c-\frac{\psi(\alpha_{c,k(n)})}{\alpha_{c,k(n)}-1}\right]\ge\lim_{n\to+\infty}(\Psi^-(c)-1/k(n))=\Psi^-(c),
\end{align}
and therefore both inequalities are equalities. If the sequence $(\alpha_{c,k(n)})_{n\in\bN}$ is monotone increasing then the first limit in \eqref{Psi approximate optimizer proof} is 
$c(\alpha_c-1)-\psi(\alpha_c^-)$, as required. If $(\alpha_{c,k(n)})_{n\in\bN}$ is monotone decreasing then the first limit in \eqref{Psi approximate optimizer proof} is 
$c(\alpha_c-1)-\psi(\alpha_c^+)$. This shows that 
$\psi(\alpha_c^+)=\psi(\alpha_c^-)$, since otherwise 
$\Psi^-(c)=(\alpha_c-1)\left[c-\frac{\psi(\alpha_{c}^+)}{\alpha_{c}-1}\right]
<(\alpha_c-1)\left[c-\frac{\psi(\alpha_{c}^-)}{\alpha_{c}-1}\right]\le\Psi^-(c)$, a contradiction. This proves \ref{Psi prop5}.

Let $c\in(D_1^+,D_{\infty})$, and $\alpha_c$ be as above. 
For every $c'>c$,
\begin{align*}
\Psi^-(c')&\ge 
\lim_{\alpha\nearrow\alpha_c}\{c'(\alpha-1)-\psi(\alpha)\}
=
c'(\alpha_{c}-1)-\psi(\alpha_{c}^-)=
c(\alpha_{c}-1)-\psi(\alpha_{c}^-)+(c'-c)(\alpha_{c}-1)\\
&=\Psi^-(c)+(c'-c)(\alpha_{c}-1)>\Psi^-(c),
\end{align*}
showing the strict monotonicity of $\Psi^-$ on $(D_1^+,D_{\infty}]$, as stated in 
\ref{Psi prop6}.
\end{proof}

Consider now the Legendre-Fenchel transform of $\tilde\psi$, 
\begin{align}
\tilde\Psi(r):=\sup_{0\le u\le 1}\{ur-\tilde\psi(u)\}
=\sup_{0\le u< 1}\{ur-\tilde\psi(u)\}
=
\sup_{1\le\alpha}\left[\frac{\alpha-1}{\alpha}r-\frac{\psi(\alpha)}{\alpha}\right],
\end{align}
where the first equality follows as 
$r-\tilde\psi(1)=\lim_{u\nearrow 1}\{ur-\tilde\psi(u)\}$, according to \eqref{Dinfty},
and the second equality follows by the change of variables $u=(\alpha-1)/\alpha$.

By the monotonicity of $\Psi$, (\ref{Psi prop1} of Lemma \ref{lemma:Psi prop}), 
the limits 
\begin{align}\label{r1 plus}
r_{1}^+:=\lim_{c\searrow D_1^+}\Psi(c),
\ds\ds\ds
\rmax:=\lim_{c\nearrow D_{\infty}}\Psi(c),
\end{align}
exist. If $-\infty<D_1^+$, then 
\ref{Psi prop1} and \ref{Psi prop2} of Lemma \ref{lemma:Psi prop} imply
\begin{align}\label{r1 plus 2}
r_{1}^+&=\Psi(D_1^+)=
D_1^+=\partial^+\tilde\psi(0),
\end{align}
If $D_{\infty}<+\infty$ then 
\begin{align}
\rmax&=\Psi(D_{\infty})\ge
\lim_{\alpha\to+\infty}\{\alpha D_{\infty}-\psi(\alpha)\}
=
\lim_{u\nearrow 1}\frac{\tilde\psi(u)-D_{\infty}}{u-1}
=
\partial^-\tilde\psi(1).\label{r infty 2}
\end{align}
It is easy to see that if $\psi$ is convex then the inequality in \eqref{r infty 2} is an equality, i.e., 
$\rmax=\partial^-\tilde\psi(1)$, in complete analogy with 
$r_1^+=\partial^+\tilde\psi(0)$ in \eqref{r1 plus 2}; see also \eqref{psi tilde derivative lim}.

\begin{prop}\label{prop:Hanti LF}
In the above setting assume that $D_1^+>-\infty$.
\begin{enumerate}
\item\label{Hanti LF1}
$\tilde\Psi$ is a finite-valued, monotone increasing, convex, and continuous function on $\bR$. 
\item\label{Psi tilde pos}
$\tilde\Psi(r)\ge 0$, and $\tilde\Psi(r)=0$ $\iff$ $r\le r_1^+$. 
\item
The Legendre-Fenchel transform of $\tilde\psi$ can be expressed as
\begin{align}\label{Hanti LF2}
\tilde\Psi(r)
=u_rr-\tilde\psi(u_r^-)
=\begin{cases}
0,&r\le r_1^+,\\
r-\Psi\inv(r)=\Psi^-\bz\Psi\inv(r)\jz,&r\in(r_1^+,\rmax),\\
r-D_{\infty},&r\ge \rmax,
\end{cases}
\end{align}
where 
$u_r:=\frac{\alpha_{\Psi\inv(r)}-1}{\alpha_{\Psi\inv(r)}}\in(0,1)$ with $\alpha_{\Psi\inv(r)}\in(1,+\infty)$ as in 
\ref{Psi prop5} of Lemma \ref{lemma:Psi prop} when 
$r\in(r_1^+,\rmax)$, and $u_r:=1$ when $r\ge \rmax$.
When $r\le r_1^+$, we define $u_r:=0$ and $\tilde\psi(0^-):=\tilde\psi(0):=0$.
\item
For every $r\in(r_1^+,\rmax)$ there exist 
$0<u_{r,\min}< u_{r,\max}<1$ such that 
\begin{align}\label{Hanti LF optimal range}
\sup_{u\in[0,1]\setminus[u_{r,\min}, u_{r,\max}]}\{ur-\tilde\psi(u)\}<\tilde\Psi(r).
\end{align}
\item\label{Hanti LF large r}
For every $r> \rmax$, and every $u\in[0,1)$, 
$ur-\tilde\psi(u)<\tilde\Psi(r)$.
\end{enumerate}
\end{prop}
\begin{proof}
The assertion in \ref{Hanti LF1} is immediate from the definition. 

Clearly, $\tilde\Psi(r)\ge 0\cdot r-\tilde\psi(0)=0$. For any $u\in(0,1]$, 
$ur-\tilde\psi(u)=u(r-\tilde\psi(u)/u)\le 0$ if $r\le r_1^+=D_1^+=\inf_{u\in(0,1]}\frac{\tilde\psi(u)}{u}$, showing that $\tilde\Psi(r)=0$. By the same reasoning, 
$ur-\tilde\psi(u)>0$ for some $u\in(0,1]$ if $r>r_1^+$, showing that $\tilde\Psi(r)>0$. 
This proves \ref{Psi tilde pos} and the first case in \eqref{Hanti LF2}.

Assume next that $r\in(r_1^+,\rmax)$, and let $c_r:=\Psi\inv(r)$.
By the assumption on $r$, $c_r\in(D_1^+,D_{\infty})$. 
Let 
$\alpha_{c_r}$ be as in \ref{Psi prop5} of Lemma \ref{lemma:Psi prop}. For any 
$\alpha\ge 1$, 
\begin{align}\label{Hanti LF proof1}
\Psi^-(c_r)-\left[\frac{\alpha-1}{\alpha}r-\frac{\psi(\alpha)}{\alpha}\right]
=
\frac{1}{\alpha}\left[\Psi^-(c_r)-\bz c_r(\alpha-1)-\psi(\alpha)\jz\right]\ge 0,
\end{align}
where the equality follows by substituting $r=\Psi(c_r)=\Psi^-(c_r)+c_r$.
Taking the limit $\alpha\nearrow\alpha_{c_r}$ yields
\begin{align*}
\Psi^-(c_r)-\left[\frac{\alpha_{c_r}-1}{\alpha_{c_r}}r-\frac{\psi(\alpha_{c_r}^-)}{\alpha_{c_r}}\right]
=
\frac{1}{\alpha_{c_r}}\left[\Psi^-(c_r)-\bz c_r(\alpha_{c_r}-1)-\psi(\alpha_{c_r}^-)\jz\right]=0,
\end{align*}
where the second equality is due to \ref{Psi prop5} of Lemma \ref{lemma:Psi prop}.
This shows that 
\begin{align*}
\Psi^-(c_r)&=\sup_{\alpha\ge 1}
\left[\frac{\alpha-1}{\alpha}r-\frac{\psi(\alpha)}{\alpha}\right]
=\left[\frac{\alpha_{c_r}-1}{\alpha_{c_r}}r-\frac{\psi(\alpha_{c_r}^-)}{\alpha_{c_r}}\right]\\
&=\sup_{0\le  u<1}\{ur-\tilde\psi(u)\}
=
u_rr-\tilde\psi(u_r^-),
\end{align*}
where the expressions in the second line follow by the change of variables
$u:=(\alpha-1)/\alpha$. 
This proves the second case in \eqref{Hanti LF2}.

Let $u_{r,\min}:=(\alpha_{c_r,\min}-1)/\alpha_{c_r,\min}$, 
with $\alpha_{c_r,\min}$ as in \ref{Psi prop5} of Lemma \ref{lemma:Psi prop}. Combining \eqref{Hanti LF proof1} and \eqref{Psi restricted opt2} yields
that for every $u<u_{r,\min}$ and corresponding $\alpha=1/(1-u)<\alpha_{c_r,\min}$,
\begin{align*}
\tilde\Psi(r)-(ur-\tilde\psi(u))=\Psi^-(c_r)-\left[\frac{\alpha-1}{\alpha}r-\frac{\psi(\alpha)}{\alpha}\right]
=
\frac{1}{\alpha}\Big[\Psi^-(c_r)-
\underbrace{\bz c_r(\alpha-1)-\psi(\alpha)\jz}_{<\Psi^-(c_r)/2}\Big]
>
\frac{\Psi^-(c_r)}{2\alpha_{c_r,\min}}
>0.
\end{align*}
On the other hand, let $\delta>0$ be such that $c_r+\delta<D_{\infty}$, and define
$\alpha_{c_r,\delta}:=\sup\{\alpha\ge 1:\,\frac{\psi(\alpha)}{\alpha-1}\le c_r+\delta\}$,
$u_{r,\max}:=(\alpha_{c_r,\delta}-1)/\alpha_{c_r,\delta}$.
Then $1<\alpha_{c_r,\min}<\alpha_{c_r,\delta}<+\infty$ and thus 
$0<u_{r,\min}<u_{r,\max}<1$. 
By \eqref{Hanti LF proof1}, for any $u>u_{r,\max}$, and corresponding $\alpha=1/(1-u)>\alpha_{c_r,\delta}$,
\begin{align*}
\tilde\Psi(r)-\{ur-\tilde\psi(u)\}=
\frac{1}{\alpha}\Big[\Psi^-(c_r)-(\alpha-1)\Big( c_r-\underbrace{\frac{\psi(\alpha)}{\alpha-1}}_{>c_r+\delta}\Big)\Big]
>\frac{1}{\alpha}\Psi^-(c_r)+\frac{\alpha-1}{\alpha}\delta>\bz 1-\frac{1}{\alpha_{c_r,\delta}}\jz\delta>0.
\end{align*}
This proves \eqref{Hanti LF optimal range}.

The third case in \eqref{Hanti LF2} is only interesting when $\rmax<+\infty$ (since otherwise there is nothing to prove), in which case we have 
$D_{\infty}=\lim_{c\to D_{\infty}}c\le\lim_{c\to D_{\infty}}\Psi(c)=\rmax<+\infty$, and thus $\rmax=\Psi(D_{\infty})$, according to Lemma \ref{lemma:Psi prop}.
This yields that for any $\alpha\ge 1$, $r\ge \Psi(D_{\infty})\ge\alpha D_{\infty}-\psi(\alpha)$, whence $D_{\infty}\le (r+\psi(\alpha))/\alpha$, and thus,
\begin{align*}
r-D_{\infty}\ge\sup_{\alpha\ge 1}\left[ \frac{\alpha-1}{\alpha}r-\frac{\psi(\alpha)}{\alpha}\right]=\sup_{0\le u<1}\{ur-\tilde\psi(u)\}.
\end{align*}
Obviously, $r-D_{\infty}=\lim_{u\nearrow 1}\{ur-\tilde\psi(u)\}\le\sup_{0\le u<1}\{ur-\tilde\psi(u)\}$, completing the proof in the third case in \eqref{Hanti LF2}.
When $r>\rmax$ then we have the strict inequality 
$r>\Psi(D_{\infty})\ge\alpha D_{\infty}-\psi(\alpha)$, which yields the strict inequality 
$r-D_{\infty}>\frac{\alpha-1}{\alpha}r-\frac{\psi(\alpha)}{\alpha}$ for any 
$\alpha\ge 1$, proving \ref{Hanti LF large r}. 
\end{proof}

\begin{cor}\label{cor:Hanti composite}
Let $\hil$ be a finite-dimensional Hilbert space, and 
$\N,\A\subseteq\S(\hil)$ be compact subsets such that 
there exist $\rho\in\N$ and $\sigma\in\A$ for which $\rho^0\le\sigma^0$.
For every $\alpha\in[1,+\infty)$ and $c\in\bR$, let 
\begin{align*}
\psi(\alpha):=\inf_{\rho\in\N,\sigma\in\A}\psi\nw(\rho\|\sigma|\alpha),
\ds\ds\ds
\Psi(c):=\sup_{\alpha\in[1,+\infty)}\{c\alpha-\psi(\alpha)\}.
\end{align*}
The following hold:
\begin{enumerate}
\item
$\Psi$ is an increasing convex and continuous function, 
which is strictly increasing on $(-\infty,D_{\infty}\nw(\N\|\A)]$.
\item
The Hoeffding anti-divergence $r\mapsto H_r\nw(\N\|\A)$ is convex, continuous and strictly increasing on $[D(\N\|\A),+\infty)$, and it can be expressed as 
\begin{align}\label{Hanti composite LF}
H_r\nw(\N\|\A)=u_rr-\tilde\psi(\rho_r\|\sigma_r|u_r)
=
\begin{cases}
0,&r\le D(\N\|\A),\\
r-\Psi\inv(r)=\Psi^-(\Psi\inv(r)),&r\in(D(\N\|\A),\rmax(\N\|\A)),\\
r-D_{\infty}\nw(\N\|\A),&r\ge \rmax(\N\|\A),
\end{cases}
\end{align}
for some $\rho_r\in\N$, $\sigma_r\in\A$ and $u_r\in[0,1]$, where
$\rmax(\N\|\A):=\Psi(D_{\infty}\nw(\N\|\A))$. 
\item
For any
$u_r$ satisfying the first equality in \eqref{Hanti composite LF},
\begin{align*}
r\le D(\N\|\A)\imp u_r=0,\ds\ds\ds
r\in(D(\N\|\A),\rmax(\N\|\A))\imp u_r\in(0,1),\ds\ds\ds
r>\rmax(\N\|\A)\imp u_r=1.
\end{align*}
\end{enumerate}
\end{cor}
\begin{proof}
It is clear that $\psi$ is finite-valued, 
$\psi(0)=0$, and 
$\alpha\mapsto \frac{\psi(\alpha)}{\alpha-1}=
\inf_{\rho\in\N,\sigma\in\A}\frac{\psi(\rho\|\sigma|\alpha)}{\alpha-1}=
D_{\alpha}\nw(\N\|\A)$ is monotone increasing. 
We have 
\begin{align*}
D_1^+
=\inf_{\alpha>1}\frac{\psi(\alpha)}{\alpha-1}
=D(\N\|\A)\ge 0,\ds\ds\ds
D_{\infty}
=\sup_{\alpha>1}\frac{\psi(\alpha)}{\alpha-1}
=D_{\infty}\nw(\N\|\A)<+\infty,
\end{align*}
according to \eqref{composite D1} and Remark \ref{rem:composite D infty}, and hence the conditions in Proposition \ref{prop:Hanti LF} are satisfied. 
The existence of $\rho_r,\sigma_r$ and $u_r$ with the given properties is immediate from 
Corollary \ref{cor:Hanti usc2}, and the rest of the statements follow from 
Proposition \ref{prop:Hanti LF}.
\end{proof}

\subsection{Classical systems}
\label{sec:classical sc}

\subsubsection{Two alternative hypotheses in a $2$-dimensional classical system: Equality might fail}
\label{sec:sc two alt}

In this section we demonstrate by an explicit example that the equality 
$\sc_r(\rho\|\A)=\sup_{\sigma\in\A}\sc_r(\rho\|\sigma)$ may fail to hold 
in the simplest possible setting, where the system is classical and $2$-dimensional, and 
$\A$ has only two elements. 
The example may be considered as a hypothesis testing between a fair coin and two biased coins, given by 
\begin{align*}
\rho,\sigma_1,\sigma_2:\,\{h=\text{heads},t=\text{tails}\}\to[0,1],\ds\ds\ds
\rho(h):=\frac{1}{2},\ds
\sigma_1(h):=\frac{1}{4},\ds
\sigma_2(h):=\frac{3}{4}.
\end{align*}

One might suspect that the failure of equality is a consequence of the fact that 
$\rho\in\co(\{\sigma_1,\sigma_2\})$. However, as we show below, equality fails even if 
we replace all states with the same states on some fixed tensor power $k$, while 
$\rho^{\otimes k}\notin\co(\{\sigma_1^{\otimes k},\sigma_2^{\otimes k}\})$ for any 
$k\ge 2$, as one can easily verify.

\begin{prop}
For $\rho,\sigma_1,\sigma_2$ as above, and arbitrary $k\in\bN$,
\begin{align*}
\sc_r(\rho^{\otimes k}\|\{\sigma_1^{\otimes k},\sigma_2^{\otimes k}\})-\max_{j\in\{1,2\}}\sc_r(\rho^{\otimes k}\|\sigma_j^{\otimes k})
\begin{cases}
>0,&r>D(\rho^{\otimes k}\|\sigma_j^{\otimes k})=k\log\frac{2}{\sqrt{3}},\\
\ge k\log\sqrt{3},&r\ge k\log 4.
\end{cases}
\end{align*}
\end{prop}
\begin{proof}
Let $\hat\rho:=\rho^{\otimes k}$, 
$\hat\sigma_1:=\sigma_1^{\otimes k}$,
$\hat\sigma_2:=\sigma_2^{\otimes k}$.
By Lemma \ref{lemma:classical strict convexity}, $\psi(\hat\rho\|\hat\sigma_j|\valt)$
is strictly convex, and hence
\begin{align*}
k\log\frac{2}{\sqrt{3}}=D(\hat\rho\|\hat\sigma_j)<D_{\alpha}\nw(\hat\rho\|\hat\sigma_j)<D_{\infty}\nw(\hat\rho\|\hat\sigma_j)=k\log 2<\rmax(\hat\rho\|\hat\sigma_j)=k\log 4,
\end{align*}
for $j=1,2$, and any $\alpha\in(1,+\infty)$, where the explicit values follow by a straightforward computation (using \eqref{psi tilde derivative lim} for $\rmax$).
In particular, 
\begin{align}\label{sc 2 class alt proof}
\sc_r(\hat\rho\|\hat\sigma_j)=H_r\nw(\hat\rho\|\hat\sigma_j)=\max_{\alpha\in[1,+\infty]}\frac{\alpha-1}{\alpha}\left[r-D_{\alpha}\nw(\hat\rho\|\hat\sigma_j)\right]<r-D(\hat\rho\|\hat\sigma_j)
=r-k\log\frac{2}{\sqrt{3}}
\end{align}
for any $r>D(\hat\rho\|\hat\sigma_j)$, where the first equality is due to 
\eqref{lemma:simple sc exp}, and the strict inequality is straightforward to verify. 

For any sequence of tests $(T_n)_{n\in\bN}$, we have 
\begin{align}
1-\alpha_n(H_0|T_n)&=
\sum_{\vecc{x}\in\{h,t\}^{kn}} \hat\rho^{\otimes n}(\vecc{x}) T_n (\vecc{x})
=
\sum_{\{h,t\}^{kn}} \frac{\hat\rho^{\otimes n}(\vecc{x})}{\hat\sigma_1^{\otimes n}(\vecc{x})+\hat\sigma_2^{\otimes n}(\vecc{x})}\left(\hat\sigma_1^{\otimes n}(\vecc{x})+\hat\sigma_2^{\otimes n}(\vecc{x})\right) T_n (\vecc{x})
\nonumber
\\
&\le
\max_{\vecc{x}\in\{h,t\}^{kn}}\!\left(
\frac{\hat\rho^{\otimes n}(\vecc{x})}{\hat\sigma_1^{\otimes n}(\vecc{x})+\hat\sigma_2^{\otimes n}(\vecc{x})}\right) \,
\sum_{\{h,t\}^{kn}} \left(\hat\sigma_1^{\otimes n}(\vecc{x})+\hat\sigma_2^{\otimes n}(\vecc{x})\right) T_n(\vecc{x})\nonumber
\\
& \le
\max_{\vecc{x}\in\{h,t\}^{kn}}\!\left(
\frac{\hat\rho^{\otimes n}(\vecc{x})}{\hat\sigma_1^{\otimes n}(\vecc{x})+\hat\sigma_2^{\otimes n}(\vecc{x})}\right) 2 \beta_n(H_1|T_n)\nonumber
\\
&\le
\frac{1}{2}\left(\frac{2}{\sqrt{3}}\right)^{kn} 2 \beta_n(H_1|T_n),
\label{sc 2 class alt proof3}
\end{align}
where the first two inequalities are obvious, and the last inequality follows 
by $\hat\rho^{\otimes n}(\vecc{x})=1/2^{kn}$, $\vecc{x}\in\{h,t\}^{kn}$, and
an application of the inequality between the arithmetic and geometric means, as
\begin{align*}
\hat\sigma_1^{\otimes n}(\underline{x})+\hat\sigma_2^{\otimes n}(\underline{x})=
\left(\frac{1}{4}\right)^{m}
\left(\frac{3}{4}\right)^{kn-m}
+
\left(\frac{3}{4}\right)^m
\left(\frac{1}{4}\right)^{kn-m}
\geq 2\,\sqrt{
\left(\frac{1}{4}\right)^m
\left(\frac{3}{4}\right)^{kn-m}
\left(\frac{3}{4}\right)^m
\left(\frac{1}{4}\right)^{kn-m}
}
= 2 \left(\frac{\sqrt{3}}{4}\right)^{kn},
\end{align*}
where $m:=|\{i:\,x_i=h\}|$.
Hence, if $\liminf_{n\to+\infty}-\frac{1}{n}\log\beta_n(H_1|T_n)\ge r$ then 
\eqref{sc 2 class alt proof3} yields
\begin{align}\label{sc 2 class alt proof2}
\liminf_n-\frac{1}{n}\log(1-\alpha_n(H_0|T_n))
\ge k\log\frac{\sqrt{3}}{2}+\liminf_n-\frac{1}{n}\log\beta_n(H_1|T_n)
\ge k\log\frac{\sqrt{3}}{2}+r=r-D(\hat\rho\|\hat\sigma_j).
\end{align}
Hence,  
\begin{align*}
\sc_r(\hat\rho\|\{\hat\sigma_1,\hat\sigma_2\})\ge r-D(\hat\rho\|\hat\sigma_j)>
H_r\nw(\hat\rho\|\hat\sigma_j)=\sc_r(\hat\rho\|\hat\sigma_j),
\end{align*}
where the second inequality holds for $r>D(\hat\rho\|\hat\sigma_j)$, according to
\eqref{sc 2 class alt proof}.
Finally, when 
$r\ge k\log 4=\rmax(\hat\rho\|\{\hat\sigma_1,\hat\sigma_2\})$, we have 
\begin{align*}
\sc_r(\hat\rho\|\{\hat\sigma_1,\hat\sigma_2\})-\max_{j\in\{1,2\}}\sc_r(\hat\rho\|\hat\sigma_j)
\ge r-D(\hat\rho\|\hat\sigma_j)-\bz r-D_{\infty}\nw(\hat\rho\|\hat\sigma_j)\jz
=
D_{\infty}\nw(\hat\rho\|\hat\sigma_j)-D(\hat\rho\|\hat\sigma_j)=k\log\sqrt{3},
\end{align*}
where the inequality is again due to \eqref{sc 2 class alt proof2}, and we used 
Corollary \ref{cor:Hanti composite}.
\end{proof}

\subsubsection{Composite iid vs adversarial: Equality for the relaxed exponent}
\label{sec:sc adv}

Below we consider classical state discrimination problems where the alternative hypothesis may be given in the composite i.i.d., the arbitrarily varying, or in the adversarial setting, and use the notations $\sc_r$, $\sc_r^{\av}$, $\sc_r^{\adv}$ for the strong converse exponents, similarly to the notations introduced in Section \ref{sec:classical exponents} for the direct exponents. 
Theorem \ref{thm:classical convex hyp sc} below is an analogy of 
Theorem \ref{thm:classical convex hyp} in the case where the null hypothesis is simple.

\begin{theorem}\label{thm:classical convex hyp sc}
Let $\X$ be a finite set, 
$\rho\in\S(\X)$, and $\A\subseteq\S(\X)$. 
Assume that $D(\rho\|\A)<D_{\infty}\nw(\rho\|\A)$.
Then for any $r>0$,
\begin{align}
H_r\nw(\rho\|\cco(\A))&=\sup_{\sigma\in\cco(\A)}\sc_r(\rho\|\sigma)
\label{classical adversarial sc0}\\
&\ge
\sc_r^{\adv}(\rho\|\cco(\A))
\ge
\sc_r^{\adv}(\rho\|\A)\label{classical adversarial sc1}\\
&
\ge \sc_r^{\av}(\rho\|\A)=
\sc_r^{\av}(\rho\|\cco(\A))\label{classical adversarial sc2}\\
&
\ge
\sc_r(\rho\|\cco(\A))
\ge
\sc_r^{\{1\}}(\rho\|\cco(\A))\\
&
\ge
\sup_{\sigma\in\cco(\A)}\sc_r(\rho\|\sigma),
\label{classical adversarial sc3}
\end{align}
and hence all the inequalities above hold as equalities.
\end{theorem}
\begin{proof}
All the inequalities are obvious, except for the first one in 
\eqref{classical adversarial sc1}, the equality in 
\eqref{classical adversarial sc2} holds for the same reason as in Theorem \ref{thm:classical convex hyp}, and the equality in \eqref{classical adversarial sc0} is due to Lemma \ref{lemma:simple sc exp}.
Hence, we only have to prove the first inequality in 
\eqref{classical adversarial sc1}, and therefore we may assume that 
$\A$ is convex and compact, so that $\cco(\A)=\A$, to simplify notation. 
Also, we may assume that $r>D(\rho\|\A)$, since otherwise 
$\sc_r^{\adv}(\rho\|\A)=0$, according to Theorem \ref{thm:classical convex hyp}.
Finally, we may assume that $\rho\notin\A$, since otherwise 
$H_r\nw(\rho\|\A)$ is easily seen to be equal to $r$, which is indeed an upper bound to 
$\sc_r^{\adv}(\rho\|\cco(\A))$, as demonstrated by the test sequence
$T_n:=e^{-nr}I$, $n\in\bN$. 

Consider first the case $r\in(D(\rho\|\A),\rmax(\rho\|\A))$,
where we use the notations of Corollary \ref{cor:Hanti composite}. 
Let $\rho_r=\rho,\sigma_r$, and $u_r\in(0,1)$ be as in Corollary \ref{cor:Hanti usc2}
and Corollary \ref{cor:Hanti composite}.
$r>D(\rho\|\A)$ implies that 
$0<H_r\nw(\rho\|\A)=H_r\nw(\rho\|\sigma_r)$, and hence, in turn, 
$D(\rho\|\sigma_r)<r$, according to \eqref{Hanti composite pos}.
In particular, $\rho^0\le\sigma_r^0$.
Moreover, $u_r<1$ implies that $r\le \rmax(\rho\|\sigma_r)$, according to 
Corollary \ref{cor:Hanti composite} applied to the sets
$\{\rho\}$ and $\{\sigma_r\}$.
Thus, $D(\rho\|\sigma_r)<r\le \rmax(\rho\|\sigma_r)$. 
Therefore, by Lemma \ref{lemma:classical strict convexity} and Remark \ref{rem:affine psi},
$\tilde\psi(\rho\|\sigma_r|\valt)$ is strictly convex, and thus 
$\hat\psi(\rho\|\sigma_r|\valt):=(\valt)r-\tilde\psi(\rho\|\sigma_r|\valt)$
is strictly concave. This implies that $u_r$ is the unique maximizer of 
$\hat\psi(\rho\|\sigma_r|\valt)$, and that
for any small enough $\ep>0$, there exist 
$0<u_{r,\ep,-}<u_r<u_{r,\ep,+}<1$ such that 
$\hat\psi(\rho\|\sigma_r|u_{r,\ep,-})
=H_r\nw(\rho\|\sigma_r)-\ep=\hat\psi(\rho\|\sigma_r|u_{r,\ep,+})$, and 
\begin{align}\label{classical sc proof1}
\hat\psi(\rho\|\sigma_r|u)<H_r\nw(\rho\|\sigma_r)-\ep,\ds\ds\ds 
u\in[0,1]\setminus[u_{r,\ep,-},u_{r,\ep,+}].
\end{align}

For any fixed $\sigma\in\A$ and $t\in[0,1]$, let $\sigma^{(t)}:=(1-t)\sigma_r+t\sigma\in\A$. 
Applying Corollary \ref{cor:Hanti composite} to the sets
$\{\rho\}$ and $\{\sigma^{(t)}\}$, we get the existence of some 
$u(t)\in[0,1]$ such that $H_r\nw(\rho\|\sigma^{(t)})=ru(t)-\tilde\psi(\rho\|\sigma^{(t)}|u(t))=
\max_{u\in[0,1]}\hat\psi(\rho\|\sigma^{(t)}|u)$, where
$\hat\psi(\rho\|\sigma^{(t)}|u):=ur-\tilde\psi(\rho\|\sigma^{(t)}|u)$.
As we have seen above, $\rho^0\le\sigma_r^0$, and hence 
$\lim_{t\to 0}\hat\psi(\rho\|\sigma^{(t)}|u)=\hat\psi(\rho\|\sigma_r|u)$ for every 
$u\in [0,1]$. However, since all 
$\hat\psi(\rho\|\sigma^{(t)}|\valt)$ are concave, the convergence is actually uniform in $u$, 
i.e., for every $\ep>0$ there exists a $t_{\ep}>0$ such that 
for all $t\in[0,t_{\ep})$, 
\begin{align}\label{psi hat uniform conv}
\max_{u\in[0,1]}\abs{\hat\psi(\rho\|\sigma^{(t)}|u)-\hat\psi(\rho\|\sigma_r|u)}<\frac{\ep}{3}.
\end{align}
Combining \eqref{classical sc proof1} and \eqref{psi hat uniform conv} yields
\begin{align*}
\sup_{u\in[0,1]\setminus[u_{r,\ep,-},u_{r,\ep,+}]}\hat\psi(\rho\|\sigma^{(t)}|u)
&\le \sup_{u\in[0,1]\setminus[u_{r,\ep,-},u_{r,\ep,+}]}\hat\psi(\rho\|\sigma_r|u)+\frac{\ep}{3}\le
H_r\nw(\rho\|\sigma_r)-\frac{2\ep}{3},\\
\hat\psi(\rho\|\sigma^{(t)}|u_r)&\ge\hat\psi(\rho\|\sigma_r|u_r)-\frac{\ep}{3}=
H_r\nw(\rho\|\sigma_r)-\frac{\ep}{3}.
\end{align*}
This implies that $u(t)\in[u_{r,\ep,-},u_{r,\ep,+}]$ for every $t<t_{\ep}$. Since 
$\lim_{\ep\searrow}u_{r,\ep,-}=\lim_{\ep\searrow}u_{r,\ep,+}=u_r$, we finally obtain that 
\begin{align*}
\lim_{t\searrow 0}u(t)=u_r.
\end{align*}
This also implies that for every small enough $t$, 
$\hat\psi(\rho\|\sigma^{(t)}|1)<H_r\nw(\rho\|\sigma^{(t)})$, and hence
$r\le \rmax(\rho\|\sigma^{(t)})$, according to Corollary \ref{cor:Hanti composite}.
Using again that $\rho^0\le\sigma_r^0$, we see that 
$\lim_{t\searrow 0}D(\rho\|\sigma^{(t)})=D(\rho\|\sigma_r)<r$. 
Thus, for every all small enough $t$,
$D(\rho\|\sigma^{(t)})<r\le \rmax(\rho\|\sigma^{(t)})$, and therefore
$\frac{d^2}{du^2}\hat\psi(\rho\|\sigma^{(t)}|u)<0$ for all $u\in(0,1)$, 
according to Lemma \ref{lemma:classical strict convexity} and Remark \ref{rem:affine psi}. 

Since $u(t)$ is the unique solution of $\frac{d}{du}\hat\psi(\rho\|\sigma^{(t)}|u)=0$, 
the implicit function theorem implies that $t\mapsto u(t)$ is differentiable
on $(0,t_0)$ for some $t_0\in(0,1)$, and hence so is
$ t\mapsto \hat\psi(\rho\|\sigma^{(t)}|u(t))$.
Therefore, the same computation as in the proof of Theorem \ref{thm:classical convex hyp2} yields that 
\begin{align*}
0&\ge
\lim_{t\searrow 0}\frac{H_r\nw(\rho\|\sigma^{(t)})-H_r\nw(\rho\|\sigma_r)}{t}
=
\lim_{t\searrow 0}\frac{\hat\psi(\rho\|\sigma^{(t)}|u(t))-\hat\psi(\rho\|\sigma_r|u_r)}{t}
=
\lim_{t\searrow 0}\frac{d}{dt}\hat\psi(\rho\|\sigma^{(t)}|u(t))\\
&=
\lim_{t\searrow 0}\frac{u(t)\sum_x\rho(x)^{\frac{1}{1-u(t)}}\sigma^{(t)}(x)^{1-\frac{1}{1-u(t)}-1}
(\sigma(x)-\sigma_r(x))}{\sum_x\rho(x)^{\frac{1}{1-u(t)}}\sigma^{(t)}(x)^{1-\frac{1}{1-u(t)}}}
=
\frac{u_r\sum_x\rho(x)^{\frac{1}{1-u_r}}\sigma_r(x)^{-\frac{1}{1-u_r}}
(\sigma(x)-\sigma_r(x))}{\sum_x\rho(x)^{\frac{1}{1-u_r}}\sigma_r(x)^{1-\frac{1}{1-u_r}}},
\end{align*}
where in the last step we used that $\lim_{t\searrow 0}u(t)=u_r$, as shown above.
Rearranging, we get
\begin{align*}
\sum_x\sigma(x)\bz\frac{\rho(x)}{\sigma_r(x)}\jz^{\alpha_r}
\le
\sum_x\sigma_r(x)\bz\frac{\rho(x)}{\sigma_r(x)}\jz^{\alpha_r},
\end{align*}
where $\alpha_r=1/(1-u_r)$.
Since this holds for every $\sigma\in\A$, the same argument as in \eqref{adv beta upper bound}--\eqref{adv type II upper bound} yields that 
\begin{align}\label{sc beta upper bound}
\beta_n(T_{n,r})\le e^{-nr},
\end{align}
where the test $T_{n,r}$ is defined as in \eqref{NP test}, with $\rho_r=\rho$.

To estimate the type I success probability from below, note that the logarithmic moment generating function of $\log\frac{\rho(\cdot)}{\sigma_r(\cdot)}$ w.r.t.~$\rho$ is 
\begin{align*}
\Lambda(\alpha):=\log\sum_{x\in\X}\rho(x)e^{\alpha\log\frac{\rho(x)}{\sigma_r(x)}}
=\log\sum_{x\in\X}\rho(x)^{1+\alpha}\sigma_r(x)^{-\alpha}
=
\psi(\alpha+1),
\end{align*}
where $\psi(\alpha):=\psi(\rho\|\sigma_r|\alpha)$.
The Legendre-Fenchel transform of $\Lambda$ at $c_r=(\psi)'(\alpha_r)$ is 
\begin{align*}
\Lambda^{\circ}(c_r)
:=
\sup_{\alpha\in\bR}\{c_r\alpha-\psi(1+\alpha)\}
=
\sup_{\alpha\in\bR}\{c_r(\alpha-1)-\psi(\alpha)\}
=\Psi(c_r)-c_r
=
H_r\nw(\rho\|\sigma_r)=H_r\nw(\rho\|\A),
\end{align*}
according to \eqref{Hanti composite LF} and Lemma \ref{lemma:simple Hr}.
Hence, by Cram\'er's large deviation theorem \cite{DZ}, we get that 
\begin{align}\label{sc alpha lower bound}
\limsup_{n\to+\infty}-\frac{1}{n}\log\bz 1-\alpha_n(T_{n,r})\jz
=
\limsup_{n\to+\infty}-\frac{1}{n}\log\rho^{\otimes n}(T_{n,r})
&\le
\Psi(c_r)-c_r=
H_r\nw(\rho\|\A).
\end{align}
Putting together \eqref{sc beta upper bound} and \eqref{sc alpha lower bound} yields
\begin{align*}
\sc_{r}^{\adv}(\rho\|\A)
\le 
H_r\nw(\rho\|\A).
\end{align*}

Assume next that $r\ge \rmax(\rho\|\A)$. In this case 
the Neyman-Pearson tests may not give the optimal exponent,
and a different approach is needed, which was originally introduced in 
\cite[Theorem 4]{NH}; see \cite[Theorem 4.10]{MO} for details.
Let $\psi(\alpha):=\inf_{\sigma\in\A}\psi(\rho\|\sigma|\alpha)$, and 
$\Psi(c):=\sup_{\alpha\ge 1}\{c-\psi(\alpha)\}$, $\alpha,c\in\bR$. 
For every $c\in(D(\rho\|\A),D_{\infty}(\rho\|\A))$, let 
$r_c:=\Psi(c)$, and $T_{n,r_c}$ be the test as above. By the strict monotonicity of $\Psi$
(Corollary \ref{cor:Hanti composite}), 
$\Psi(c)<\Psi(D_{\infty}(\rho\|\A))=\rmax(\rho\|\A)\le r$, and hence
\begin{align*}
T_n(r,c):=e^{-n(r-\Psi(c))}T_{n,r_c}
\end{align*}
is a test. By \eqref{sc beta upper bound} and \eqref{sc alpha lower bound},
\begin{align*}
\beta_n(T_n(r,c))&=e^{-n(r-\Psi(c))}\beta_n(T_{n,r_c})
\le
e^{-n(r-\Psi(c))}e^{-nr_c}=e^{-nr},\ds\ds\ds n\in\bN,\\
\limsup_{n\to+\infty}-\frac{1}{n}\log\bz 1-\alpha_n(T_n(r,c))\jz
&=
r-\Psi(c)+
\limsup_{n\to+\infty}-\frac{1}{n}\log\bz 1-\alpha_n(T_{n,r_c})\jz\\
&\le
r-\Psi(c)
+(\Psi(c)-c)=r-c.
\end{align*}
Hence,
\begin{align*}
\sc_{r}^{\adv}(\rho\|\A)
\le 
\inf_{c\in(D(\rho\|\sigma_r),D_{\infty}(\rho\|\sigma_r))}\{r-c\}=r-D_{\infty}\nw(\rho\|\A)=H_r\nw(\rho\|\A),
\end{align*}
where the equality is due to Corollary \ref{cor:Hanti composite}.
\end{proof}

Theorem \ref{thm:classical convex hyp sc} and Corollary \ref{cor:general sc equality} yield immediately the following:
\begin{cor}\label{cor:adv sc}
Let $\X$ be a finite set, 
and $\N,\A\subseteq\S(\X)$ be such that 
$D(\rho\|\A)<D_{\infty}\nw(\rho\|\A)$ for every $\rho\in\N$.
Then for any $r>0$,
\begin{align}
H_r\nw(\N\|\cco(\A))&=
\sup_{\rho\in\N,\sigma\in\cco(\A)}\sc_r(\rho\|\sigma)
\label{classical adversarial sc01}\\
&\ge
\sc_r^{\{0\}}\bz\N\Big\|\bz\cco(\A)^{(n)}_{\adv}\jz_{n\in\bN}\jz
\ge
\sc_r^{\{0\}}\bz\N\Big\|\bz\A^{(n)}_{\adv}\jz_{n\in\bN}\jz
\label{classical adversarial sc11}\\
&\ge 
\sc_r^{\{0\}}\bz\N\Big\|\bz\A^{\otimes n}_{\av}\jz_{n\in\bN}\jz=
\sc_r^{\{0\}}\bz\N\Big\|\bz\cco(\A)^{\otimes n}_{\av}\jz_{n\in\bN}\jz
\label{classical adversarial sc21}\\
&\ge
\sc_r^{\{0\}}(\N\|\cco(\A))
\ge
\sc_r^{\{0,1\}}(\rho\|\cco(\A))\\
&\ge
\sup_{\rho\in\N,\sigma\in\cco(\A)}\sc_r(\rho\|\sigma),
\label{classical adversarial sc31}
\end{align}
and hence all the inequalities above hold as equalities.
\end{cor}
\begin{proof}
The equality in \eqref{classical adversarial sc01} is due to Lemma \ref{lemma:simple sc exp},
and the equality in \eqref{classical adversarial sc21} follows the same way as in 
Theorem \ref{thm:classical convex hyp}.
All the inequalities are obvious except for the first one in \eqref{classical adversarial sc11}, for which we have 
\begin{align*}
\sc_r^{\{0\}}\bz\N\Big\|\bz\cco(\A)^{(n)}_{\adv}\jz_{n\in\bN}\jz
&\le
\inf_{r'>r}\sup_{\rho\in\N}\sc_{r'}\bz\rho\Big\|\bz\cco(\A)^{(n)}_{\adv}\jz_{n\in\bN}\jz
\le
\inf_{r'>r}\sup_{\rho\in\N}H_r\nw(\rho\|\cco(\A))
=
\inf_{r'>r}H_r\nw(\N\|\cco(\A))\\
&=
H_r\nw(\N\|\cco(\A)).
\end{align*} 
The first inequality above is due to 
Corollary \ref{cor:general sc equality},
the second inequality follows from 
Theorem \ref{thm:classical convex hyp sc}, 
the first equality is by definition, 
and the last equality follows from the continuity 
of $r\mapsto H_r\nw(\N\|\cco(\A))$; see Corollary \ref{cor:Hanti composite}.
\end{proof}

\subsubsection{Composite i.i.d. versus simple i.i.d.: Equality for a restricted parameter range}
\label{sec:ball test}

For a sequence $\vecc{x}\in\X^n$, let 
$P_{\vecc{x}}$ denote its empirical distribution, or type, defined by 
$P_{\vecc{x}}(y):=\frac{1}{n}\abs{\{i:\,x_i=y\}}$, $y\in \X$. For 
any $n\in\bN$, let $\P_n:=\{P_{\vecc{x}}:\,\vecc{x}\in\X^n\}$ be the set of 
$n$-types. It is clear that any probability distribution on $\X$ can be arbitrarily well approximated by $n$-types for sufficiently large $n$. We will need the following 
refined statement, whose proof follows by a simple modification of that of 
Lemma A.2 in \cite{Hoeffding}. 
We give a detailed proof in Appendix \ref{sec:Hoeffding lemma} for readers' convenience.

\begin{lemma}\label{lemma:Hoeffding}
Let $\rho\in\S(\X)\cap A_{v,c}$, where $A_{v,c}:=\{f\in\bR^{\X}:\,\sum_{x\in\X}v(x)f(x)\ge c\}$ is a half-space, and let $r:=|\supp \rho|$. For every $n\ge r(r-1)$, there exists a $\rho_n\in\P_n\cap A_{v,c}$
such that $\rho_n^0\le \rho^0$ and $\norm{\rho-\rho_n}_1\le\frac{2(r-1)}{n}$.
\end{lemma}

We will also need the following continuity bound for the relative entropy, which is a 
simple consequence of an analogous continuity bound for the entropy
\cite{Aud:FannesIneq,FannesIneq,Petzbook}:
\begin{lemma}\label{lemma:Fannes}
Let $\rho_1,\rho_2,\sigma\in\S(\X)$ be such that 
$\supp\rho_1\cup\supp\rho_2\subseteq\supp\sigma$. Then 
\begin{align*}
\abs{D(\rho_1\|\sigma)-D(\rho_2\|\sigma)}
\le \half\norm{\rho_1-\rho_2}_1\log(|\X|-1)+h_2\bz\half\norm{\rho_1-\rho_2}_1\jz
+
\norm{\rho_1-\rho_2}_1\bz-\log\min_{x\in \supp\rho_1\cup\supp\rho_2}\sigma(x)\jz,
\end{align*}
where $h_2(x):=-x\log x-(1-x)\log(1-x)$, $x\in[0,1]$. 
\end{lemma}

\begin{theorem}\label{thm:composite classical strong sc}
Consider the classical i.i.d~state discrimination problem with composite null hypothesis 
$\N\subseteq\S(\X)$ and simple alternative hypothesis $\sigma\in \S(\X)$. 
Assume that 
\begin{align*}
\delta_{\min}(\N):=\sup_{\rho\in\N}(-\log\min_{x\in\supp\rho}\rho(x))<+\infty.
\end{align*}
Then, for any $D(\N\|\sigma)<r<\inf_{\rho\in\N}\rmax(\rho\|\sigma)$,
\begin{align}\label{HK statement}
\sc_r(\N\|\sigma)=H_r\nw(\N\|\sigma)=\sup_{\rho\in\N}\sc_r(\rho\|\sigma).
\end{align}
\end{theorem}
\begin{proof}
The second equality in \eqref{HK statement} is immediate from 
Lemma \ref{lemma:simple sc exp}, and hence we only need to prove the first one.

For a fixed $r$ as in the statement, and $n\in\bN$, consider the test
\begin{align*}
B_{n,r}:=\{\vecc{x}\in\X^n:\,D(P_{\vecc{x}}\|\sigma)\ge r\}.
\end{align*}
Then, by 
a standard type estimate (see, e.g., Eq.~(2.1.12) in \cite{DZ}),
\begin{align}\label{HK proof0}
\beta_n(B_{n,r})&\le (n+1)^{|\X|}e^{-n\inf_{\omega\in\P_n:\,D(\omega\|\sigma)\ge r}D(\omega\|\sigma)}\le (n+1)^{|\X|}e^{-nr}.
\end{align}

Assume that $\rho\in\N$ is such that $D(\rho\|\sigma)\ge r$. Since
$\{\omega\in\S(\X):\,D(\omega\|\sigma)\le r\}$ is a convex set, it can be separated from 
$\rho$ by a hyperplane, i.e., $\rho\in A_{v,c}$ for some $v\in\bR^{\X}$ and $c\in\bR$, 
and $\{\omega\in\S(\X):\,D(\omega\|\sigma)\le r\}\subseteq A_{-v,-c}$. 
For every $n\ge |\X|^2$, let $\rho_n\in\S(\X)\cap A_{v,c}$ as in Lemma \ref{lemma:Hoeffding}.
Then $D(\rho_n\|\sigma)\ge r$, and thus
\begin{align}\label{HK proof1}
\rho^{\otimes n}(B_{n,r})\ge\rho^{\otimes n}\bz\left\{\vecc{x}\in\X^n:\,P_{\vecc{x}}=\rho_n\right\}\jz\ge(n+1)^{-|\X|}e^{-nD(\rho_n\|\rho)},
\end{align}
where the last inequality follows by another well-known type estimate
\cite[Lemma 2.6]{CsiszarKorner}.
By lemma \ref{lemma:Fannes},
\begin{align}\label{HK proof2}
D(\rho_n\|\rho)\le \underbrace{D(\rho\|\rho)}_{=0=H_r\nw(\rho\|\sigma)}+\frac{|\X|^2}{n}+
h_2\bz\frac{|\X|}{n}\jz+\frac{2|\X|\delta_{\min}(\N)}{n},
\end{align}
where we used \eqref{Hanti composite pos}. Combining \eqref{HK proof1} and \eqref{HK proof2}, we get
\begin{align}\label{HK proof3}
-\frac{1}{n}\log\rho^{\otimes n}(B_{n,r})\le
H_r\nw(\rho\|\sigma)+
|\X|\frac{\log(n+1)}{n}+\frac{|\X|^2}{n}+
h_2\bz\frac{|\X|}{n}\jz+\frac{2|\X|\delta_{\min}(\N)}{n}.
\end{align}

Consider now a $\rho\in\N$ such that $D(\rho\|\sigma)<r$.
By assumption, we also have $r<\rmax(\rho\|\sigma)$, and hence, by 
Lemma \ref{lemma:simple Hr}, there exists a unique $\alpha_r\in(1,+\infty)$ such that
$D(\mu_{\alpha_r,\rho,\sigma}\|\sigma)=r$, 
$D(\mu_{\alpha_r,\rho,\sigma}\|\rho)=H_r\nw(\rho\|\sigma)$.
Again, $\mu_{\alpha_r,\rho,\sigma}$ can be separated by a hyperplane from 
$\{\omega\in\S(\X):\,D(\omega\|\sigma)\le r\}$, and Lemma \ref{lemma:Hoeffding}
yields the existence of a $\mu_n\in\P_n$ such that 
$D(\mu_n\|\sigma)\ge r$ and $\norm{\mu_n-\mu_{\alpha_r,\rho,\sigma}}_1\le\frac{2|\X|}{n}$,
whenever $n\ge |\X|^2$. By the same argument as above, 
\begin{align*}
\rho^{\otimes n}(B_{n,r})\ge\rho^{\otimes n}\bz\left\{\vecc{x}\in\X^n:\,P_{\vecc{x}}=\mu_n\right\}\jz\ge(n+1)^{-|\X|}e^{-nD(\mu_n\|\rho)},
\end{align*}
and
\begin{align*}
D(\mu_n\|\rho)&\le
\underbrace{D(\mu_{\alpha_r,\rho,\sigma}\|\rho)}_{=H_r\nw(\rho\|\sigma)}
+\frac{|\X|^2}{n}+h_2\bz\frac{|\X|}{n}\jz+\frac{2|\X|(-\log\sigma_{\min})}{n},
\end{align*}
where $\sigma_{\min}:=\min_{x\in\supp\sigma}\sigma(x)$. Thus,
\begin{align}\label{HK proof4}
-\frac{1}{n}\log\rho^{\otimes n}(B_{n,r})\le
H_r\nw(\rho\|\sigma)+
|\X|\frac{\log(n+1)}{n}+\frac{|\X|^2}{n}+
h_2\bz\frac{|\X|}{n}\jz+\frac{2|\X|(-\log\sigma_{\min})}{n}.
\end{align}

Combining \eqref{HK proof3} and \eqref{HK proof4} yields
\begin{align}\label{HK proof5}
-\frac{1}{n}\log\alpha_n(B_{n,r}|\N)
=
\sup_{\rho\in\bN}-\frac{1}{n}\log\rho^{\otimes n}(B_{n,r})
\le
H_r\nw(\N\|\sigma)+\frac{c}{n}+|\X|\frac{\log (n+1)}{n}+h_2\bz\frac{|\X|}{n}\jz
\end{align}
for some positive constant $c$. 
Finally, \eqref{HK proof0} and \eqref{HK proof5} together yield
$\sc_r(\N\|\sigma)\le H_r\nw(\N\|\sigma)$, as required.
\end{proof}

\begin{rem}
It is clear from its definition in \eqref{psi tilde derivative lim} that $\rmax(\rho\|\sigma)\ge D_{\infty}\nw(\rho\|\sigma)$, 
and thus Theorem \ref{thm:composite classical strong sc} is valid, in particular, for every 
$r\in(D(\N\|\sigma),D_{\infty}\nw(\N\|\sigma))$.  
This is typically a non-degenerate interval; an easily verifiable case where this holds is when $\N$ is compact and $\alpha\mapsto\psi(\rho\|\sigma|\alpha)$ is not affine 
for a $\rho\in\N$ such that $D_{\infty}\nw(\rho\|\sigma)=D_{\infty}\nw(\N\|\sigma)$;
see Lemma \ref{lemma:classical strict convexity}.
\end{rem}


\section{Conclusion}

While the error exponents of simple binary i.i.d.~state discrimination are by now completely 
characterized in terms of divergences of the two states representing the two hypotheses,
the composite case seems to present far more open problems than definitive solutions so far. 
The results presented in this paper clarify the picture to some extent, but of course there are many problems still left open. 
The most obvious one is to find tractable expressions for the error exponents in cases where the equality $\e(\N\|\A)=E(\N\|\A)$ does not hold, or at least find some possibly sub-optimal
but non-trivial and universallly applicable lower bounds. 
Though a complete characterization of the Stein exponent in the finite-dimensional quantum case was given in \cite{BertaBrandaoHirche2017}, and general lower bounds are available as in 
Proposition \ref{prop:general lower bounds}, these involve regularizations that are practically not feasible to solve explicitly, and call for the search of better computable lower bounds. One such bound was given in \cite{HMH09}, where it was shown that 
$\cli(\rho\|\A)\ge\half C(\rho\|\A)$ when $\A$ is finite. 
It would be interesting to find similar bounds for more general settings and for other exponents, too.

It is interesting that even the finite-dimensional classical case seems to present 
some open problems. We are not aware of the equality problem $\e(\N\|\A)=E(\N\|\A)$
having been completely solved 
for the Stein, Chernoff, and the direct exponents without some constraint (finiteness, convexity) on the sets $\N,\A$. On the other hand, it also seems to be open whether closedness and convexity 
of the sets always guarantees equality; in our classical counterexamples to equality in 
Sections \ref{sec:infinite classical} and \ref{sec:sc two alt} the composite sets are non-convex. 
On the other hand, convexity is certainly not sufficient for equality 
in the quantum setting:
an example with a closed convex alternative hypothesis $\A$ and a simple null-hypothesis 
$\rho$ was shown in \cite{BertaBrandaoHirche2017} for which $\sli(\rho\|\A)<D(\rho\|\A)$. 
It is quite likely that this example is also a counterexample to 
equality for the strong converse exponents with rates $\sli(\rho\|\A)<r<D(\rho\|\A)$, since
$H_r\nw(\rho\|\A)=0$ in this range; the missing part 
is whether $\sc_r(\rho\|\A)>0$ when $r>\sli(\rho\|\A)$, 
i.e., the exponential strong converse 
part of Stein's lemma with composite alternative hypothesis and a simple null-hypothesis. 


\section*{Acknowledgments}
This work was partially funded by the
National Research, Development and 
Innovation Office of Hungary via the research grants K124152, KH129601, and K132097, and
by the Ministry of Innovation and
Technology and the National Research, Development and Innovation
Office within the Quantum Information National Laboratory of Hungary. The work of WM was further supported 
by a Bolyai J\'anos Fellowship of the Hungarian Academy of Sciences, 
and the \'UNKP-19-4 New National Excellence Program of the 
Ministry for Innovation and Technology. 
The authors are grateful to P\'eter Vrana for the idea of the proof of Proposition \ref{prop:sc general equality}.

\appendix

\section{Proof of Lemma \ref{lemma:Hoeffding}}
\label{sec:Hoeffding lemma}

\begin{proof}
Let $x_1,\ldots,x_r$ be an ordering of $\supp \rho$ and let $p_i:=\rho(x_i)$,
$v_i:=v(x_i)$.
If $r=1$ then there is nothing to prove, and hence for the rest we assume the contrary.  Obviously, there exists an $i$ such that 
$p_i\ge 1/r$; assume without loss of generality that $p_r\ge 1/r$. For each $i<r$, 
let
\begin{align}
k_i:=\begin{cases}
\ceil{np_i},&v_i>v_r,\\
\floor{np_i},&v_i\le v_r,
\end{cases}
\ds\ds\ds\text{so that}\ds\ds\ds
(v_i-v_r)(k_i-np_i)\ge 0.\label{Hoeffding proof1}
\end{align}
Define $k_r:=n-(k_1+\ldots+k_{r-1})$. By the above, 
\begin{equation*}
k_r\ge
n-(np_1+1)-\ldots-(np_{r-1}+1)=np_r-(r-1)\ge
\frac{n}{r}-(r-1)=\frac{n-r(r-1)}{r},
\end{equation*}
and therefore $k_r\ge 0$ if $n\ge r(r-1)$. 
Thus, $\rho_n(x_i):=\frac{k_i}{n},\,i=1,\ldots, r$, 
$\rho_n(x):=0$, $x\in\X\setminus\{x_1,\ldots,x_r\}$, 
defines an element of $\P_n$, and it is clear from the definition that $\rho_n^0\le\rho^0$. 
We have
\begin{equation*}
|k_r-np_r|=|n-k_1-\ldots-k_{r-1}-n(1-p_1-\ldots-p_{r-1})|\le
\sum_{k=1}^{r-1}|k_i-np_i|\le r-1,
\end{equation*}
and therefore 
\begin{equation*}
\norm{\rho_n-\rho}_1=\sum_{k=1}^r|k_i/n-p_i|\le\frac{2(r-1)}{n}.
\end{equation*}
Finally, 
\begin{align*}
0&\le\sum_{i=1}^{r-1}(v_i-v_r)(k_i/n-p_i)=
\sum_{i=1}^{r-1}v_i(k_i/n-p_i)
-v_r\underbrace{\sum_{i=1}^{r-1}(k_i/n-p_i)}_{=1-k_r/n-1+p_r}\\
&=
\sum_{i=1}^{r}v_i(k_i/n-p_i)=
\underbrace{\sum_{i=1}^{r}v_ik_i/n}_{=\sum_x v(x)\rho_n(x)}-\underbrace{\sum_{i=1}^{r}v_ip_i}_{=\sum_xv(x)\rho(x)\ge c}
\le
\sum_x v(x)\rho_n(x)-c,
\end{align*}
where the first inequality is due to \eqref{Hoeffding proof1}.
Thus, $\rho_n\in A_{v,c}$.
\end{proof}


\bibliography{bibliography201018}

\end{document}